\documentclass[12pt,letterpaper]{article}
\pdfoutput=1
\usepackage{jheppub}
\usepackage{amsmath}
\usepackage{bm}
\usepackage{comment}

\hypersetup{
    colorlinks=true,       
    linkcolor=red,          
    citecolor=blue,        
    filecolor=magenta,      
    urlcolor=blue           
}

\def\k{{\bf k}}
\def\p{{\bf p}}

\def\x{{\bf x}}
\def\y{{\bf y}}

\def\mW{m_{\rm W}}
\def\tr{\,{\rm tr}\,}
\def\ZN{{\mathbb Z}_N}
\def\nf{n_{\rm f}}

\def\bmu{{\bm \mu}}
\def\bsigma{{\bm \sigma}}
\def\blambda{{\bm \lambda}}
\def\balpha{{\bm \alpha}}
\def\bnabla{{\bm \nabla}}
\def\bnu{{\bm \nu}}
\def\brho{{\bm \rho}}
\def\half{\tfrac{1}{2}}

\def\sumprime{\sideset{}{'}\sum}

\def\sigmap{\tilde\sigma}
\def\LambdaIR{\Lambda_{\rm IR}}
\def\eq{\,{=}\,}
\def\S{\Sigma}
\def\O{\mathcal {O}}

\def\P{\mathcal{P}}

\title		{QCD on a small circle}
\author[a]	{Kyle Aitken,}
\author[b]	{Aleksey Cherman,}
\author[c]	{Erich~Poppitz,}
\author[a]	{Laurence G.~Yaffe}
\emailAdd	{kaitken17@gmail.com}
\emailAdd	{aleksey.cherman.physics@gmail.com}
\emailAdd	{poppitz@physics.utoronto.ca}
\emailAdd	{yaffe@phys.washington.edu}
\affiliation[a]	{Department of Physics, University of Washington, Seattle, WA 98195-1560, USA}
\affiliation[b]	{Institute for Nuclear Theory, University of Washington, Seattle, WA 98195-1560, USA}
\affiliation[c]	{Department of Physics, University of Toronto, Toronto, ON M5S 1A7, Canada}

\abstract
    {%
    QCD-like theories
    can be engineered to remain in a confined phase when
    compactified on an arbitrarily small circle,
    where their features may be studied quantitatively in a controlled fashion.
    Previous work has elucidated the generation of a
    non-perturbative mass gap and the spontaneous breaking of chiral symmetry
    in this regime.
    Here, we study the rich spectrum of hadronic states,
    including glueball, meson, and baryon resonances.
    We find an exponentially growing Hagedorn density of states,
    as well as the emergence of non-perturbative energy scales given
    by iterated exponentials
    of the inverse Yang-Mills coupling $g^2$.
    }

\preprint{INT-PUB-17-026}

\begin{document}
\maketitle

\section	{Introduction}\label{sec:intro}

There are few ways to analytically study 
the low temperature and density behavior of QCD-like quantum field theories.%
\footnote
    {%
    By ``QCD-like'' we mean 4D asymptotically free $SU(N)$ gauge theories,
    possibly containing fermions but without light fundamental scalar fields.
    We assume that the fermion content is such that the theory,
    when defined on $\mathbb R^4$, has a confining phase
    characterized by some strong scale $\Lambda$.
    }
Near the chiral limit (in theories containing light fermions),
chiral perturbation theory may be used to systematically characterize
the low energy consequences of spontaneously broken chiral symmetry
using a small number of low energy parameters.
(See, e.g., Ref.~\cite{Scherer:2002tk} for a review.)
But the demonstration of chiral symmetry breaking
and determination of these low energy constants
requires other methods, such as large scale lattice gauge theory simulations
or input of experimental data.
Gauge-gravity duality \cite{Aharony:1999ti}
has provided insight into some 4D confining gauge theories
\cite{Witten:1998zw,Maldacena:2000yy,Klebanov:2000hb,Polchinski:2000uf,Sakai:2004cn,Mia:2009wj},
but is usefully applicable primarily in theories
which are strongly coupled at all scales, not asymptotically free,
and have a large number $N$ of colors.
For 4D confining, asymptotically free gauge theories, analytic methods based on controlled approximations are generally unavailable.

In this paper,
we study properties of 4D confining  QCD-like theories, at finite $N$,
in a regime which allows controlled analytic calculations.
Specifically, we consider theories on $\mathbb R^3 \times S^1$,
with one dimension compactified on a circle of circumference $L$
which is small compared to the inverse strong scale of the theory,
$L \ll \Lambda^{-1}$
(and henceforth denoted $S^1_L$).
This is a very old idea (see, e.g., Ref.~\cite{vanBaal:2000zc} for a review)
but interest has been renewed in recent years
with the realization that a wide range of QCD-like theories may be engineered
to possess a phase diagram in which the small-$L$ regime is
continuously connected to the large-$L$ or decompactified regime.
Achieving such ``adiabatic compactification'' requires non-thermal
boundary conditions and suitable matter content (or the addition of
double trace deformations) \cite{Unsal:2007vu,Unsal:2007jx,Unsal:2008ch,Shifman:2008ja,Shifman:2009tp,Unsal:2010qh}.

Compactifying one direction on a small circle does, obviously,
change properties of a theory.
Lorentz invariance is reduced from $SO(1,3)$ to $SO(1,2)$
and physical quantities will depend on the newly introduced scale $L$.
But if one can engineer compactifications where the $L$ dependence is
smooth (``adiabatic''), then studies of the small-$L$ regime
may teach one qualitative lessons which remain valid in
the large-$L$ limit.
Previous work
\cite{Unsal:2008ch,
Shifman:2008ja,Shifman:2008cx,Shifman:2009tp,
Cossu:2009sq,Myers:2009df,Simic:2010sv,Unsal:2010qh,Vairinhos:2011gv,
Thomas:2011ee,Anber:2011gn,
    Poppitz:2012sw,
    Poppitz:2012nz,Argyres:2012ka,Argyres:2012vv,
    Anber:2013doa,Cossu:2013ora,
    Anber:2014lba,Bergner:2014dua,Li:2014lza,
    Anber:2015kea,Anber:2015wha,Misumi:2014raa,
    Cherman:2016hcd}
has examined symmetry realizations at small $L$ and studied
the properties of the very lightest excitations.
One finds that it is possible to prevent the spontaneous breaking of
the $\mathbb Z_N$ center symmetry of pure Yang-Mills (YM) theory,
which would signal a deconfinement transition.
With massless quarks present, one finds that
chiral symmetry is spontaneously broken.
The mechanism of confinement,
the generation of a non-perturbative mass gap
(without massless quarks),
and the spontaneous breaking of chiral symmetry
(with massless quarks)
all may be nicely understood in the small-$L$ regime using
semiclassical methods.
All evidence supports the view that these center-stabilized
compactifications are, indeed, adiabatic.%
\footnote
    {%
    Consistency of symmetry realizations between small and large $L$ is,
    of course, necessary but not sufficient for physics to be smooth in $L$.
    Phase transitions not involving any change in symmetry realization
    could always be present at some intermediate value of $L$.
    For center-stabilized QCD, with light quarks, the
    careful lattice studies which would be needed to
    rule out this possibility are not yet available.
    In the absence of any evidence to the contrary,
    we proceed assuming that for the compactifications we study below,
    physical properties are smooth in $L$.
    }

Given the weight of evidence that adiabatic compactifications exist,
it is interesting to use these calculable settings to
explore properties of QCD-like theories in more detail.
In this paper we initiate efforts in this direction by
investigating qualitatively, and where possible quantitatively,
the spectrum and properties of glueballs, mesons, and baryons
in the small-$L$ regime of adiabatically compactified theories.
Some of the hadronic states we find are stable, but naturally most
are resonances.
In the weakly coupled small-$L$ regime, hadronic resonances are narrow
with parametrically small decay widths.
Portions of the spectrum have interesting parallels with what one obtains from
naive quark models, but in a context where the dynamics of the quantum field
theory are under systematic theoretical control.

\begin{figure}[t]
\begin{center}
\vspace*{-7pt}
\includegraphics[scale=0.33]{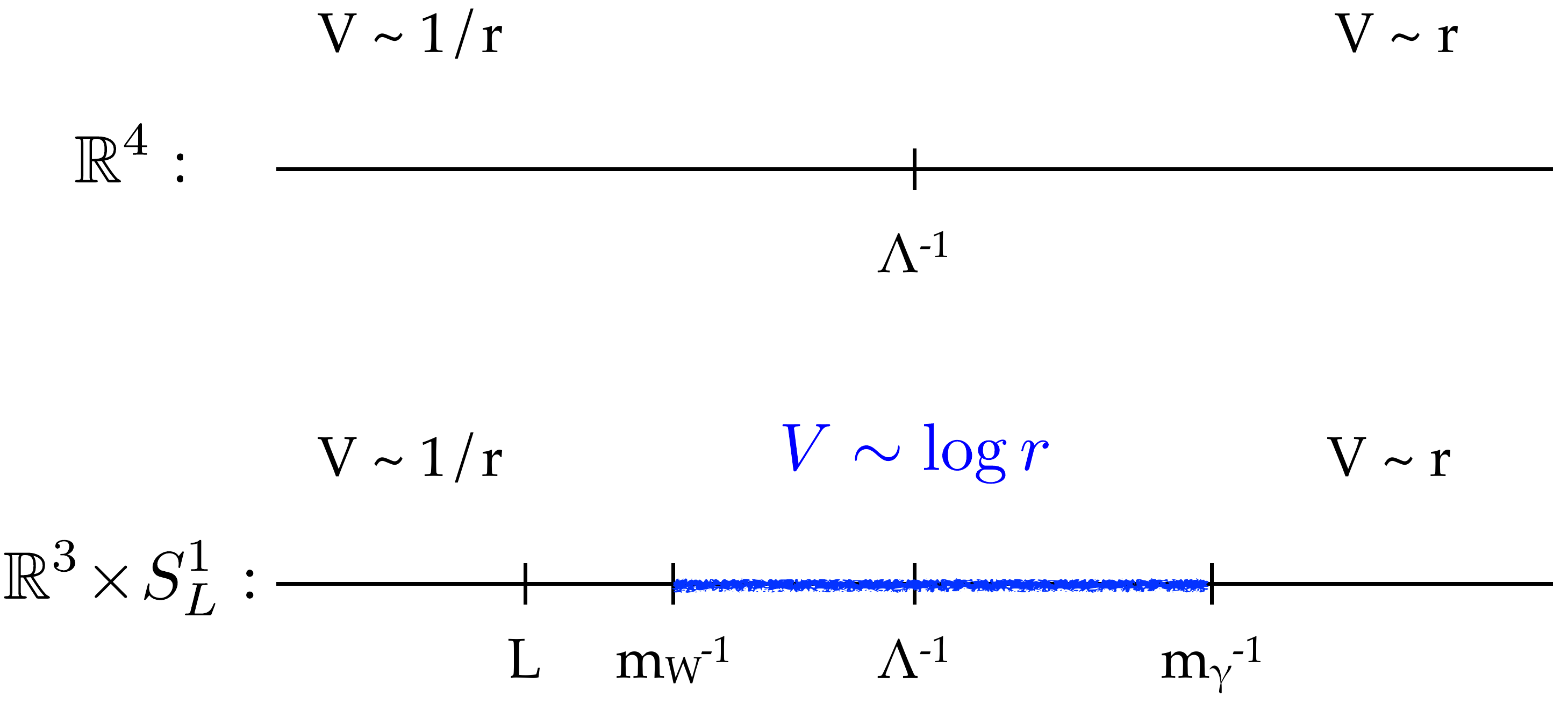}\hspace*{1cm}
\vspace*{-12pt}
\end{center}
\caption
    {%
    Characteristic length scales in the static potential $V(r)$
    for a heavy quark and antiquark separated by a distance $r$
    in Yang-Mills theory on $\mathbb{R}^4$ (top), and in
    adiabatically compactified YM theory on $\mathbb{R}^3 \times S^1_L$
    (bottom).
    On $\mathbb{R}^4$ there is only one intrinsic length scale
    $\Lambda^{-1}$ which separates the short ($V \sim 1/r$)
    and long ($V \sim r$) distance regimes.
    In the small-$L$ regime of adiabatically compactified
    YM theory on $\mathbb{R}^3 \times S^1_L$,
    there is a parametrically large intermediate regime,
    $\mW^{-1} \ll r \ll m_\gamma^{-1}$,
    in which the potential is logarithmic, $V \sim \ln r$.
    Here $\mW^{-1} \sim NL$ and
    $m_{\gamma}^{-1} \sim NL \, \eta^{-11/6}$ with
    $\eta \equiv N L \Lambda \ll 1$.
    (See Eq.~(\ref{eq:dYMphoton_mass}) below for details.)
    \label{fig:potential_scales}
    }
\end{figure}

We mention here two especially curious aspects of our results.
First, we find that the lightest glueballs (or dual photons in the
small-$L$ description) form bound states whose binding energies
are given by iterated exponentials of the Yang-Mills coupling,
$\Delta E \sim \exp (-A g^{k} \exp(B/g^2))$.
Second, we find that the density of states of both glueballs and mesons
exhibits Hagedorn (or exponential) growth with energy,
but this growth has an unusual origin.
Hagedorn scaling of the density of mesonic states is typically
attributed to the fluctuations of a long, highly excited confining
string, and can only be established systematically in the large $N$
limit where mesons cannot decay.
The origin of Hagedorn scaling in our context is quite different.
The extra scale $L$ introduced
by the adiabatic compactification modifies the potential
experienced by heavy test quarks separated by a distance $r$, and
introduces a parametrically large regime where the potential is
logarithmic, as illustrated in Fig.~\ref{fig:potential_scales}.
The compactified theory has many narrow resonances which can be
described using non-relativistic quantum mechanics with this logarithmic
potential, leading to a Hagedorn spectrum.
The fact that stringy dynamics are not the only way to obtain
a Hagedorn spectrum, and in particular that such a spectrum arises
in ordinary quantum mechanics with logarithmic potentials
does not seem to be widely appreciated.%
\footnote
    {%
    However, the notion of a limiting temperature for systems
    with exponential densities of states was first introduced by
    Rumer in 1960 \cite{Rumer:1960}, precisely in quantum mechanics
    with a logarithmic potential,
    several years before Hagedorn's suggestion \cite{Hagedorn:1965st}
    that such a density of states may arise in hadronic physics.
    }

To make our presentation reasonably self-contained, we begin
in Section~\ref{sec:adiabatic_compactification}
with a summary of center-stabilized adiabatic compactifications.
Section~\ref{sec:NR_EFT_light} discusses the light sector of the
compactified theory, with a focus on the spectrum of bound states.
In Section~\ref{sec:NR_EFT}, we formulate the 3D non-relativistic
effective field theory (EFT) which efficiently describes the dynamics
of heavy quark and gluon degrees of freedom.
Section~\ref{sec:symmetries} describes how the various symmetries of the
underlying 4D gauge theory act within our 3D effective field theory.
In Section~\ref{sec:bound_states} we examine the resulting spectrum of
heavy bound states, while Section~\ref{sec:decays} discusses decay
processes.
We summarize our findings in Section~\ref{sec:discussion}
and discuss some of their consequences,
including large $N$ scaling relations and implications
for the thermodynamics of QCD-like theories.
Several appendices contain technical details.

\section{Adiabatic compactification}
\label{sec:adiabatic_compactification}

Consider $SU(N)$ Yang-Mills theory compactified on
$\mathbb R^3 \times S^1$,
with the spatial circle having circumference $L$,
\begin{equation}
S_{\text{YM}}=\frac{1}{4g^{2}}\int_{\mathbb{R}^3\times S^{1}}d^{4}x\,\left(F_{\mu\nu}^{a}\right)^{2}.
\label{eq:YM Lag}
\end{equation}
If all matter fields added to the theory transform in the
adjoint representation of the gauge group,
then the theory has a $\mathbb{Z}_N$ center symmetry.
(We discuss below the addition of fundamental representation fermions.)
Order parameters for center symmetry are built from
the holonomy of the gauge field in the compact direction
(or ``Polyakov loop''),
\begin{equation}
    \Omega
    \equiv
    \mathcal P \> e^{i \int_0^L dx_3 \> A_3} \,.
\end{equation}
Center symmetry transformations multiply the (fundamental representation)
trace of the holonomy by a phase factor equal to an $N$'th root of unity.
The defining transformation is
\begin{equation}
   \tr \Omega \to \omega \, \tr \Omega \,,\qquad
    \omega \equiv e^{2\pi i/N} \,.
\label{eq:centersym}
\end{equation}
At large $L$ center symmetry is unbroken, implying
that $\langle \tr \Omega^n \rangle = 0$ for all integer $n \neq 0 \mod N$.
This is a hallmark of a confining phase.
At small $L$ the realization of center symmetry is analytically
calculable \cite{Gross:1980br,Weiss:1981ev}.
We require that the theory is engineered to prevent
spontaneous breaking of the $\mathbb Z_N$ center symmetry
in the $L \to 0$ limit,
so that the theory is \emph{not} in a deconfined plasma phase at small $L$.
This can be achieved by adding suitable double trace deformations
of the form $|\!\tr \Omega|^2$ (plus higher windings)
to the action of pure Yang-Mills theory \cite{Unsal:2008ch,Myers:2007vc}.
Alternatively, the center symmetry at small $L$ can be stabilized by
the addition of massless or sufficiently light adjoint representation fermions
\cite{Kovtun:2007py,
    Myers:2009df,
    Unsal:2010qh,
    Azeyanagi:2010ne,
    Catterall:2010gx}.%
\footnote
    {%
    If center symmetry is stabilized with adjoint fermions,
    we assume that $2 \le n_{\rm adj} \le 5$ species of adjoint Majorana fermions
    are added, so the theory is asymptotically free but non-supersymmetric
    (in the massless limit).
    We also take the adjoint fermion mass $m_{\rm adj}$ to be large compared
    to the mass gap scale $m_\gamma$ discussed below.
    }
If the adjoint fermions are massive,
center stabilization for small $L$ requires that their mass $m_{\rm adj}$
satisfy the constraint $m_{\rm adj} \lesssim 2\pi/NL$
\cite{Unsal:2010qh}.

With center symmetry stabilized,
the one-loop effective potential $V_{\rm eff}(\Omega)$ for the holonomy,
obtained by integrating out field modes with non-zero
Kaluza-Klein (KK)  momentum in the compact direction,
has a unique (up to gauge equivalence)
$\mathbb Z_N$ symmetric minimum,
\begin{equation}
    \Omega
    =
    \omega^{-(N-1)/2} \>
    \mathrm{diag}\,(1, \omega, \omega^2, {\cdots}, \omega^{N-1}) \,.
\label{eq:Omega}
\end{equation}
For sufficiently small $L$, the gauge coupling at the compactification
scale is weak and quantum fluctuations are suppressed.
Hence, one may regard the holonomy $\Omega$ as a nearly constant
$SU(N)$ matrix with eigenvalues
which are all $N$'th roots of unity for $N$ odd,
and all $N$'th roots of $-1$ for $N$ even.
The holonomy acts like an adjoint representation Higgs field,
``breaking'' the non-Abelian gauge symmetry
(using typical sloppy perturbative language)
down to the $U(1)^{N-1}$ Cartan subgroup.
We will refer to the $N{-}1$ diagonal Cartan components of
the gauge field as ``photons.''
The off-diagonal components
of gauge field (charged under the Cartan subgroup)
will be termed ``$W$-bosons'' and receive masses given by
positive integer multiples of
\begin{equation}
    \mW \equiv 2\pi/(NL) \,.
\label{eq:mW}
\end{equation}

Fluctuations in the eigenvalues of the holonomy will have an effective mass
$m_\Omega$ whose value depends on the details of the center symmetry
stabilization.
One may regard $m_\Omega \sim \sqrt\lambda\, \mW$ as a characteristic
fiducial value,
with $\lambda \equiv g^2 N$ the usual 't Hooft coupling.
This is the typical size resulting from modifications to the
one-loop effective potential for the holonomy,
unless one fine-tunes the stabilization mechanism,
for instance by considering a nearly supersymmetric limit of the theory.

The dynamical Higgs mechanism and resulting Abelianization
induced by the center-symmetric holonomy is the key feature responsible for the
analytic tractability of the theory at small $L$.
All charged degrees of freedom have masses of order $\mW$ or more,
so the 4D 't Hooft coupling $\lambda$ does not continue to run
below the scale $\mW$.
If $\mW \gg \Lambda$, or equivalently
\begin{align}
\eta \equiv N L \Lambda \ll 1 \,,
\label{eq:etadef}
\end{align}
then the long-distance value of the 't Hooft coupling
will be small, $\lambda(\mW) \ll 1$.
We focus on this regime in what follows and,
unless stated otherwise,
the value of $g^2$ is taken at the scale $\mW$.

Previous work on adiabatically compactified QCD-like theories has focused
exclusively on the lightest subsector in the small $L$ limit, with
characteristic energies and momenta much less than $\mW$ and $m_\Omega$.
On these scales, the physics can be described by an effective field
theory of $N{-}1$ Abelian photons living in three dimensions.
Non-perturbative monopole-instanton effects generate
small but relevant interactions between the photons.
The Euclidean action for the diagonal components
of the gauge field
has the schematic form%
\footnote
    {%
    Perturbative corrections
    generate photon mixing terms (as well as higher derivative
    terms which are irrelevant at long distances).
    The photon mixing matrix has been calculated in $\mathcal{N}=1$
    SYM theory to first order in $\lambda$ \cite{Anber:2014lba}.
    This photon mixing is diagonalized by the same $\mathbb Z_N$
    Fourier transform mentioned below, and does not affect the following
    discussion.
    \label{fn:mixing}
    }
\begin{equation}
    S_{\rm light}
    =
    L \int d^3x \, \left[ \tfrac 1{4  g^2} (F_{\mu\nu}^a)^2
    + \mathcal L^{\rm monopole}_{\rm int} \right] .
\label{eq:Slight0}
\end{equation}
A three-dimensional Abelian duality transformation leads to the Coulomb gas
representation,%
\footnote
    {%
    A redundant field component has been introduced in this representation,
    as if the original gauge group were $U(N)$ instead of $SU(N)$.
    The unphysical components,
    $\sum_{a=1}^N F_{\mu\nu}^a$ and $\sum_{a=1}^N \sigma^a$,
    exactly decouple and can be ignored.
    See, e.g., Ref.~\cite{Unsal:2008ch} for more detailed discussion.
    Appendix \ref{app:lightsector} contains
    details of our conventions, normalizations, and duality transformation.
    }
\begin{equation}
    S_{\rm light}
    =
    \int d^3x \,
    \biggl[
	\frac{\lambda \mW}{16\pi^3} \, (\nabla \bsigma)^2
	-
	\zeta \> \sum_{i=1}^N
	\cos(\balpha_i \cdot \bsigma + \theta/N)
    \biggr] .
\label{eq:Slight}
\end{equation}
The field
$\bsigma = \{ \sigma^i \} $ is an $N$-component
compact scalar field; in our basis it is independently periodic
in every component with period $2\pi$.
The fundamental domain of $\bsigma$ is the unit cell of the weight lattice,
generated by the shifts
$
    \bsigma \to \bsigma + 2\pi \bmu_i~
$
where $\{ \bmu_i \}$ are the fundamental weight vectors of $SU(N)$
and $\{ \balpha_i \}$ are the corresponding root vectors.
The ``fugacity''
\begin{equation}
    \zeta = A \, \mW^3 \, \lambda^{-2} \, e^{-8\pi^2 /\lambda} \,,
    \label{eq:monopole_fugacity}
\end{equation}
where $A$ is an $\O(1)$ coefficient which depends on the choice of regularization scheme.
Although not immediately apparent, the action (\ref{eq:Slight})
is invariant, as it must be,
under shifts in the QCD $\theta$ angle by multiples of $2\pi$.

To obtain an expression for the masses of the dual photons,
note that the potential
$
    V = - \zeta \sum_{i=1}^N
    \cos(\balpha_i \cdot \bsigma + \theta/N)
$
has $N$ extrema in the unit cell of the weight lattice
located at $\langle \bsigma\rangle_k = \frac{2 \pi k}{N} \, \brho$
for $k=0,{\cdots},N{-}1$,
where $\brho = \sum_{i=1}^{N-1} \bmu_i$ is the Weyl vector.%
\footnote
    {%
    To see this, use $\balpha_i \cdot \brho = 1$ for
    $i=1,{\cdots}, N{-}1$, together with $\balpha_N \cdot \brho = 1{-}N$.
    }
For $\theta =0$ the minimum lies at  $k=0$. For
general $\theta$, the vacuum energy density is given by
\begin{equation}
    V_0 = - N \zeta\;
    {\max_k} \Big(  \cos \frac{2 \pi k +\theta }{ N} \,\Big) .
\label{eq:vacenergy}
\end{equation}
Expanding
the  potential around each of the $N$ extrema and diagonalizing the
curvature (via a $\ZN$ Fourier transform)
yields the $\theta$-dependent mass spectrum in each of the $N$ extrema
(not all of which are minima).
At the lowest-energy minimum, which determines the physical mass spectrum,
one finds
\begin{equation}
\label{massspectrum}
    m_p^2 = m_\gamma^2 \, \sin^2 \left(\frac{\pi p}{N} \right)\;
    {\max_k}\Big(\cos \frac{2 \pi k + \theta }{ N}\, \Big) ,
\end{equation}
for $p = 1, 2, {\cdots}, N{-}1$, with
\begin{equation}
    m_\gamma
    \equiv
    C \, \mW \, \lambda^{-3/2} \, e^{-4\pi^2/\lambda} \,.
\label{eq:mgamma}
\end{equation}
The $\O(1)$ coefficient $C$ is determined in terms of the coefficient
$A$ in the fugacity \eqref{eq:monopole_fugacity}.
The label $p$ can be viewed as the charge under $\mathbb Z_N$
center symmetry transformations; this is discussed
in Sec.~\ref{sec:symmetries}.
One may also show that expectation values of
large fundamental representation Wilson loops
(not wrapping the compactified direction)
have area law behavior, with a string tension \cite{Unsal:2008ch}
\begin{equation}
    T = C' \, \lambda \, \mW \, m_\gamma \,,
\label{eq:stringtension}
\end{equation}
with $C'$ another $\O(1)$ coefficient.

The dual photon mass $m_\gamma$ can be expressed in terms of the strong scale
$\Lambda$ by using the renormalization group to relate
$\lambda$ at the scale of $\mW$ to $\Lambda$.
The specific form of this relation depends on the value of the
beta function, and hence on whether center symmetry is stabilized
by double trace deformations, or by the addition of adjoint fermions.
If center symmetry is stabilized by a double trace deformation,  then
parametrically \cite{Unsal:2008ch}
\begin{align}
  m_{\gamma}
  \sim \Lambda (NL\Lambda)^{5/6}
  = \O(\Lambda \, \eta^{5/6}) \, ,
  \label{eq:dYMphoton_mass}
\end{align}
and $m_\gamma/\mW = \O(\eta^{11/6})$.%
\footnote
    {%
    If center symmetry is stabilized by the addition of $n_{\rm adj}$
    light adjoint Majorana fermions with mass comparable to $\mW$,
    then $m_\gamma/\mW = \O(\eta^{(11 - 2 n_{\rm adj})/6})$.
    }

\subsection{Addition of fundamental quarks}\label{sec:quarks}

We will consider center-stabilized adiabatically compactified QCD
in addition to pure Yang-Mills theory.
This entails adding $\nf$ flavors of quarks
--- fundamental representation Dirac fermions.
We restrict our discussion to $\nf \le N$ and,
for simplicity, focus on the massless quark limit,
\begin{align}
m_q = 0 \,,
\end{align}
where the uncompactified theory has an
$SU(\nf)_L \times SU(\nf)_R \times U(1)_V$ continuous
chiral symmetry.%
\footnote
    {%
    For $\nf > N$, it is not currently known how to ensure that chiral symmetry
    realizations coincide at large and small $L$.
    }
When compactifying the theory on $\mathbb R^3 \times S^1$,
one must specify the boundary conditions on the quark fields.
Instead of simply choosing periodic, or antiperiodic, boundary conditions
for all quark flavors, we will consider flavor-twisted boundary conditions,
or equivalently introduce a non-dynamical flavor holonomy
$\Omega_F \in U(\nf)_V$.
If one regards the quark fields $q$ as an $N \times \nf$ matrix of spinors,
then in $A_3 = 0$ gauge
(where the gauge holonomy becomes encoded in boundary conditions),
the boundary conditions on quarks are
\begin{equation}
    q(t,\x,L) = \Omega \, q(t, \x, 0) \, \Omega_F^\dagger \,.
\label{eq:twistbcA}
\end{equation}
We specifically choose the flavor holonomy $\Omega_F$ to have
a set of eigenvalues which are invariant under
$\mathbb Z_{\nf}$ cyclic permutations.  The symmetry structure of QCD with such boundary conditions was discussed in Ref.~\cite{Cherman:2017tey}
(see also Refs.~\cite{Kouno:2012zz,
    Sakai:2012ika,
    Kouno:2013zr,
    Kouno:2013mma,
    Iritani:2015ara,
    Kouno:2015sja,
    Hirakida:2016rqd,
    Hirakida:2017bye,
    Shimizu:2017asf}).
To preserve reflection (in the compactified direction) and
charge conjugation symmetries, we also require
that complex conjugation leave this set of eigenvalues unchanged.
These two conditions imply that the eigenvalues of $\Omega_F$ are either
given by all $\nf$'th roots of $+1$, or by all $\nf$'th roots of $-1$.
Finally, to simplify our discussion and leave unchanged the relevant degrees
of freedom in the non-perturbative analysis of the light sector,
we want all flavors of quarks to receive non-zero effective masses
from the compactification.
This requires that no eigenvalue of the gauge holonomy coincide with
an eigenvalue of the flavor holonomy.

Solutions to these just-stated constraints depend on the values of
$N$ and $\nf$, in particular whether $N$ is even or odd
and (when $N$ is even) whether $N$ and $\nf$ have common divisors.
For simplicity of exposition we will henceforth assume that $N$ is odd,
unless stated otherwise,
so that the eigenvalues (\ref{eq:Omega}) of the gauge holonomy $\Omega$
are $N$'th roots of unity.
To avoid coinciding gauge and flavor eigenvalues,
this implies that the flavor holonomy eigenvalues must equal $\nf$'th
roots of $-1$.
Consequently, we choose
\begin{equation}
    \Omega_F =
    \mathrm{diag}\,(\xi^{\frac 12}, \xi^{\frac 32}, {\cdots}, \xi^{\nf-\frac 12}) \,,
    \qquad
    \xi \equiv e^{2\pi i/\nf} \,.
\label{eq:OmegaF}
\end{equation}
When the gauge holonomy is encoded in a non-zero value of $A_3$
(so that the gauge field satisfies simple periodic boundary conditions),
the resulting quark boundary conditions are
\begin{equation}
    q^A(t,\x,L) = \xi^{\frac 12 -A} \, q^A(t,\x,0) \,,
\label{eq:twistbc}
\end{equation}
where $A = 1,{\cdots},\nf$ is a flavor index.
The effect of these boundary conditions is to shift the moding
(i.e., the allowed values of the momentum in the compact direction),
in a flavor-dependent fashion
which is detailed below.
The boundary conditions \eqref{eq:twistbc}
reduce the non-Abelian flavor symmetry to the Abelian subgroup%
\footnote
    {%
    More precisely, the unbroken subgroup is
    $
    U(1)^{\nf-1}_L \times U(1)^{\nf-1}_R \times U(1)_V \big/ \mathbb Z_{\nf}
    $.
    Henceforth,
    we will not be explicit with the discrete identification
    needed to avoid double counting $\mathbb Z_{\nf}$ phase rotations.
    }
\begin{equation}
    U(1)^{\nf-1}_L \times U(1)^{\nf-1}_R \times U(1)_V \,.
\label{eq:chiralsym}
\end{equation}
Note that this residual flavor symmetry of our compactified theory
contains the axial subgroup $U(1)_A^{\nf-1}$ which
differentially rotates the phases of left and right handed quarks
in a flavor-dependent fashion.

In the center-stabilized regime of YM theory,
the addition of massless quarks with the boundary conditions \eqref{eq:twistbc}
produces fermion zero modes localized on the monopole-instantons.
The presence of these zero modes modifies the
non-perturbative long distance dynamics.
After a 3D duality transformation,
one may show that $\nf{-}1$
of the dual scalar fields remain
exactly massless \cite{Cherman:2016hcd}, while the remaining $N {-} \nf$
dual scalar fields develop non-perturbative masses just
as in center-stabilized YM theory without fundamental quarks.
The mechanism causing $\nf{-}1$ dual scalars to become massless in
the presence of fermion zero modes involves their acquisition
of non-trivial transformation properties under the anomaly-free
$U(1)^{\nf-1}_A$ axial symmetry, as explained in Ref.~\cite{Cherman:2016hcd}.
Consequently, these exactly-massless fields are precisely the
expected Nambu-Goldstone bosons (or `neutral pions')
produced by spontaneous breaking of the chiral symmetry
\eqref{eq:chiralsym}
down to the diagonal vector-like $U(1)_V^{\nf}$ subgroup
\cite{Cherman:2016hcd}.

If a small quark mass $m_q$ is added to the theory,
then some of the dual photons, or neutral pions, become massive.
For example, when $\nf \eq N$ one finds \cite{Cherman:2016hcd}
(at $\theta = 0$) that
\begin{align}
    m_{p} = C \sqrt{\mW m_q} \,
    e^{-4\pi^2/\lambda} \sin \frac{\pi p}{N} \,.
\label{eq:pNGBmass0}
\end{align}
(Here $p$ is the charge of the pion under cyclic flavor
permutations.)
One may again relate $m_{p}$ to the strong scale $\Lambda$ by
taking into account the contribution of the fundamental
fermions to the running of the coupling at the scale $\mW$.
With the pure-YM center symmetry stabilized via double trace deformations
and $\nf \eq N$, one finds
\begin{align}
    m_{p}
    =
    \O \big(
	\eta
	\sqrt{m_q \Lambda}
    \big) \,,
\label{eq:pNGBmass}
\end{align}
where, once again, $\eta \equiv NL \Lambda$.

\section {\boldmath Light sector bound states}
\label{sec:NR_EFT_light}
As noted in the introduction, when the color holonomy has the
center symmetric form (\ref{eq:Omega}),
a rich spectrum of hadronic states is present in the
small-$L$ regime of the compactified theory.
These states fall into two categories based on the scale of
their rest masses.  One set of states have rest masses of order of the
light scale $m_{\gamma}$, while the other set has rest masses of
order of the heavy scale $\mW$.  As will be shown below, in both sectors
the binding momenta are small compared to the rest masses of constituents,
so the most efficient way to describe each sector of the theory involves
constructing an appropriate non-relativistic effective field theory.
In this section we describe the effective field theory for the
light `dual photon' sector and discuss the resulting light bound state
spectrum.

\subsection{$N=2$ bound states}
To illustrate the relevant physics in the simplest setting, consider
adiabatically compactified pure YM theory with $N\,{=}\,2$ and
$\theta \,{=}\, 0$.
The relativistic 3D effective theory describing interactions
of the single (physical) dual photon field
$\sigma \equiv \sigma_1{-}\sigma_2$,
to leading non-trivial order in the semiclassical expansion,
is
\begin{align}
    S_{\rm{3D}, \rm{rel}}
    =
    \int d^3{x} \, \left[
	\frac{\lambda \mW}{32\pi^3} \,
	(\partial_{\mu} \sigma)^2 - 2 \zeta \cos(\sigma)
    \right] .
\label{eq:photon_2color}
\end{align}
Introducing a canonically normalized field
$
\sigmap \equiv \sigma \, \big(\frac{ \lambda \mW}{16 \pi^3 }\big)^{1/2}
$,
and expanding the potential, one finds
\begin{align}
    S_{\rm{3D}, \rm{rel}} =
    \int d^3{x} \,
    \left[
	\tfrac{1}{2} (\partial_{\mu}\sigmap)^2
	+ \tfrac{1}{2} m \, _{\gamma}^2 \, \sigmap^2
	- \tfrac{2}{3}\epsilon\,  m_{\gamma}\, \sigmap^4
	+ \tfrac{16}{45}  \epsilon^2\, \sigmap^6
	- \tfrac{32}{315} \epsilon ^3\, m_{\gamma}^{-1} \,  \sigmap^8
	+ \cdots
    \right] ,
\label{eq:photon_rel_EFT}
\end{align}
where
\begin{align}
    \epsilon \equiv \frac{  \pi^3 m_{\gamma}}{   \lambda m_{W}}
    = \O\big( \lambda^{-5/2} \, e^{-4\pi^2/\lambda}\big)
    \,.
\label{eq:bare_epsilon}
\end{align}
At first glance it is tempting to assume that the interaction terms
in \eqref{eq:photon_rel_EFT} have negligible consequences.
To our knowledge, effects of these weak interactions have not previously
been considered, either in the literature on adiabatically compactified
4D theories starting with Ref.~\cite{Unsal:2008ch}, or in the original
literature on the Polyakov model in three dimensions \cite{Polyakov:1976fu}.
As we now discuss, this presumption overlooks interesting physics.

The $\sigmap^{8}$ and higher terms in the action \eqref{eq:photon_rel_EFT}
are irrelevant and
can be ignored when focusing on the long distance behavior of the
theory.    The $\sigmap^4$ coupling is relevant, but its coefficient
is exponentially small in units of the $\sigma$ mass.
The $\sigmap^6$ coupling is marginal and infrared-free
\cite{PhysRevB.7.248,PhysRevB.12.256,PISARSKI1981356,Pisarski:1982vz}.
It is also exponentially small and stops running below the mass gap
scale $m_{\gamma}$.
These considerations might naively be interpreted to imply that all interaction effects
in the low energy theory \eqref{eq:photon_rel_EFT} are tiny.  But
consider interactions of $\sigmap$ modes with low momenta $p \ll m_{\gamma}$.
Such interactions can be described by a non-relativistic
effective field theory.
Writing
$
    \tilde\sigma =
    (2m_{\gamma})^{-1/2} e^{- i m_{\gamma} t} \, \Sigma + (\mathrm{h.c.})
$, where $\Sigma$ is the non-relativistic field,
and integrating out rapidly oscillating terms leads to
the non-relativistic description,%
\footnote
    {%
    Here and in Eq.~(\ref{eq:Nphoton_NR_EFT}) below, we flip the overall
    sign so that the nonerelativistic action $S_{\rm 3D,NR}$ has the
    conventional $T{-}V$ form.}
\begin{align}
    S_{\rm{3D}, \rm{NR}}
    =
    \int dt\,d^2{x} \,
    \bigg[
	\Sigma^{\dag} \bigg(i\partial_{t} + \frac{{\nabla}^2} {2m_\gamma} \bigg) \Sigma
	+ \frac{  \epsilon}{  m_{\gamma}} \, (\Sigma^{\dag})^2 \,\Sigma^2
	- \frac{8 \epsilon^2}{ 9 m_{\gamma}^3} \, (\Sigma^{\dag})^3 \,\Sigma^3
	+ \cdots
    \bigg] .
\label{eq:photon_NR_EFT}
\end{align}
The scaling dimension assignments appropriate to non-relativistic theories
in spacetime dimension $d$ are $[t] = -2$,
$[x]=-1$, $[\Sigma] = \tfrac{d-1}{2}$, and $[m_{\gamma}] = 0$.
This implies that the coefficients of the $(\Sigma^{\dag}\Sigma)^2$ and
$(\Sigma^{\dag}\Sigma)^3$ interactions have dimensions $d{-}3$ and $2(d{-}2)$,
respectively.
In $d{=}3$, this shows that the two particle $(\Sigma^\dagger \Sigma)^2$ interaction
becomes \emph{marginal} in non-relativistic dynamics,
while the three particle $(\Sigma^{\dag}\Sigma)^3$ interaction becomes irrelevant.
In fact, the $(\Sigma^{\dag}\Sigma)^2$ coupling $\epsilon$ runs logarithmically
with scale \cite{Hammer:2004as,Kaplan:2005es}, as may be seen
(for example) by calculating the two particle scattering amplitude.
Consequently, the definition \eqref{eq:bare_epsilon} should be
interpreted as the value of the running interaction strength
$\epsilon$ at the UV momentum cutoff $\mu_{\rm UV} \sim m_{\gamma}$.
In the non-relativistic limit the only diagrams which contribute
to the renormalization group (RG) evolution of $\epsilon$ beyond
tree level are iterated bubble diagrams.
Summing them yields the exact beta function for $\epsilon$.
Using dimensional regularization, one simply finds \cite{Kaplan:2005es}
\begin{align}
    \mu \frac{d\,  \epsilon(\mu)}{d \mu} = -\frac{1}{\pi}\, \epsilon(\mu)^2 \,.
\end{align}
When the initial coupling $\epsilon(\mu_{\rm UV})$ is positive,
corresponding to an attractive interaction,
$\epsilon(\mu)$ diverges at the momentum scale
$
    \LambdaIR  =
    \mu_{\rm UV}  \exp\left[-{{ \pi}/{(  \epsilon(\mu_{\rm UV}))}}\right]
$.
As a function of momentum, the two particle
scattering amplitude $\mathcal{A}(k)$ becomes singular at
$k^2 = -\LambdaIR^2$.
A pole develops at this position, indicating that
$\LambdaIR$ can be interpreted as the
binding momentum for a two-body bound state of dual photons.%
\footnote
    {%
    One may also directly solve the quantum mechanical problem a particle of
    reduced mass $\half m_\gamma$ moving in the attractive
    potential $- \frac{ 2 \epsilon }{  m_\gamma} \, \delta^{(2)} ({\x})$.
    The bound state wave function equals $K_0(r/r_B)$,
    with the bound state size $r_B = |m_\gamma \, \Delta E_2|^{-1/2}$
    and $\Delta E_2$ equaling the binding energy (\ref{eq:B2}).
    }
The two particle binding energy is thus
\begin{equation}
    \Delta E_2
    = -\frac{k^2}{m_{\gamma}}
    = - \frac{\mu_{\rm UV}^2 }{ m_\gamma} \, e^{-2 \pi/  \epsilon(\mu_{\rm UV})}
  = -\tfrac {1}{4}\, {c^2}\, m_\gamma \, e^{-2 \lambda \mW / \pi^2 m_\gamma}
    \,.
\label{eq:B2}
\end{equation}
In the final form we used the bare value (\ref{eq:bare_epsilon}) of $\epsilon$
and set the ultraviolet cutoff to the reduced mass $\half m_\gamma$ times an
$\O(1)$ coefficient $c$, whose determination  requires
a more careful matching calculation and is left for future work.
The two dual photon bound state has a rest mass
\begin{align}
    m_{2}
    &=
    2 m_\gamma + \Delta E_2
    =
    m_{\gamma} \,
    \big( 2 - \tfrac {1}{4} \, {c^2}\,e^{-2 \lambda m_{W} /\pi^2 m_{\gamma}} \big)  .
\end{align}
Expressed in terms of the original gauge coupling,
the fractional binding energy involves
a non-perturbative double exponential,
\begin{align}
    \frac{\Delta E_2}{2m_{\gamma}} =
	- \tfrac{1}{4} \, c^2 \, \exp\big(
	    -\tfrac{2}{\pi^2 C} \, \lambda^{5/2} \, e^{4\pi^2/\lambda}
	    \big) ,
\label{eq:B2rel}
\end{align}
whose appearance is quite peculiar in the context of the 4D gauge theories.%
\footnote
    {%
    However,
    the existence of double-exponential non-perturbative scales
    in gauge theory has been previously advocated \cite{Shifman:1994yf},
    based on quite different considerations from those discussed here.
    }

In addition to a two particle bound state, an
attractive two-body interaction in two space dimensions
also binds higher multi-body bound states.
(See, for example, Refs.~\cite{Hammer:2004as,Lee:2005nm}.)
The magnitude of the $k$-body binding energy $\Delta E_k$ increases exponentially with $k$,
with $\Delta E_{k+1}/\Delta E_{k} \sim 8.6$ for large $k$ \cite{Hammer:2004as}.
In our context,
we thus deduce the presence of a very large number of
bound states of dual photons,
one slightly below each $k$-particle threshold at $E = k m_\gamma$
for $k = 2, 3, \cdots$,
with fractional binding energies proportional to the
non-perturbative double exponential (\ref{eq:B2rel}).%
\footnote
    {%
    This weak coupling non-relativistic description breaks down
    when $k \, (\ln 8.6)$ becomes exponentially large and comparable to
    $2 \lambda \mW / \pi^2 m_\gamma \sim \lambda^{5/2} e^{+4\pi^2/\lambda}$.
    }

\subsection{$N>2$ bound states}

We now briefly consider the generalization to arbitrary $N$,
still with $\theta = 0$.
Using a $\mathbb Z_N$ Fourier transform to
diagonalize the mass terms,
$
    \sigma_i \equiv \big(\frac{\lambda\mW}{8\pi^3}\big)^{-1/2}
    \sum_{p=1}^{N-1} \omega^{ i p} \, \tilde\sigma_p/\sqrt{N}
$
(with $\tilde\sigma_p^* = \tilde\sigma_{N-p}$),
the generalization of the action \eqref{eq:photon_rel_EFT} is
\begin{align}
    S_{\rm 3D}
    =
    \!\int d^3x
    \sum_{p=1}^{N-1}
    \half \left(
	|\partial_\mu \tilde \sigma_p|^2 + m_p^2 |\tilde \sigma_p|^2
    \right)
    &-
  \frac{4  \epsilon  \, m_{\gamma} }{ 3 N} \!
    \sum_{p_1\cdots p_4 = 1}^{N-1} \!
	\delta_{p_1+p_2+p_3+p_4,0} \>
	e^{i{\pi}(p_1+p_2+p_3+p_4)/N} \,
\nonumber
\\ &{} \times
	\biggl[\,
	    \prod_{i=1}^4 \sin\left(\frac{\pi p_{i}}{N}\right)
	\biggr] \,
	\sigmap_{p_1}\sigmap_{p_2}\sigmap_{p_3}\sigmap_{p_4}
    + \O(\tilde\sigma^6) \,,
\label{eq:quarticphoton}
\end{align}
where all center charges $\{p_k\}$ are understood to be defined modulo $N$.
The masses $\{ m_p \}$ and coupling $\epsilon$ are given by
Eqs.~(\ref{massspectrum}) and (\ref{eq:bare_epsilon}), respectively.
[Recall that the field $\tilde\sigma_0 \propto \sum_i \sigma_i$ decouples,
and is omitted.
Expression (\ref{eq:quarticphoton}) reduces to
the earlier form (\ref{eq:photon_rel_EFT}) for $N\eq 2$, as it should.]

The sign of the quartic interaction depends on the values of
the center charges of the particles under consideration.
For elastic scattering of dual photons with arbitrary charges $p_1$ and $p_2$,
the relevant piece of the quartic interaction has an overall minus sign,
which corresponds to attraction.
The effective theory which follows from
a non-relativistic reduction of the action (\ref{eq:quarticphoton}),
and generalizes the earlier form (\ref{eq:photon_NR_EFT}) to arbitrary $N$,
is
\begin{align}
\label{eq:Nphoton_NR_EFT}
    S_{\rm{3D}, \rm{NR}}
    =
    \int dt\,d^2{x} & \,
    \Bigg[ \sum_{p=1}^{N-1} \>
	\S^{\dag}_p \left(i\partial_{t} + \frac{{\nabla}^2}{2 m_p} \right) \S_p
    +
    \frac{2 \epsilon}{N} \, \frac{ m_p ^2 }{ m_\gamma^3}\;
    (\S^{\dag}_p)^2 \, \S_p^2
\nonumber
\\ &{}
    +
    \sum_{p_1 < p_2} \>
    \frac{8 \epsilon}{N} \, \frac{ m_{p_1} m_{p_2} }{ m_\gamma^3} \;
    \S^{\dag}_{p_1} \S^{\dag}_{p_2} \S_{p_2} \S_{p_1}
    + \cdots
    \Bigg] ,
\end{align}
where we have included only those terms contributing to elastic
$2 \leftrightarrow 2$ scattering.%
\footnote
    {%
    The interaction (\ref{eq:quarticphoton}) also includes
    charge exchange processes which lead to mixing among
    bound states with differing constituents but the same total
    center charge.  For generic values of $N$ and choices of $p_1$ and $p_2$
     the effects of such interaction
    terms on binding energies are suppressed in the non-relativistic limit,
    because the masses of the dual photons depend on their center charge.
    Charge exchange processes can only become relevant if states with
    differing constituents and the same total charge also have the same
    total constituent mass. Such mixing will deepen the binding of the
    lowest energy bound states of a given total charge.
    We defer a complete multi-channel treatment to future work.
    }
Note the factor of 4 difference in the coefficients of the quartic terms
responsible for scattering of identical vs.~non-identical particles.

Applying the earlier analysis
(either solving the two-particle Schr\"odinger equation with a delta function
potential, or resumming bubble diagrams and locating the resulting pole in
the scattering amplitude)
to states containing particles of center charge $p_1$ and $p_2$,
one finds the binding energy
\begin{align}
    \Delta E_2^{p_1 \ne p_2}
    =
    - 2c^2 \, m \,
    \exp \left(
	- \frac{\pi N }{ 4 \epsilon} \,
	\frac { m_\gamma^3 }{ m_{p_1} m_{p_2} m}
    \right)   ,
\label{eq:non-id binding}
\end{align}
if $p_1 \ne p_2$.
Here $m \equiv (m_{p_1}^{-1} + m_{p_2}^{-1})^{-1}$ is the reduced mass
of the two constituents.
If the two constituents are identical, then the result is
\begin{align}
    \Delta E_2^{p_1 = p_2}
    =
    - c^2 \, m_{p_1} \,
    \exp \left(
	- \frac{\pi N }{ \epsilon} \,
	\frac { m_\gamma^3 }{ m_{p_1}^3 }
    \right)  .
\label{eq:id binding}
\end{align}

Bound states composed of equal mass constituents can
have either equal or opposite charge constituents.
For the first case, with charges $p_1 = p_2 = p$, the
identical particle binding energy
(\ref{eq:id binding}) gives a total mass
\begin{equation}
    m_{2}^{p,p}
    =
    m_{p} \left[
	2 -  c^2 \, e^{- \frac{\pi N }{\epsilon} (m_\gamma/m_{p})^3}
	\right] .
\label{eq:M2_eq}
\end{equation}
For opposite charges, $p$ and $N{-}p$,
the non-identical binding energy (\ref{eq:non-id binding})
with $m_{p_1} = m_{p_2} = 2m = m_p$ gives total mass
\begin{equation}
    m_{2}^{p,N-p}
    =
    m_p \left[
	2 -  c^2 \, e^{- \frac{\pi N }{ 2 \epsilon} (m_\gamma/m_p)^3}
	\right]
\label{eq:M2_opp}
\end{equation}
(except for the special case of $p \eq N/2$ with $N$ even,
where the first result (\ref{eq:M2_eq}) applies).
In other words,
the fractional binding energy for non-identical particles is
$
    \O\big( e^{- \frac{\pi N }{ 2 \epsilon} (m_\gamma/m_{p})^3} \big)
    =
    \O\big( e^{- \frac{\pi N }{ 2 \epsilon}
    |\sin \frac {\pi p}{N}|^{-3} } \big)
$,
while bound states of identical constituents have twice the exponential
suppression in their binding energy.

\section {\boldmath Heavy sector effective field theory}
\label{sec:NR_EFT}
We now consider states with rest masses of order $\mW$ and above,
and characteristic binding momenta $p$ in the range
\begin{equation}
    m_\gamma \ll p \ll \mW \,.
\label{eq:NRrange}
\end{equation}
This section describes the construction of a non-relativistic effective theory suitable for the description of such states.
We begin with the effective theory characterizing pure gauge, or glueball,
dynamics, and then discuss the addition of fundamental representation quarks.

\subsection {Gauge field contributions}

The center-symmetric holonomy (\ref{eq:Omega}) may equivalently be
regarded as a non-vanishing constant diagonal gauge field in the
compact direction, $A_3$,
together with conventional periodic boundary conditions.
The $\tr [A_3, {\bm A}]^2$ term in the classical Yang-Mills action
generates tree-level masses of order $\mW$ for the charged $W$-bosons.
The efficient description of the interactions of these massive charged
degrees of freedom with the Cartan photons is provided by
a non-relativistic effective field theory with action:
\begin{align}
    S_{\textrm{heavy}}
    &=
    \sum_{{a,b=1}}^{N} \> \sumprime_{n=-\infty}^\infty \>
    \int dt \, d^2x
	\left[
	(\vec\phi_{n}^{\,ab})^\dagger \,i\partial_t \, \vec\phi_{n}^{\,ab}
	- M_{n}^{ab} \, |\vec\phi_{n}^{\,ab}|^2
	- \frac{|\bnabla \vec\phi_{n}^{\,ab}|^2}{2 m_n^{ab}}
	\right]
\nonumber
\\ & \quad{}
    +
    \frac {\lambda \mW}{4\pi} \sum_{a=1}^{N} \int dt\, d^2x \, d^2y \>
	\rho^a(t,\x) \, G (\x{-}\y) \, \rho^a(t,\y) \,,
\label{eq:Sheavy}
\end{align}
where
\begin{equation}
    G(\x{-}\y) \equiv \tfrac 1{2\pi} \, \ln (\mu |\x{-}\y|)
\label{eq:G}
\end{equation}
is the two dimensional Laplacian Green's function.
The derivation of this effective theory is detailed in appendix
\ref{app:NReft}.
Higher order (in $\lambda$) corrections, such as magnetic moment interactions,
are omitted for simplicity.

The two-dimensional vector fields $\vec\phi^{\,ab}_{n}$
are the non-relativistic reduction of
the $n$'th Fourier component (in the compact direction)
of the $(ab)$ component of the $SU(N)$ gauge field,
viewed as an $N \times N$ Hermitian matrix.
The color (or `Cartan') indices $a,b$ run from 1 to $N$,
and the Kaluza-Klein index $n$ is an arbitrary integer.
In the action (\ref{eq:Sheavy}),
the prime on the sum over $n$ is an indication to omit the $n=0$ term
when $a = b$, but not otherwise.
The vector field $\vec\phi_n^{\,ab}$ annihilates $W$-bosons with
charges $(+1,-1)$ with respect to the
$a$'th and $b$'th unbroken $U(1)$ gauge groups.
The spatial gradient $\bnabla$ is a two-dimensional
$U(1)^N$ covariant derivative defined by
\begin{equation}
    (\bnabla)_i (\phi^{ab}_n)_j
    \equiv
    \left[\nabla_i - i g_3 (A_i^a {-} A_i^b) \right] (\phi^{ab}_n)_j \,.
\end{equation}
Here $i,j = 1,2$ label the two non-compact spatial directions and
$\{ \vec A^{\, a} \}$ are $N$ independent spatial gauge fields.
We have introduced $N$ Abelian gauge fields, instead of $N{-}1$,
as if the original gauge group were $U(N)$ instead of $SU(N)$.
This simplifies notation, and makes no difference
as the unphysical extra photon,
$\bar A_i \equiv \sum_a A^a_i$,
will exactly decouple from all physical states.
We have also reverted to a perturbative normalization for the
gauge fields, with a dimensionless gauge coupling $g_3$
appearing inside the covariant derivative,
and a corresponding 3D Maxwell action given by
$
    L \int d^3x \> \frac 14 (F^a_{ij})^2
$.
The 3D gauge coupling is, to lowest order,
just the 4D gauge coupling evaluated at the scale $\mW$,
\begin{equation}\label{eq:coupling3}
    g_3^2 \equiv g^2(\mW) \,.
\end{equation}

Due to the non-trivial holonomy $\Omega$,
momentum in the compact direction carried by individual field components
is quantized in units of $\mW$, not $N \mW = 2\pi/L$.
The Kaluza-Klein reduction of the $(ab)$ component of the gauge field yields
a sum of modes with momentum
\begin{subequations}%
\label{eq:k}%
\begin{align}
    p_3 &= \mW \, k \,,
\\
\noalign{\hbox{where}}
    k &= a-b + n N \,, \qquad n \in \mathbb Z.
\end{align}
\end{subequations}
For any given value of $a = 1,{\cdots},N$
specifying a row of the $SU(N)$ gauge field,
there is a one-to-one mapping between the
momentum index $k$ and the corresponding values
of the column $b$ and KK index $n$,
\begin{equation}
    b-1 = (k-a+1) \bmod N \,,\qquad n = (k-a+b)/N \,.
\label{eq:kinv}
\end{equation}
In the following,
we will sometimes write expressions involving the relabeled field
\begin{equation}
    \vec\phi^{\,a}_k
    \equiv
    \vec\phi^{\,ab}_n \,,
\end{equation}
with the implicit understanding that momentum index $k$ is
related to the (antifundamental) column and KK indices $\{b,n\}$
via relations (\ref{eq:kinv}).
The momentum index $k$ may take on any integer value other than zero.
For charged $W$-bosons, $k \bmod N \ne 0$.
The ``diagonal'' operators $\vec\phi_n^{\,aa}$ with $n\ne 0$
annihilate the neutral (uncharged under $U(1)^N$) gauge bosons
carrying non-zero KK momentum.
These gauge bosons form the Kaluza-Klein tower whose $n=0$ modes
(excluded from $S_{\rm heavy}$)
are the $U(1)^N$ light Abelian photons.

The rest and kinetic mass parameters appearing in the effective theory
(\ref{eq:Sheavy}) only depend on the Cartan and KK indices via the
combination $k$, and equal the magnitude of  the compact momentum $p_3$,
up to higher order radiative corrections.
In other words,%
\begin{subequations}\label{eq:W-masses}%
\begin{align}%
    M_n^{ab} = M_k
    &\equiv \mW \, ( |k| + \O(\lambda) )
    = \mW |a - b + nN| + \O(\lambda\mW) \,,
\\
    m_n^{ab} = m_k
    &\equiv \mW \, ( |k| + \O(\lambda) )
    = \mW |a - b + nN| + \O(\lambda\mW) \,.
\end{align}
\end{subequations}
Although they coincide at lowest order,
the kinetic and rest masses appearing as parameters in
our 3D non-relativistic effective field theory (\ref{eq:Sheavy}),
or any other non-relativistic EFT,
may differ when subleading corrections are included,
even when the underlying theory retains full $2{+}1$ dimensional
Lorentz invariance.

In the effective action (\ref{eq:Sheavy}),
the time components of the $U(1)^N$ Abelian gauge fields
have been integrated out, producing non-local Coulomb interactions.
The operators
\begin{equation}
    \rho^a \equiv \sum_{{b = 1}}^{N} \> \sumprime_{n=-\infty}^\infty
    \left[
    (\vec\phi^{\,ab}_n)^\dagger \cdot \vec\phi_n^{\, ab}
    -
    (\vec\phi^{\,ba}_n)^\dagger \cdot \vec\phi_n^{\, ba}
    \right],
\label{eq:na}
\end{equation}
are the $U(1)^N$ charge densities.
(Note that $\bar \rho \equiv \sum_a \rho^a$ vanishes identically.)
The conserved charges defined by
spatial integrals of these charge densities must vanish,
\begin{equation}
    Q^a \equiv \int d^2x \> \rho^a(\x) = 0 \,,
\end{equation}
when acting on any physical, gauge invariant state.
Because of this, the dependence of the
2D Laplacian Green's function (\ref{eq:G}) on the
arbitrary scale $\mu$ inside the logarithm cancels
in any physical state,
since the variation of the Lagrangian with respect to $\mu$
is proportional to $(Q^{a})^2$.

The non-relativistic effective theory (\ref{eq:Sheavy})
describes the dynamics of all modes of the non-Abelian
gauge field which are charged under the $U(1)^N$ Cartan subgroup,
namely $W$-bosons,
plus the uncharged gauge field modes which carry non-zero KK
momentum, which we will term ``heavy photons.''
However, we have not included any fields describing fluctuations of the eigenvalues of the holonomy in the effective field theory.
These could easily be included as $N{-}1$ additional neutral scalar fields
(not 2D vectors like $\vec\phi_n^{\,ab}$) with $\O(\sqrt\lambda\mW)$ masses
whose precise values depend on the matter content or double trace
deformations used to stabilize the center symmetry.
These scalar fields only interact with $\vec\phi_n^{\,ab}$ via
higher dimension local operators, suppressed by powers of $\lambda$.
For the physics we choose to focus on,
holonomy fluctuation fields will not play
any significant role and may be neglected.
If adjoint fermions are used to stabilize the center symmetry,
then these fermions are also missing from our non-relativistic
effective theory.
They could be easily included but, for simplicity,
we will limit our attention to states where adjoint fermions
(and eigenvalue fluctuations) play no significant role.

Reading off the quantum Hamiltonian from the effective action
(\ref{eq:Sheavy}) is trivial, except for one UV subtlety.
The Hamiltonian of the second quantized non-relativistic theory
(with rest energies included) is
\begin{align}
    \hat H
    &=
    \sum_{{a,b=1}}^{N} \>
    \sumprime_{n=-\infty}^\infty \>
    \int d^2x  \;
	\phi_n^{ab}(\x)_i^\dagger \!
	\left[ -\frac{\bnabla^2}{2m_k} + M_k(\mu) \right]
	\phi_n^{ab}(\x)_i
\nonumber
\\ &-
	\sum_{a,b,c=1}^{N} \>
	\sumprime_{m,n=-\infty}^\infty \>
	\int d^2x \, d^2y \;
	\frac{\lambda \mW}{8\pi^2}
	\ln (\mu|\x{-}\y|)  \times {}
\nonumber
\\ &\qquad\qquad \times
	\left[
	\phi_n^{\,ab} (\x)_i^\dagger
	\left(
	\phi_m^{\,ac} (\y)_j^\dagger \,
	\phi_{m}^{\,ac}(\y)_j
	-
	\phi_m^{\,ca} (\y)_j^\dagger \,
	\phi_{m}^{\,ca}(\y)_j
	\right)
	\phi_{n}^{\,ab}(\x)_i
	\right.
\nonumber
\\ &\qquad\qquad\; -
	\left.
	\phi_n^{\,ba} (\x)_i^\dagger
	\left(
	\phi_m^{\,ac} (\y)_j^\dagger \,
	\phi_{m}^{\,ac}(\y)_j
	-
	\phi_m^{\,ca} (\y)_j^\dagger \,
	\phi_{m}^{\,ca}(\y)_j
	\right)
	\phi_{n}^{\,ba}(\x)_i
	\right] .
\label{eq:HNR}
\end{align}
where the field operators satisfy canonical commutation relations,
\begin{equation}
    \left[ \phi_n^{\,ab}(\x)_i^{\vphantom{\dagger}} ,\,
	    \phi_{n'}^{\,cd}(\y)_j \right] = 0
    \,,\quad
    \left[ \phi_n^{\,ab}(\x)_i ,
	    \phi_{n'}^{\,cd}(\y)_j^\dagger \right] =
    \delta^{ac} \, \delta^{bd} \, \delta_{nn'} \, \delta_{ij} \,
    \delta^2 (\x{-}\y) \,.
\end{equation}
In the Hamiltonian (\ref{eq:HNR})
we have written out the charge densities $\rho^a$ explicitly and
normal ordered the results.
In the quartic terms, normal ordering removes the UV sensitive
self-energy of each charged $W$-boson.
The price of that removal is that the $\mu$ dependence of the
Coulomb interaction terms no longer vanishes identically.
Instead, this unphysical dependence on the scale $\mu$ is canceled
by explicit dependence on $\mu$ which has been introduced
into the bare rest masses (of charged $W$'s only),
\begin{equation}
    \mu \frac d{d\mu} \, M_k(\mu)
    =
    - \frac {\lambda \mW}{4\pi^2} \, (1-\delta^0_{k \bmod N}) \,.
\label{eq:mw-running}
\end{equation}

The effective action (\ref{eq:Sheavy}), and corresponding
Hamiltonian (\ref{eq:HNR}), depend on the 3D gauge coupling $g_3$,
or equivalently the 't Hooft coupling $\lambda$,
both in the coefficient of the Coulomb interactions and inside
the spatial covariant derivatives.
But when considering phenomena for which the coupling to the
transverse Cartan gauge fields $\{ \vec A^a \}$ may be neglected,
the remaining dependence on $\lambda$ takes a very simple form.
To see this, rescale all spatial coordinates,
$\x \to \x'/s$, $\y \to \y'/s$,
and then redefine
$
    \vec\phi_k^{\,a}(\x'/s) = s \, \vec\varphi_k^{\,a}(\x')
$.
This is a unitary transformation;
the rescaled operators $\{\vec\varphi_k^{\,a}(\x)\}$ satisfy
the same canonical commutation relations as the original operators
$\{\vec\phi_k^{\,a}(\x)\}$.
In the Hamiltonian, the effect of this rescaling is to change
the relative coefficients of the kinetic and Coulomb energy terms.
Let
\begin{equation}
    \hat N_n^{ab}
    \equiv
    \int d^2x  \; \vec\phi_n^{\,ab}(\x)^\dagger \cdot \vec\phi_{n}^{\,ab}(\x)
\end{equation}
denote the number operator which counts the number of constituents
of the indicated type,
and define
\begin{equation}
    \hat H_{\rm NR} (\lambda;\mu)
    \equiv
    \left. \hat H \right|_{\vec A^a = 0} -
    \sum_{{a,b=1}}^{N} \>
    \sumprime_{n=-\infty}^\infty \>
    M_{k}(\mu) \, \hat N_{n}^{\,ab}
\end{equation}
as the non-relativistic Hamiltonian with rest energy contributions removed,
the spatial Abelian gauge fields set to zero,
and dependence on $\lambda$ and the scale $\mu$ made explicit.
If one chooses $s = \sqrt\lambda$, then a short exercise shows that
\begin{align}
    \hat H_{\rm NR}(\lambda;\mu)
    &\cong
    \lambda \, \hat H_{\rm NR}(1,\mu/\sqrt\lambda)
    =
    \lambda \, \hat H_{\rm NR}(1,\mu)
    -
    \frac{\lambda \ln \lambda}{8\pi^2} \, \mW \hat N_{\rm W} \,,
\label{eq:rescale}
\end{align}
where $\cong$ denotes unitary equivalence and
\begin{equation}
    \hat N_{\rm W} \equiv
    \sum_{\substack{a,b=1\\a\ne b}}^{N} \>
    \sumprime_{n=-\infty}^\infty \>
    \hat N_n^{ab}
\end{equation}
is the total number of charged $W$-bosons.
The scaling relation (\ref{eq:rescale}) shows that the
spectrum of the 2D Coulomb Hamiltonian $\hat H_{\rm NR}(\lambda;\mu)$
is simply proportional to the 't Hooft coupling $\lambda$,
up to an overall additive shift proportional to $\lambda \ln\lambda$
times the number of charged constituents.
This relation may equivalently be expressed as
\begin{equation}
    \frac 1{\lambda} \,
    \hat H_{\rm NR}(\lambda;\mu)
    \cong
    \frac 1{\lambda'} \,
    \hat H_{\rm NR}(\lambda';\mu)
    -
    \frac {\mW}{8\pi^2} \, \ln (\lambda/\lambda') \,
    \hat N_W \,.
\label{eq:rescale2}
\end{equation}

\subsection{Quark contributions}

The quark fields modify the light and heavy sectors of the
theory in several ways.
In addition to their effects on the non-perturbative large distance dynamics,
already mentioned in the previous section,
the compactified quark fields contain
massive degrees of freedom which play a role in physics on the
scale of $\mW$ and above.
Specifically, every flavor and color component of a fundamental
representation Dirac fermion leads, in a non-relativistic description,
to a pair of two-component spinor fields which we will denote
as $\psi_n^{aA}$ and $\chi_n^{aA}$.
The field $\psi_n^{aA}$ annihilates quarks of flavor $A$ which have
charge $+1$ under the $a$'th $U(1)$ gauge group
(and are neutral with respect to all other $U(1)$ gauge group factors).
The field $\chi_n^{aA}$ annihilates antiquarks of flavor $A$
and charge $-1$ under the $a$'th $U(1)$ gauge group
(and are neutral with respect to the other $U(1)$ gauge group factors).
It will be convenient to define quark KK indices as
half-integers, $n \in \mathbb Z+\half$.
These fields satisfy canonical anticommutation relations,
\begin{align}
    \left\{
	\psi_n^{aA}(\x)_s,\,
	\psi_{n'}^{bB}(\y)^\dagger_{s'}
    \right\}
    &=
    \left\{
	\chi_n^{aA}(\x)_s,\,
	\chi_{n'}^{bB}(\y)^\dagger_{s'}
    \right\}
	=
	\delta^{ab} \, \delta^{AB} \, \delta_{nn'} \, \delta_{ss'} \,
	\delta^2(\x{-}\y) \,,
\end{align}
where $s,s' = \pm$ are spin-1/2 spinor indices.
All other anticommutators vanish.
To describe the dynamics of the quarks,
one must add another set of terms to the effective theory
\eqref{eq:Sheavy} describing $W$-bosons, namely
\begin{align}
    S_{\rm quark}
    =
    \sum_{a=1}^{N}
    \sum_{A=1}^{\nf}
    \sum_{n\in \mathbb Z + \frac 12}
    \int dt \, d^2x
    &\left[
	(\psi_n^{aA})^\dagger \, i\partial_t \, \psi_n^{aA}
	- M_n^{aA} |\psi_n^{aA}|^2
	-\frac{|\bnabla\psi_n^{aA}|^2}{2m_n^{aA}}
    \right.
\nonumber
\\
    &\!\!+\left.
	(\chi_n^{aA})^\dagger \, i\partial_t \, \chi_n^{aA}
	- M_n^{aA} |\chi_n^{aA}|^2
	-\frac{|\bnabla\chi_n^{aA}|^2}{2m_n^{aA}}
    \right] ,
\label{eq:Squark}
\end{align}
where the covariant spatial gradients acting on fermions are defined by
\begin{equation}
    (\bnabla)_i \, \psi^{aA}_n
    \equiv
    \left[\nabla_i - i g_3 A_i^a \right] \psi^{aA}_n \,,\qquad
    (\bnabla)_i \, \chi^{aA}_n
    \equiv
    \left[\nabla_i + i g_3 A_i^a \right] \chi^{aA}_n \,.
\end{equation}
The compact momentum $p_3$ carried by
a quark created by $(\psi_n^{aA})^\dagger$
is
\begin{equation}
    p_3 = \mW \big[ (a{-}\half) - (A{-}\half) \, N/\nf + n N \big] \,,
\label{eq:quark_p3}
\end{equation}
while the antiquark created by $(\chi_n^{aA})^\dagger$
carries the opposite momentum $-p_3$.
The rest and kinetic quark masses equal $|p_3|$,
the magnitude of the compact momentum,
up to higher order radiative corrections,
\begin{align}
    M_n^{aA} &= |p_3| \, (1 + \O(\lambda))
    \,,\quad
    m_n^{aA} = |p_3| \, (1 + \O(\lambda)) \,.
\label{eq:quarkmasses}
\end{align}
Note that these fermion masses in the effective theory
have nothing to do with chiral symmetry breaking quark masses in the
underlying 4D theory, which we have assumed vanish.
Our EFT fully respects the chiral symmetry \eqref{eq:chiralsym}
of the compactified theory.
Nevertheless, the non-relativistic quark masses
\eqref{eq:quarkmasses} are non-vanishing for all values of
$n \in \mathbb Z+\half$, $a = 1,{\cdots},N$, and $A=1,{\cdots},\nf$.
(Recall that we have assumed that $N$ is odd.)
Our explicit calculations in Sec.~\ref{sec:bound_states}
will focus on the special case of $\nf = N$, for which
the allowed values of the compact momentum of a quark
become half-integers (times $\mW$),
\begin{equation}
    p_3 = \mW \, k \,,\quad \mbox { with }
    k \equiv a-A + n N \,.
\label{eq:k-quark}
\end{equation}
For a given Cartan index $a$,
relation (\ref{eq:k-quark}) gives a one-to-one mapping between
the flavor and KK indices $\{A,n\}$ and the
quantized momentum index $k$.
When discussing the $\nf\,{=}\,N$ theory, it will often be convenient to use
the momentum index $k \in \mathbb Z + \half$
in place of the (equivalent) values of the
the flavor and KK indices and relabel the quark fields as
\begin{equation}
    \psi^a_k \equiv \psi^{aA}_n \,,\quad
    \chi^a_{k} \equiv \chi^{aA}_n \,,
\end{equation}
with the implicit understanding that the flavor, KK and momentum
indices are connected via relation (\ref{eq:k-quark}).
In other words, $\psi^a_k$ annihilates a quark with compact momentum
$p_3 = \mW k$ and charge $+1$ under the $a$'th $U(1)$ gauge group,
while $\chi^a_k$ annihilates an antiquark with compact momentum
$p_3 = -\mW k$ and charge $-1$ under the $a$'th $U(1)$ group.

In addition to the above quark kinetic terms,
the Abelian charge densities $\rho^a$
appearing in the Coulomb interactions of the effective theory (\ref{eq:Sheavy})
must be augmented to include the quark contributions,
\begin{equation}
    \rho^a \equiv
    \sum_{b = 1}^{N} \> \sumprime_{n\in\mathbb Z}
    \left[
    (\vec\phi^{\,ab}_n)^\dagger \cdot \vec\phi_n^{\, ab}
    -
    (\vec\phi^{\,ba}_n)^\dagger \cdot \vec\phi_n^{\, ba}
    \right]
    +
    \sum_{A=1}^{\nf}
    \sum_{n\in \mathbb Z+\frac 12}
    \left[
    (\psi^{aA}_n)^\dagger \psi_n^{aA}
    -
    (\chi^{aA}_n)^\dagger \chi_n^{aA}
    \right] ,
\label{eq:na2}
\end{equation}
and the form of the Coulomb interactions appearing in the
action (\ref{eq:Sheavy}) must now have the contribution from
the unwanted extra $U(1)$ gauge group removed,
\begin{equation}
    S_{\rm Coulomb}
    =
    \frac{\lambda \mW}{4\pi} \!
    \int dt \, d^2x \, d^2y  \; G(\x{-}\y) \!
    \left[
	\sum_{a=1}^N \rho^a(t,\x) \, \rho^a(t,\y)
	-
	\frac 1N
	\sum_{a,b=1}^N \rho^a(t,\x) \, \rho^b(t,\y)
    \right] .
\label{eq:SCoulomb}
\end{equation}
(Without the subtraction of the second term in this expression,
the Coulomb energy would be that of a $U(N)$ gauge theory instead of $SU(N)$.)
With quarks added to the theory, all the conserved Abelian charges $Q^a$,
when acting on physical states, equal the baryon number,
\begin{equation}
    Q^a = N_B \equiv
    \frac 1N
    \sum_{a,A,n}
    \int d^2x
    \left[
    (\psi^{aA}_n)^\dagger \psi_n^{aA}
    -
    (\chi^{aA}_n)^\dagger \chi_n^{aA}
    \right] .
\label{eq:N_B}
\end{equation}

Conversion of the effective action for quarks (\ref{eq:Squark})
to the corresponding quark contribution of the non-relativistic
Hamiltonian proceeds as described earlier.
As with the $W$-bosons,
normal ordering the Coulomb interactions induces
logarithmic dependence on the scale $\mu$ in
the quark rest masses,
\begin{equation}
    \mu \frac d{d\mu} \, M_n^{aA}(\mu)
    =
    -\frac {\lambda\mW}{8\pi^2} \, (1{-}\tfrac 1N) \,.
\label{eq:mq-running}
\end{equation}
In the presence of quarks
the rescaling relation (\ref{eq:rescale2}) becomes
\begin{align}
    \frac1{\lambda} \,
    \hat H_{\rm NR}(\lambda;\mu)
    &\cong
    \frac1{\lambda'} \,
    \hat H_{\rm NR}(\lambda',\mu)
    - \frac{\mW}{16\pi^2} \ln(\lambda/\lambda')
    \left[
	2 \hat N_{\rm W}
	+ (1{-}\tfrac 1N) \, \hat N_{\rm q{+}\bar q}
    \right] ,
\label{eq:rescale3}
\end{align}
where
\begin{align}
    \hat H_{\rm NR} (\lambda;\mu)
    \equiv
    \left. \hat H \right|_{\vec A^a = 0}
    &- \sum_{{a, b=1}}^{N} \> \sumprime_{n\in\mathbb Z} \>
	M_{n}^{ab}(\mu) \, \hat N_{n}^{\,ab}
    - \sum_{a=1}^{N} \> \sum_{A=1}^{\nf} \sum_{n\in\mathbb Z+\frac 12} \>
	M_{n}^{aA}(\mu) \, \hat N_{n}^{\,aA}
\end{align}
is the non-relativistic Hamiltonian with all rest energies removed,
\begin{equation}
    \hat N_n^{aA}
    \equiv
    \int d^2x  \,
    \left[
    \psi_n^{\,ab}(\x)^\dagger \psi_{n}^{\,ab}(\x) +
    \chi_n^{\,ab}(\x)^\dagger \chi_{n}^{\,ab}(\x)
    \right]
\end{equation}
counts the number of quarks plus antiquarks of the specified type,
and the operator
$
    \hat N_{\rm q {+}\bar q}
    \equiv
    \sum_{A=1}^{\nf}
    \sum_{a=1}^N
    \sum_{n\in\mathbb Z+\frac 12}
    \hat N_n^{aA}
$
is the total number of quarks plus antiquarks.

\section {Symmetries}
\label{sec:symmetries}

As already noted, physical states in an $SU(N)$ gauge theory must be
gauge invariant.
In the compactified theory, this is trivially enforced dynamically:
gauge invariant states are those which do not have
divergent Coulomb energies.
This is equivalent to the just-stated condition (\ref{eq:N_B}) that
all $U(1)$ changes equal the baryon number,
$Q^a = N_B$.
To see this connection more explicitly, it may be helpful to note
that our effective $W$-boson fields, $\vec\phi_n^{\,ab}$, which were
 described earlier in a basis-dependent fashion as coming from
a specified row and column of the 4D gauge field
--- when the holonomy has the specific form (\ref{eq:Omega}) ---
could have been introduced in a manifestly basis-independent fashion by
first defining the operators
\begin{equation}
    \P_a
    \equiv
    \frac 1N \sum_{n=0}^{N-1}
    \omega^{-(a-\frac 12 (N{+}1))n} \> \Omega^n \,,
    \qquad \mbox{$a = 1,{\cdots},N$.}
\label{eq:p_a}
\end{equation}
The operators (\ref{eq:p_a})
are mutually orthogonal Hermitian projection operators,
$
    \P_a \, \P_b = \delta_{ab} \, \P_a
$,
when $\Omega$ lies at the center-symmetric minimum (\ref{eq:Omega})
and the eigenvalues of $\Omega$ are all $N$'th roots of $-1$ or $+1$.
Our effective 3D fields correspond to pieces of
the original 4D fields extracted by these projection operators,%
\footnote
    {%
    These are leading order relations.
    As with any effective field theory,
    field redefinitions and matching corrections complicate
    higher order relations between fields in the effective and original
    theories.
    }
\begin{equation}
    F^a_{\mu\nu} \propto
    \tr (\P_a F_{\mu\nu}) \,,\quad
    \vec \phi^{ab}_n \propto
    \P_a \vec D \, \P_b \,,\quad
    \psi^{aA}_n \propto \P_a \, q^A \,,\quad
    \chi^{aA}_n \propto \bar q^A \P_a \,,
\end{equation}
(neglecting details of the KK decomposition, spinor structure,
etc.).  This highlights the point that the Cartan gauge fields are
associated with manifestly gauge invariant 4D operators, while the
$W$-boson and quark fields are gauge covariant, as one would
expect.  With the aid of such expressions, it is easy
to see that composite operators in the 3D theory which map onto
manifestly gauge invariant 4D operators are precisely those
satisfying the condition $Q^a = N_B$.
As examples, the operators
\begin{subequations}\label{eq:examples}%
\begin{align}
    G^{ab}
    &\equiv
    \vec \phi_0^{\,ab} \cdot \vec \phi_0^{\,ba}
    \hspace{1.6cm}\sim
    \tr (D_i \, \P_b \, D_i \, \P_a)\,,
\\
    M^{a}_{AB}
    &\equiv
    \chi_{1/2}^{aA} \, \psi_{1/2}^{aB}
    \hspace{1.65cm}\sim
    \bar q^B \P_a \,  q^A \,,
\\
    B_A
    &\equiv
    \psi_{1/2}^{1,A} \,
    \psi_{1/2}^{2,A}
    \cdots
    \psi_{1/2}^{N,A}
    \sim
    (\P_1  q^A) \,
    (\P_2  q^A) \cdots
    (\P_N  q^A) \,,
\end{align}
\end{subequations}
(with no implied sums on Cartan indices,
and extraneous structure suppressed)
are prototypical glueball, meson, and baryon operators, respectively.

The global symmetries which are respected by our compactification
and under which eigenstates of the Hamiltonian may be classified
include the spacetime symmetries of 2+1 dimensional Minkowski space,
leading to conserved total 2D spatial momentum ($\vec P$)
and angular momentum ($J_z$).
States with vanishing $J_z$ may be further classified by their
behavior under 2D spatial reflections.%
\footnote
    {%
    Reflections are only a symmetry of the theory when $\theta = 0$
    (or $\pi$), but the violation of reflection symmetry induced
    by a non-zero $\theta$ only affects the long distance non-perturbative
    physics.
    For a more thorough discussion of the action of various symmetry
    transformations in the 3D effective theory,
    refer to Appendix \ref{app:symms}.
    }
Translation invariance in the compactified direction implies
conservation of the total compact momentum,
\begin{align}
    P_3 &\equiv
    \! \int \! d^2x \>
    \biggl\{
    \sum_{a,b=1}^N
    \sum_{n\in\mathbb Z}
    \mW (a-b+nN) \>
    (\vec\phi_n^{\,ab})^\dagger \vec\phi_n^{\,ab}
\nonumber\\ &\;{}
    +
    \sum_{a=1}^N
    \sum_{A=1}^{\nf}
    \sum_{n\in\mathbb Z+\frac 12}
    \mW\big( (a{-}\half)-\tfrac N{\nf}(A{-}\half) + nN\big)
    \>
    \big[
    (\psi_n^{\,aA})^\dagger \psi_n^{\,aA}
    -
    (\chi_n^{\,aA})^\dagger \chi_n^{\,aA}
    \big]
    \biggr\} .
\label{eq:P3}
\end{align}
As discussed earlier, our individual fields carry compact momentum
quantized in units of $\mW$ (for $\vec\phi^{\,ab}_n$)
or linear combinations of $\mW$ and $(N/\nf) \, \mW$
(for $\psi^{aA}_n$ and $\chi^{aA}_n$).
Physical glueball and flavor singlet mesons states 
must have total compact momentum equal to an integer
multiple of $2\pi/L = N \mW$, as these states remain invariant
when translated once around the compact dimension.
Due to our flavor-twisted boundary conditions for quarks,
flavor non-singlet mesons can have $P_3$ equal to
integer multiples of $2\pi/(\nf L)$.
The allowed values of $P_3$ for flavor singlet (non-singlet)
baryons are integer or half-integer multiples of $2\pi/L$
(or $2\pi/(\nf L)$) depending on whether $N$ is even or odd.

When quarks are present, the unbroken
$U(1)^{\nf}_V$ flavor symmetry transformations
are generated by the conserved flavor charges
\begin{equation}
    N^A \equiv
    \int d^2x
    \sum_{a=1}^N
    \sum_{n\in\mathbb Z+\frac 12}
    \left[
    (\psi^{aA}_n)^\dagger \psi_n^{aA}
    -
    (\chi^{aA}_n)^\dagger \chi_n^{aA}
    \right] .
\label{eq:N^A}
\end{equation}
The sum of these flavor charges equals the total number of quarks
minus antiquarks, or $N$ times the baryon number $N_B$.

Axial $U(1)^{\nf}_A$ flavor symmetry transformations
act as spin rotations on the EFT fermions and
are generated by the axial charges
\begin{equation}
    N^A_5 \equiv
    \int d^2x
    \sum_{a=1}^N
    \sum_{n\in\mathbb Z+\frac 12}
    \left[
    (\psi^{aA}_n)^\dagger \sigma_3 \, \psi_n^{aA}
    +
    (\chi^{aA}_n)^\dagger \sigma_3 \, \chi_n^{aA}
    \right] .
\label{eq:N5^A}
\end{equation}
The perturbative dynamics conserves these charges
but the long range non-perturbative dynamics violates conservation
of $\overline N_5 \equiv \sum_A N^A_5$
(and the non-perturbative vacuum is not annihilated by
the other axial charges).

In the absence of quarks, the compactified theory is
invariant under the $\mathbb Z_N$ center symmetry which,
by construction, remains unbroken.
The defining center symmetry transformation (\ref{eq:centersym})
multiplies the holonomy by an $N$'th
root of unity,
$
    \Omega \to \omega \, \Omega
$.
This permutes the projection operators (\ref{eq:p_a}),
$
    \P_a \to \P_{a-1}
$
(with $\P_0 \equiv \P_N$),
and also acts as a cyclic permutation on our 3D fields,
\begin{equation}
    \sigma^a \to \sigma^{a-1} \,,\quad
    \vec \phi_k^{\,a} \to \vec\phi_{k}^{\,a-1} \,.
\label{eq:ZNa}
\end{equation}
Here, Cartan indices are to be understood to be defined modulo $N$
(so $a{-}1\equiv N$ when $a\eq 1$).
Glueball operators such as
$
    G^a_k \equiv \vec\phi^{\,a}_k \cdot \vec\phi^{\,a-q}_{-k}
$
(with $k \bmod N \equiv q$)
are likewise cyclically permuted by
center symmetry transformations.
To diagonalize center symmetry,
one must perform a discrete $\mathbb Z_N$ Fourier transform and define,
for example,
\begin{equation}
    \tilde \sigma^p
    \equiv \frac{1}{\sqrt{N}} \sum_{a=1}^{N} \> \omega^{a p} \, \sigma^a \,,\quad
    \widetilde G^{p}_k
    \equiv \frac{1}{\sqrt{N}} \sum_{a=1}^{N} \> \omega^{a p} \, G^{a}_k \,.
\label{eq:diag-center}
\end{equation}
These operators now have definite center symmetry charge $p = 0,{\cdots},N{-}1$,
meaning that
under the center symmetry transformation (\ref{eq:centersym})
they transform into themselves multiplied by
the eigenvalue $\omega^p = e^{2\pi i p/N}$.

Adding fundamental representation quarks to the theory generally
breaks the $\mathbb Z_N$ center symmetry.
However, in the special case of $\nf \eq N$, the theory retains
an intertwined $\mathbb Z_N$ color-flavor center symmetry
(see, e.g., Refs.~\cite{Kouno:2012zz,Cherman:2017tey}).%
\footnote
    {%
    More generally,
    if $d \equiv \mathrm{gcd}(\nf,N) > 1$,
    then a $\mathbb Z_d$ color-flavor center symmetry remains
    \cite{Cherman:2017tey}.
    For simplicity, we will focus on the case of $\nf \eq N$.
    }
This symmetry combines the usual center transformation (\ref{eq:centersym})
with a cyclic permutation of quark flavors.
In terms of our 3D fields, this flavor-intertwined center symmetry acts as
\begin{equation}
    \sigma^a \to \sigma^{a-1} \,,\quad
    \vec \phi_k^{\,a} \to \vec\phi_{k}^{\,a-1} \,,\quad
    \psi_k^{a} \to \psi_{k}^{a-1} \,,\quad
    \chi_{k}^{a} \to \chi_{k}^{a-1} \,,
\label{eq:ZNb}
\end{equation}
and again may be diagonalized by a discrete $\mathbb Z_N$ Fourier transform.

Because the sets of eigenvalues (\ref{eq:Omega}) and (\ref{eq:OmegaF})
of the gauge holonomy $\Omega$ and our chosen flavor holonomy $\Omega_F$
are invariant under complex conjugation,
both charge conjugation and reflection of the compactified dimension
($x_3 \to -x_3$) remain symmetries of theory provided they are combined
with global gauge and flavor transformations which suitably permute
the Cartan and flavor indices.
The ordering (\ref{eq:Omega}) of the eigenvalues of the gauge holonomy was
chosen so that the required global gauge transformation $V$ is just a
permutation which flips Cartan indices, $a \to N{+}1 - a$, reflecting the
fact that
\begin{equation}
    \Omega^* = V \, \Omega \, V^\dagger \,,
\end{equation}
with $V \equiv \| \delta_{a+b,N+1} \|$ an anti-diagonal transposition.
Similarly, given the order (\ref{eq:OmegaF}) of the flavor holonomy eigenvalues,
the required flavor transformation $V_F$ also corresponds to a simple
flip of flavor indices, $A \to \nf{+}1 - A$, since
\begin{equation}
    \Omega_F^* = V_F \, \Omega_F \, V_F^\dagger \,,
\end{equation}
with $V_F \equiv \| \delta_{A+B,\nf+1} \|$.
This redefined charge conjugation symmetry acts on the fields of our
our dimensionally reduced EFT as
\begin{subequations}
\begin{align}
    \sigma^a &\to -\sigma^{\bar a} \,,
    &\psi^{aA}_n &\to \chi^{\bar a \bar A}_{-n} \,,
\\
    \vec\phi^{\,ab}_n &\to -\vec\phi^{\;\bar b \bar a}_{n} \,,
    &\chi^{aA}_n &\to \psi^{\bar a \bar A}_{-n} \,,
\end{align}
\end{subequations}
where $\bar a \equiv N{+}1-a$, $\bar A \equiv \nf{+}1-A$.%
\footnote
    {%
    The form of this transformation
    relies on our simplifying assumption that $N$ is odd,
    so that eigenvalues  of $\Omega$ are roots of $+1$ and
    $\Omega_F$ eigenvalues are roots of $-1$.
    If $N$ is even then
    both $\pm 1$ can be eigenvalues of the flavor holonomy $\Omega_F$
    for some values of $\nf \le N$.
    When two eigenvalues of $\Omega_F$ are real,
    the required flavor transformation $V_F$ which must be combined
    with the naive action of charge conjugation no longer corresponds
    to the simple flip $A \to \bar A$ of flavor indices.
    }
Note that center symmetry does not commute with charge conjugation.
In choosing a basis for degenerate levels of the Hamiltonian, one must
choose between specifying center symmetry charge, or the sign under
the (appropriately redefined) charge conjugation symmetry;
we will generally opt for the former.

Finally, reflection in the compact direction,
$x_3 \ to -x_3$, when combined with the same global gauge
and flavor transformations $V$ and $V_F$, remains a symmetry.
This redefined reflection symmetry acts on our 3D EFT fields as
\begin{subequations}
\begin{align}
    \sigma^a &\to \sigma^{\bar a} \,,
    &\psi^{aA}_n &\to  -i\sigma_2 \, \psi^{\bar a \bar A}_{-n} \,,
\\
    \vec\phi^{\,ab}_n &\to \vec\phi^{\,\bar a \bar b}_{-n} \,,
    &\chi^{aA}_n &\to i\sigma_2 \, \chi^{\bar a \bar A}_{-n} \,.
\end{align}
\end{subequations}
The combined symmetry of charge conjugation times $x_3$ reflection
does not involve any global gauge or flavor transformations and acts as
\begin{subequations}
\begin{align}
    \sigma^a &\to -\sigma^{a} \,,
    &\psi^{aA}_n &\to  i\sigma_2 \, \chi^{aA}_{n} \,,
\\
    \vec\phi^{\,ab}_n &\to -\vec\phi^{\,ba}_{-n} \,,
    &\chi^{aA}_n &\to -i\sigma_2 \, \psi^{aA}_{n} \,.
\end{align}
\end{subequations}
This is the same as a CP transformation times a 180$^\circ$
rotation in the uncompactified directions.

\section {Heavy sector spectrum}
\label{sec:bound_states}

\subsection {Overview}

Three basic types of bound states can be formed
from the constituents of our non-relativistic effective theory:
glueballs, mesons, and baryons.
Here, ``bound state'' means either
a genuine single particle eigenstate of the full theory,
or a narrow resonance whose fractional decay width vanishes
in the $L \to 0$ (and correspondingly $\lambda\to 0$) limit.
In this section, we neglect the coupling to the
Abelian gauge fields contained in the spatial covariant derivatives,
as well as higher dimension operators not shown explicitly in
our effective theories (\ref{eq:Sheavy}) and (\ref{eq:Squark}).
Effects of these terms are discussed in Sec.~\ref{sec:decays}
which discusses decay processes.

By glueballs we mean bound states of two or more charged $W$-bosons,
and no quarks or antiquarks.
Mesons are, of course, bound states of a quark and antiquark,
possibly containing additional $W$-bosons,
while baryons are bound states of $N$ quarks
(perhaps with additional charged $W$-bosons).
In our weakly coupled small-$L$ regime, mixing between glueballs and
flavor singlet mesons is suppressed, so they are clearly distinguishable.
Manifestly gauge invariant interpolating operators for
simple examples of such states were shown in Eq.~\eqref{eq:examples}.
Further possibilities, which we will not focus on in this paper,
include multi-meson or multi-glueball ``molecules''
and multi-baryon bound states.

\begin{figure}[t]
\begin{center}
\vspace*{-3.5cm}
\includegraphics[scale=0.5]{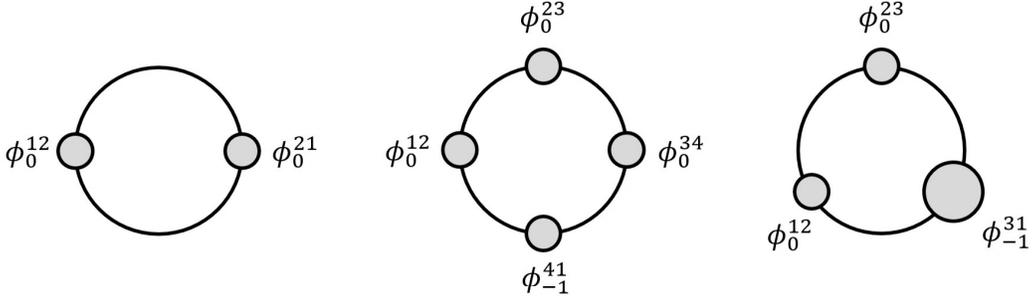}
\vspace*{-4cm}
\end{center}
\caption
    {%
    Examples of glueball states when $N \eq 4$.
    Filled circles represent the charged $W$-bosons,
    with larger circles indicating more massive constituents.
    Lines connecting the constituents indicate attractive
    logarithmic interactions (of relative strength 1).
    \label{fig:glueballs}
    }
\end{figure}

As discussed above, all physical (gauge invariant) states must satisfy
$Q^a = N_B$.
Hence, glueballs and mesons must be composed of combinations
of constituents for which all $U(1)^N$ charges sum to zero.
The simplest glueballs are two-body bound states of a $W$-boson
and its oppositely charged antiparticle, created by operators
such as
\begin{equation}
    (\vec \phi_0^{\,ab})^\dagger \cdot (\vec\phi_0^{\,ba})^\dagger \,,
\end{equation}
with $a \ne b$.
Two different $U(1)$ gauge group factors contribute to the
logarithmic interaction between these constituents,
giving an attractive interaction of relative strength 2.
The explicit two-body Hamiltonian, and its spectrum, is examined
in Sec.~\ref{sec:glueballs} below.
Bound states of more than two $W$-bosons can also form.
States of this type which cannot be decomposed into two or more
separately gauge invariant glueballs consist of $W$-bosons whose
charge assignments lead to a ring-like color structure with nearest-neighbor
logarithmic interactions.
Examples of operators creating such states are
\begin{equation}
    (\phi_0^{\,ab})^\dagger_i
    (\phi_0^{\,bc})^\dagger_j
    (\phi_0^{\,ca})^\dagger_k \,,
    \quad
    (\phi_0^{\,ab})^\dagger_i
    (\phi_0^{\,bc})^\dagger_j
    (\phi_0^{\,cd})^\dagger_k
    (\phi_0^{\,da})^\dagger_l \,,
\label{eq:gluerings}
\end{equation}
etc., with up to $N$ constituents and Cartan indices
$a,b,c,{\cdots}$ all distinct.
We will refer to these as ``closed string'' glueballs.
These are all single trace operators when
expressed in terms of the original 4D fields (as in Eq.~(\ref{eq:examples})).
In these multi-body states,
a single $U(1)$ factor generates an attractive
logarithmic interaction (of relative strength 1)
between each pair of neighboring constituents in the cyclic list.
This is illustrated schematically in Fig.~\ref{fig:glueballs}.  We note that there is an amusing similarity between these states and the picture advocated long ago in Ref.~\cite{Thorn:1978kf}.

\begin{figure}
\begin{center}
\vspace*{-4.5cm}
\includegraphics[scale=0.5]{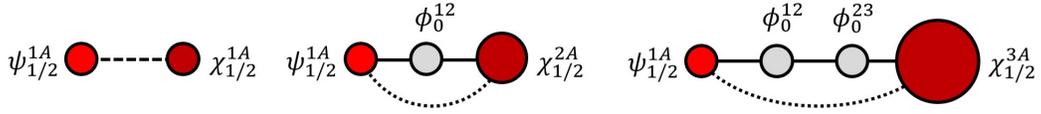}
\vspace*{-5cm}
\end{center}
\caption
    {%
    Examples of meson states (with $N \ge 3$).
    Filled circles represent the charged constituents.
    Solid lines connecting constituents indicate attractive
    logarithmic interactions of relative strength 1,
    dashed lines represent attractive interactions of strength
    $1{-}\frac 1N$, and dotted lines represent repulsive logarithmic
    interactions of strength $1/N$.
    \label{fig:mesons}
    }
\end{figure}

\begin{figure}
\begin{center}
\vspace*{-3.5cm}
\includegraphics[scale=0.5]{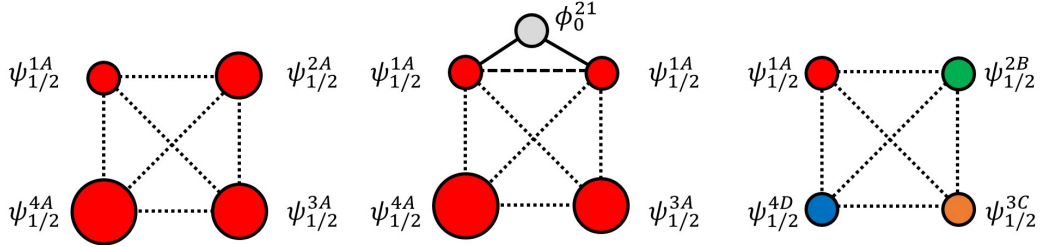}
\vspace*{-3.5cm}
\end{center}
\caption
    {%
    Examples of baryon states when $N \eq 4$.
    Filled circles represent the charged constituents,
    with larger circles indicating more massive constituents.
    Dotted lines represent attractive logarithmic
    interactions of strength $1/N$,
    dashed lines represent repulsive interactions of
    strength $1{-}\frac 1N$, and solid lines show
    attractive interactions of strength 1.
    In the single flavor example (left),
    each quark constituent has a different mass due to their differing
    Cartan indices.
    The multi-flavor example (right) shows the special case with
    $\nf \eq 4$ where all constituents have equal mass.\label{fig:baryons}
    }
\end{figure}

The situation with mesons is similar.
The simplest mesons are two-body bound states,
created by operators such as
\begin{equation}
    (\chi_{1/2}^{aA})^\dagger (\psi_{1/2}^{aB})^\dagger \,.
\end{equation}
The attractive logarithmic interaction between the quark and antiquark
has relative strength of $(1{-}\frac 1N)$,
with the reduction from 1 coming from the subtraction of the unwanted
``extra'' $U(1)$ contribution in the Coulomb interaction (\ref{eq:SCoulomb}).
There are also mesons in which one or more additional
$W$-bosons are present.
States of this type which cannot be decomposed into meson-glueball products
have charge assignments implying an ``open string'' color structure.
Examples of operators creating such states include
\begin{equation}
    (\chi_{1/2}^{aA})^\dagger (\phi_0^{ab})^\dagger_i (\psi_{1/2}^{bB})^\dagger \,,
    \quad
    (\chi_{1/2}^{aA})^\dagger
    (\phi_0^{ab})^\dagger_i
    (\phi_0^{bc})^\dagger_j
    (\psi_{1/2}^{cB})^\dagger \,,
\end{equation}
etc, with up to $N{-}1$ $W$-bosons inserted between the quark and antiquark
and Cartan indices $a,b,c,{\cdots}$ all distinct.
There are attractive logarithmic interactions of relative strength 1
between each pair of neighboring constituents,
along with a repulsive logarithmic
interaction of strength $1/N$ between the quark and antiquark
(with differing Cartan charges).
This is illustrated schematically in Fig.~\ref{fig:mesons}.

Finally, baryons containing $N$ quarks,
potentially with additional $W$-bosons as well,
are present as finite energy bound states because
our gauge group is $SU(N)$, not $U(N)$.
The simplest non-exotic baryons are created by operators like
\begin{equation}
    (\psi_{1/2}^{1,A})^\dagger
    (\psi_{1/2}^{2,B})^\dagger
    (\psi_{1/2}^{3,C})^\dagger
    \cdots
    (\psi_{1/2}^{N,Z})^\dagger \,.
\end{equation}
In such states, every pair of quarks has an attractive logarithmic interaction
of relative strength $1/N$.
Two such baryon states,
as well as a baryon state containing an additional $W$-boson,
are illustrated schematically in Fig.~\ref{fig:baryons}.

The stability of these various hadronic states will depend on
their relative energy differences and the resulting radiative
transition and short distance annihilation rates.
These are discussed below in Sec.~\ref{sec:decays}.

\subsection {Two-body states}\label{sec:2body}

Neglecting couplings to the spatial Abelian gauge fields
(which are relevant for radiative decays but not the leading order spectrum),
the dynamics of all two-body sectors of our effective theory \eqref{eq:HNR},
namely glueballs composed of oppositely charged $W$-bosons,
quark-antiquark mesons, and diquark baryons in the special case
of $N \eq 2$,
are described by a common first-quantized two-dimensional
non-relativistic Hamiltonian,
\begin{equation}\label{eq:pair}
    \hat H
    =
    \frac{\p_1^2}{2m_1} + \frac{\p_2^2}{2m_2}
    + \kappa \, \ln(\mu|\x_1{-}\x_2|) \,,
\end{equation}
with a logarithmic potential
and positive interaction strength, $\kappa>0$.
Before discussing our specific application to glueball, meson,
and $N\eq2$ baryons in compactified QCD,
we first summarize properties of the spectrum
of this quantum theory.

\subsubsection {2D logarithmic QM}

Starting with the two particle Hamiltonian \eqref{eq:pair},
separating the center of mass motion
and working in the center-of-mass frame
leads to a one-body Hamiltonian for the relative motion,
\begin{equation}
    \hat H_{\rm relative}
    =
    \frac{\p^2}{2m}
    + \kappa \, \ln(\mu|\x|) \,,
\label{eq:Hrelative}
\end{equation}
where $m \equiv m_1 m_2 / (m_1+m_2)$ is the reduced mass.
Non-relativistic dimensional analysis (with $\hbar \equiv 1$)
shows that $\kappa m/\mu^2$ is the only dimensionless combination
of parameters appearing in the Hamiltonian \eqref{eq:Hrelative},
so its eigenvalues must have the form
$E = \kappa \, f(\kappa m/\mu^2)$ for some univariate function $f$.
The manifestly trivial $\mu$ dependence,
$\partial E/\partial \mu = \kappa/\mu$,
then implies that the energy eigenvalues of $\hat H_{\rm relative}$
are given by
\begin{equation}
\label{eq:logE}
    E = \kappa \Big[ \epsilon - \half \ln \frac{\kappa m}{\mu^2} \Big] ,
\end{equation}
where $\epsilon$ is an eigenvalue of the theory with
$\kappa = m = \mu \equiv 1$.
Introducing a dimensionless radial variable
$r = \sqrt{\kappa m}\, |\mathbf x|$,
eigenstates with orbital angular momentum
$L_z \equiv \ell = 0, \pm1, \pm2, \cdots$
satisfy the one-dimensional radial Schr\"odinger equation,
\begin{equation}
    \left[ -\half \frac {d^2}{dr^2} + V_\ell(r) \right] \chi(r)
    =
    \epsilon \, \chi(r) \,,
\label{eq:radial}
\end{equation}
with effective radial potential
\begin{equation}
    V_\ell(r) \equiv \frac {\ell^2{-}\tfrac 14}{2r^2} + \ln r \,.
\label{eq:V_l}
\end{equation}

\begin{table}[tp]
\begin{center}
\footnotesize
\vspace*{-0.5cm}
\hspace*{-1.0cm}%
\begin{tabular}{c|c|c|c|c|c|c|c|c|c|c}
$n$ & $|\ell|=0$ & 1 & 2 & 3 & 4 & 5 & 6 & 7 & 8 & 9
\tabularnewline
\hline
$0^{\strut}$
& 0.179935 & 1.03961 & 1.49780 & 1.81127 & 2.04971 & 2.24214 & 2.40348 & 2.54238 & 2.66432 & 2.77301
\tabularnewline
$1$
& 1.31468 & 1.66290 & 1.92929 & 2.14154 & 2.31731 & 2.46710 & 2.59753 & 2.71299 & 2.81656 & 2.91044
\tabularnewline
$2$
& 1.83061 & 2.04777 & 2.23348 & 2.39248 & 2.53070 & 2.65265 & 2.76163 & 2.86008 & 2.94982 & 3.03224
\tabularnewline
$3$
& 2.16887 & 2.32609 & 2.46790 & 2.59439 & 2.70781 & 2.81028 & 2.90360 & 2.98920 & 3.06819 & 3.14152
\tabularnewline
$4$
& 2.42105 & 2.54403 & 2.65839 & 2.76311 & 2.85901 & 2.94717 & 3.02859 & 3.10416 & 3.17460 & 3.24054
\tabularnewline
$5$
& 2.62222 & 2.72309 & 2.81873 & 2.90790 & 2.99083 & 3.06805 & 3.14015 & 3.20770 & 3.27118 & 3.33102
\tabularnewline
$6$
& 2.78959 & 2.87502 & 2.95712 & 3.03466 & 3.10761 & 3.17622 & 3.24085 & 3.30185 & 3.35957 & 3.41429
\tabularnewline
$7$
& 2.93290 & 3.00696 & 3.07882 & 3.14735 & 3.21239 & 3.27407 & 3.33257 & 3.38814 & 3.44101 & 3.49138
\tabularnewline
$8$
& 3.05822 & 3.12356 & 3.18740 & 3.24875 & 3.30740 & 3.36337 & 3.41677 & 3.46776 & 3.51650 & 3.56314
\tabularnewline
$9$
& 3.16956 & 3.22799 & 3.28541 & 3.34092 & 3.39428 & 3.44548 & 3.49457 & 3.54165 & 3.58684 & 3.63024
\end{tabular}%
\hspace*{-1.0cm}
\end{center}
\caption
    {%
    The first ten eigenvalues $\epsilon_{n,\ell}$
    of the radial Schr\"odinger equation
    (\ref{eq:radial}),
    for $|\ell| = 0, 1, {\cdots}, 9$.
    All digits shown are accurate.
    \label{tab:Energy-levels-of}
    }
\end{table}

\begin{figure}[tp]
\begin{center}
\vspace*{-1cm}%
\includegraphics[scale=0.55]{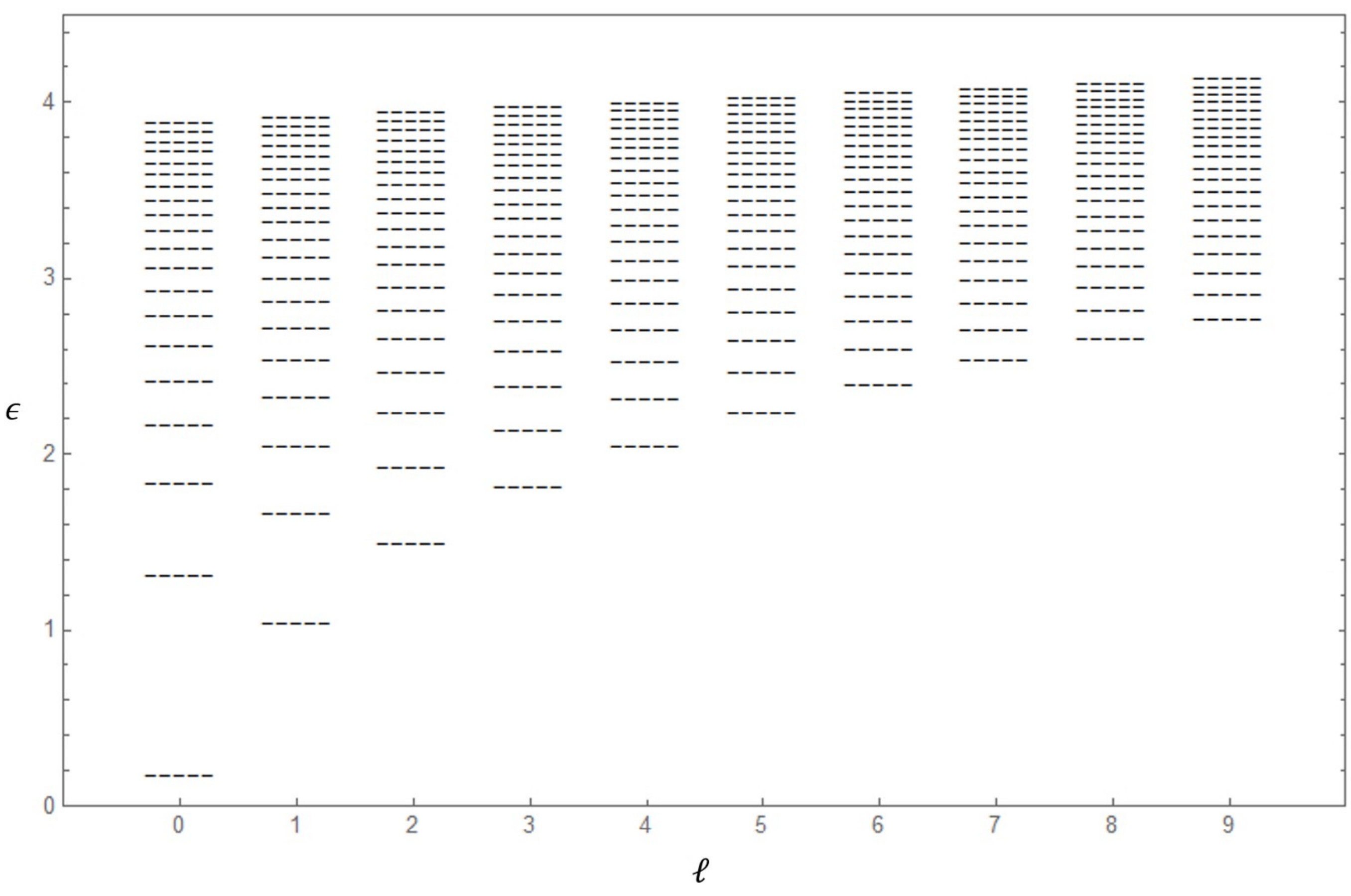}
\vspace*{-1.0cm}
\end{center}
\caption
    {%
    Energy spectrum of the radial Schr\"odinger equation (\ref{eq:radial}).
    \label{fig:Energy-spectrum-of 2-body}
    }
\end{figure}

Solutions to the Schr\"odinger equation (\ref{eq:radial})
are not expressible in terms of familiar special functions.
The equation was analyzed numerically over 40 years ago
\cite{PhysRevA.9.2617}
(see also Refs.~\cite{Rumer:1960,Quigg:1979vr}),
but we will present our own more accurate and extensive results.
Calculations of low-lying energy levels are fairly
straightforward using variational methods and a suitable basis set,
or alternatively using pseudo-spectral methods \cite{boyd2001chebyshev} with a Gauss-Laguerre
grid for the semi-infinite radial domain.%
\footnote
    {%
    A simple choice of basis for a variational
    calculation consists of 2D harmonic oscillator eigenstates
    with definite angular momentum $\ell$.
    Given a suitable adjustment of the scale of the harmonic oscillator
    basis functions, a truncated basis of 40 harmonic oscillator states
    is sufficient to find the lowest energy level
    of the logarithmic Hamiltonian (\ref{eq:Hrelative})
    to an accuracy of a few parts in $10^4$.
    However,
    pseudo-spectral discretization using a Laguerre
    grid turns out to provide significantly better accuracy for a
    given basis size.
    (This is because harmonic oscillator wavefunctions with their
    Gaussian envelope decrease too rapidly at large $r$;
    as discussed below eigenstate wavefunctions in a logarithmic
    potential decrease much more slowly.)
    To obtain the eigenvalues shown in Table \ref{tab:Energy-levels-of}
    and compute transition matrix elements for radiative decays,
    discussed in Sec.~\ref{sec:decays}, we used Gauss-Laguerre
    grids with 100--200 points.
    To avoid excessive precision loss in the evaluation of the spectral
    differentiation matrices and the resulting eigenvalue computation,
    we used extended precision arithmetic with slightly over twice
    as many digits as the number of grid points.
    \label{fn:numerics}
    }
The first ten levels, for each $|\ell| = 0, {\cdots}, 9$,
are listed in Table~\ref{tab:Energy-levels-of}.
The spectrum is shown graphically
in Fig.~\ref{fig:Energy-spectrum-of 2-body}.
Notice that levels at neighboring values of $\ell$ are interleaved,
$
    \epsilon_{n,|\ell|} < \epsilon_{n,|\ell|+1} < \epsilon_{n+1,|\ell|}
$.

As $|\ell|$ increases, the minimum of the potential moves
to larger values, with $r_{\rm min} \sim |\ell| + \O(\ell^{-2})$.
When $|\ell| \gg 1$, a quadratic approximation to the potential
is sufficient to find low-lying states.
For fixed level number $n$ (starting from 0),
\begin{equation}
    \epsilon_{n,\ell}
    =
    \ln (|\ell|) + \tfrac 12
    + \frac{2n{+}1}{\sqrt 2 \, |\ell|}
    + \O (\ell^{-2}) \,.
\label{eq:2bodylargel}
\end{equation}

Standard WKB methods may be used
to study more highly excited states.
When the energy $\epsilon$ is large compared to $\max(1,\ln |\ell|)$,
the classically allowed region
of the Schr\"odinger equation (\ref{eq:radial})
extends out to a turning point at
$r_* \equiv \exp(\epsilon)$.
For $r > r_*$, the WKB solution which decays as $r \to \infty$ is
\begin{equation}
    f_{\rm I}(r)
    =
    \big[\!\ln(r)/\epsilon{-}1\big]^{-1/4} \>
    \exp\big[ -\sqrt{2\epsilon} \, |Q_0(r)| + \O(\epsilon^{-1/2}) \big] ,
\label{eq:WKBI}
\end{equation}
where
\begin{equation}
    Q_0(r) \equiv \int_r^{r_*} dr' \> \sqrt{1{-}\ln(r')/\epsilon} \,.
\end{equation}
The usual Airy function matching across the turning point
(or analytic continuation around the turning point) shows that
this solution matches onto the allowed region WKB solution
\begin{equation}
    f_{\rm II}(r)
    =
    [1-\ln(r)/\epsilon]^{-1/4} \>
    \cos \big[
	\sqrt{2\epsilon} \, Q_0(r) - \tfrac \pi4 + \O(\epsilon^{-1/2})
	\big] .
\label{eq:WKBII}
\end{equation}
This WKB approximation is valid down to $r = \O(1)$, where
\begin{equation}
    f_{\rm II}(r)
    \sim
    \cos \big[
	\sqrt {2\epsilon} \, r - I(\epsilon)
	+ \tfrac \pi4 + \O(\epsilon^{-1/2})
	\big] \times (1 + \O(\epsilon^{-1})) \,,
\end{equation}
with
\begin{equation}
    I(\epsilon) \equiv \sqrt{2\epsilon} \, Q_0(0)
    = \sqrt{\tfrac \pi 2} \, \exp(\epsilon) \,.
\label{eq:I}
\end{equation}
For parametrically small values of $r$,
the centrifugal term in the potential cannot be
neglected but the logarithmic term is subdominant.
In this region, the appropriate solution satisfying
regularity at the origin is
\begin{equation}
    f_{{\rm III}}(r)
    = (\half\epsilon)^{1/4} \sqrt{\pi r} \> J_{|\ell|}(\sqrt {2\epsilon} \, r) \,.
\end{equation}
When $r \gg \epsilon^{-1/2}$,
$
    f_{{\rm III}}(r) \sim
    \cos(\sqrt{2\epsilon}\, r - \frac \pi 2 \, |\ell| - \frac \pi 4)
    + \O\big((\sqrt\epsilon r)^{-1}\big)
$.
For $\O(1)$ values of $r$, this matches onto the
the classically allowed WKB solution (\ref{eq:WKBII}) provided
\begin{equation}
    I(\epsilon) = \half (2n {+} |\ell| {+} 1) \, \pi + \O(\epsilon^{-1/2})\,,
\label{eq:WKBmatch}
\end{equation}
for some integer $n$.
Inserting the result (\ref{eq:I}), one finds that
eigenvalues $\epsilon_{n,\ell}$ of the radial
Schr\"odinger equation (\ref{eq:radial}) are given by
\begin{equation}
    \epsilon_{n,\ell}
    =
    \ln ( 2n {+} |\ell| {+} 1 ) + \half \ln \tfrac \pi 2 \,,
\label{eq:wkb energy}
\end{equation}
up to corrections vanishing faster than $\O(1/n)$.
One may verify that $n$ equals the number of nodes in this solution,
so $n$ is level number when counting from 0.

Numerically, the accuracy of the WKB approximation 
(\ref{eq:wkb energy}) to energy levels
is surprisingly good for modest values of the level number $n$.
For $\ell \eq 0$ and $n \eq 10$,
the difference between our numerical and WKB results is less than
2 parts in $10^4$.
The relative deviation grows with increasing $\ell$ at fixed $n$,
reaching 2\% for $\ell \eq n \eq 10$.

The WKB result (\ref{eq:wkb energy}) shows that the level spacing
(at fixed $\ell$)
decreases with increasing level number,
$d\epsilon/dn = 2/(2n{+}|\ell|{+}1)$.
Inverting this relation, one finds that
the asymptotic density of states with fixed orbital angular momentum $\ell$
rises exponentially with energy,
\begin{equation}
    \frac{\partial n_\ell}{\partial \epsilon}
    \sim
    \frac {e^\epsilon} {\sqrt{2\pi}} \,.
\label{eq:hagedorn2}
\end{equation}
(This neglects any spin degeneracy of the constituents.)
The integral of this density of states gives
the total number of quantum states, with fixed $\ell$,
below a given energy, and asymptotically equals
the area of the classically allowed region
in phase space (in units of $2\pi \hbar$),
\begin{align}
    n_\ell(\epsilon)
    &=
	\int \frac {dp}{2\pi} \, dr \>
	\Theta(\epsilon {-} \half p^2 {-} V_\ell(r))
    =
	\frac{\sqrt{2}}{\pi}
	\int_{r_{\rm min}}^{r_{\rm max}} dr \>
	\sqrt{\epsilon {-} V_\ell(r)}
\nonumber
\\ &
    =
	\frac{\sqrt{2}}{\pi}
	\left[
	\int_0^{e^\epsilon} dr \>
	\sqrt{\epsilon {-} \ln r}
	\right]
	+ \O(|\ell|{-}\half)
    =
	\frac{e^\epsilon}{\sqrt{2\pi}} + \O(|\ell|{-}\half) \,.
\label{eq:nl}
\end{align}
The total number of states
below energy $\epsilon$
(with vanishing total momentum, but no projection onto definite $\ell$),
$N(\epsilon) = \sum_\ell n_\ell(\epsilon)$,
coincides asymptotically with
the classically allowed phase space volume
of the 2D relative dynamics.
This grows exponentially at twice the rate
of the fixed-$\ell$ result,
\begin{equation}
    N(\epsilon)
    =
	\int \frac {d^2p}{(2\pi)^2} \, d^2r \>
	\Theta(\epsilon {-} \half p^2 {-}\ln r)
    =
	\int_0^{e^\epsilon} r \, dr \> (\epsilon {-} \ln r)
    =
	\tfrac 14 \, e^{2\epsilon} \,.
\label{eq:N}
\end{equation}
This exponential growth is a direct consequence of the
slow increase of the confining logarithmic potential with distance.
Bound states spread over rapidly growing
spatial regions as their energy increases.
The exponential behavior (\ref{eq:nl}) of the fixed-$\ell$ number
of states is nothing but linear dependence on the turning point
radius $r_*$, while the total number of states (\ref{eq:N}) is,
up to a factor of $1/(4\pi)$, just the spatial area of the allowed region,
$\pi r_*^2$.

\subsubsection{Glueballs\label{sec:glueballs}}

For every pair of oppositely charged $W$-bosons there is a manifold
of bound states described by the two-body logarithmic interaction Hamiltonian
(\ref{eq:pair}) with interaction strength
\begin{equation}
    \kappa = \frac{\lambda \mW}{2\pi^2} \,.
\label{eq:kappaW}
\end{equation}
This is analogous to the ro-vibrational states associated with each
electronic level in molecular spectroscopy.
For a pair of $W$-bosons with compact momentum indices $k$ and $k'$
(defined by the relation (\ref{eq:k}) and satisfying the constraint
$k+k'= 0 \bmod N$ so that the $W$-bosons have opposite Cartan charges),
the resulting bound state energies are given by
\begin{equation}
    E_{WW} = M_k(\mW) + M_{k'}(\mW)
    + \frac{\lambda \mW}{2\pi^2}
    \Bigl(
	\epsilon_{n,\ell}
	-\half \ln \frac {\lambda \, m_{kk'}}{2\pi^2\mW}
    \Bigr) ,
\label{eq:2bodyglue}
\end{equation}
where the reduced mass
$m_{kk'} \equiv m_k \, m_{k'} /(m_k {+} m_{k'})$,
and we have chosen to set the arbitrary scale $\mu$ equal to $\mW$.
The lightest glueballs are composed of $W$-bosons with one unit of
compact momentum, $|k| = |k'| = 1$,
and tree-level constituent mass $\mW$,
leading to glueball energies
\begin{equation}
    E = 2 M_1(\mW)
    + \frac{\lambda \mW}{2\pi^2}
    \Bigl(
	\epsilon_{n,\ell}
	-\half \ln \frac {\lambda}{4\pi^2}
    \Bigr) .
\label{eq:lightglue}
\end{equation}

Neglecting higher order relativistic corrections, as well as
non-perturbative physics on the scale of $m_\gamma$,
two-body glueball states have a degeneracy of $4N$
if they are $\ell = 0$ and CP self-conjugate.
(Center symmetry gives a factor of $N$, and
there is a spin degeneracy of 4 since each
massive $W$-boson has two spin states.)
There is an additional factor of 2 degeneracy for
states with non-zero orbital angular momentum
(corresponding to positive and negative values of $\ell$,
which are exchanged by 2D spatial reflections),
and a separate additional factor of 2 degeneracy for
states which are not CP self-conjugate.
The lightest glueball level (\ref{eq:lightglue})
contains CP self-conjugate $\ell=0$ states,
and hence has the minimal degeneracy of $4N$.

Relativistic corrections to the above results contribute
$\O(\lambda^2 \mW)$ energy shifts,
or relative $\O(\lambda)$ corrections to binding energies.
Spin-orbit corrections give an energy shift proportional to
$\ell \, S_z$
(where $S_z \equiv s^{(1)}_z + s^{(2)}_z$),
with a positive coefficient.
In our dimensionally reduced effective theory,
spin-spin (or hyperfine) interactions are local and
proportional to $s_z^{(1)} s_z^{(2)} \, \delta^2 (\mathbf x)$,
also with a positive coefficient.%
\footnote
    {%
    In two spatial dimensions, spin-spin interactions do not have a
    long range dipolar form since the magnetic field produced by
    a current loop is localized inside the loop.
    }
This spin-spin correction only has a non-zero expectation value in
$\ell = 0$ states.
Hence, first order relativistic corrections produce an energy shift
of the form
\begin{equation}
    \Delta E_{\rm fine\mbox{-}structure}
    =
    \lambda^2 \mW
    \left[ A \, \ell \, S_z + B \, \delta_\ell^0 \, (S_z^2 {-} 2)
    \right],
\end{equation}
where $A$ and $B$ are positive $\O(1)$ coefficients
(depending on $n$ and $|\ell|$).
For a given $n$ and $\ell \ne 0$,
the spin-orbit correction splits the four possible spin states,
$\{
    \mid\uparrow\uparrow\rangle$, $
    \mid\uparrow\downarrow{\pm}\downarrow\uparrow\rangle$, $
    \mid\downarrow\downarrow\rangle
\}$,
into three sublevels
with the $S_z = -2 \ell/|\ell|$ state moving lower in energy,
the $S_z = +2 \ell/|\ell|$ state moving higher, and
the two $S_z = 0$ states unchanged.
For $\ell = 0$ levels,
the spin-spin interaction produces two sublevels, with
the energy of the $S_z = \pm 2$ states shifted upward,
and the $S_z = 0$ states downward.
The degeneracy between the spin symmetric and antisymmetric
$S_z = 0$ states,
$
    \mid\uparrow\downarrow{\pm}\downarrow\uparrow\rangle
$,
is not lifted by these leading relativistic corrections,
but should be removed at higher orders.

Short distance effects will also induce higher order
corrections to the rest and kinetic masses,
leading to further spin-independent $\O(\lambda^2\mW)$ energy shifts.
Operators producing $\O(\lambda^2\mW)$ corrections are listed
in Appendix \ref{app:NReft},
which discusses the relevant power counting rules.
The structure of higher dimensional operators that appear
in our non-relativistic EFT follow the same pattern known, for example,
from studies of hydrogenic spectra or heavy quark physics in QCD
\cite{Lepage:1992tx},
but quantitative evaluation of these higher order effects
is left to future work.

The factor of $N$ degeneracy associated with center symmetry would be
lifted by the non-perturbative long distance physics on the scale of
$m_\gamma$ but, more importantly,
this degeneracy is first lifted by one loop perturbative
corrections which generate photon mixing terms
(mentioned earlier in footnote \ref{fn:mixing}).
Such mixing arises from vacuum polarization corrections
which are sensitive to the differing masses $M_n^{ab}$ of the
charged virtual $W$-bosons.
This mixing (when rediagonalized) induces $\O(\lambda)$
variations in the coupling strengths of different light photons.
Eigenstates of bound $W$-bosons will have definite center charge and
are constructed by a $\mathbb Z_N$ Fourier transform,
as in Eq.~(\ref{eq:diag-center}).
The energies of states with differing values of center charge will be
split by $\O(\lambda^2 \mW)$, or in other words additional
$\O(\lambda)$ relative corrections to binding energies.

\subsubsection{Mesons}\label{sec:mesons}

Differences between the two-body meson and glueball
spectra arise from the differing constituent masses and
the strength of the logarithmic interaction.
For an oppositely charged quark-antiquark pair, the
interaction strength is given by
\begin{equation}
    \kappa = (1{-}\tfrac 1N)\, \frac{\lambda\mW}{4\pi^2} \,.
\label{eq:kappaq}
\end{equation}
The allowed values of compact momentum (\ref{eq:quark_p3}) depend on both $N$
and $\nf$.
As mentioned earlier, a particularly simple case which we will focus on
is $\nf = N$.
For this number of flavors the
tree-level constituent quark masses (\ref{eq:quarkmasses}) become
half-integers times $\mW$,
\begin{equation}
    M_n^{aA} = M_k \equiv \mW (|k| + \O(\lambda))\,, \qquad
    m_n^{aA} = m_k \equiv \mW (|k| + \O(\lambda))\,,
\end{equation}
with $k = a - A + nN$ and $n \in \mathbb Z+\half$.
The resulting bound state energies are given by
\begin{equation}
    E_{\bar q q} = M_k(\mW) + M_{k'}(\mW)
    + (1{-}\tfrac 1N) \frac{\lambda \mW}{4\pi^2}
    \Bigl(
	\epsilon_{n,\ell}
	-\half \ln \frac {(1{-}\frac 1N)\lambda \, m_{kk'}}{4\pi^2\mW}
    \Bigr) ,
\label{eq:mesonmass}
\end{equation}
where, once again, $m_{kk'}$ is the reduced mass.
The lightest mesons have $|k|=|k'| = \half$, leading to
\begin{equation}
    E_{\bar q q} = 2M_{1/2}(\mW)
    + (1{-}\tfrac 1N) \frac{\lambda \mW}{4\pi^2}
    \Bigl(
	\epsilon_{n,\ell}
	-\half \ln \frac {(1{-}\frac 1N)\lambda}{16\pi^2}
    \Bigr) .
\label{eq:lightestmeson}
\end{equation}
Neglecting higher order relativistic corrections,
the lightest two-body meson levels (\ref{eq:lightestmeson})
have a degeneracy of $16N$ if they have $\ell \eq 0$, with
an additional factor of 2 if $\ell \ne 0$.
(Four factors of 2 coming from the choice of spin for quark
and antiquark, plus the choice of sign of each momentum index,
and a factor of $N$ from one choice of flavor, or equivalently
from the choice of which $U(1)$ photon provides the binding.)
Higher order spin-orbit, spin-spin and other radiative effects
partially lift this degeneracy in the same manner discussed above
for glueballs.


\subsubsection{$N\,{=}\,2$ baryons}

Finally, in the special case of two-color QCD,
the simplest baryons are bound states of two quarks
(with no additional $W$-bosons).
The interaction strength
$\kappa$ equals $\frac 1N \lambda \mW/(4\pi^2)$ which, for $N\,{=}\,2$,
coincides with the quark-antiquark interaction strength.
Consequently, the resulting diquark baryon spectrum is identical to
the meson spectrum (\ref{eq:mesonmass}) and (\ref{eq:lightestmeson})
given above, when specialized to $N\,{=}\,2$.
The degeneracy of the lightest baryon levels
(neglecting relativistic corrections)
is $16$ for $\ell = 0$ states, with an additional factor
of two for $\ell \ne 0$.

\subsection {Multi-body states}

\subsubsection{Glueballs}\label{sec:multiglue}

As noted in the overview,
in addition to two-body $W$-boson bound states,
multi-body bound states containing three or more $W$-bosons
with a ring-like color structure can also form,
such as those illustrated in Fig.~\ref{fig:glueballs}.
The spectrum of such ``closed string'' states is quite rich.

The rest mass of $W$-bosons is given by Eq.~(\ref{eq:W-masses}),
reproduced here for convenience,
\begin{equation}
    M_{n}^{ab}
    = M_k
    \equiv
    \mW |k|
    =
    \mW \,|a-b+nN|,
\end{equation}
up to $\O(\lambda \mW)$ corrections.
To form a physical (gauge invariant) bound state, the $U(1)^N$ Cartan charges
of all $W$-bosons in the bound state must sum to zero.
For closed-string glueball states
which are not decomposable into multiple separate glueballs,
this means that each neighboring pair of $W$'s in the ring is bound together by
a distinct Abelian gauge interaction.
Bound states containing $3 \le P \le N $ constituents having
compact momentum indices
$\{ k_1,\, k_2,\,{\cdots}, k_P \}$ exist,
consistent with this constraint, provided that
\begin{equation}
    \sum_{i=1}^P k_i = 0 \bmod N \,.
\label{eq:ring-constraint}
\end{equation}
For this state to be non-decomposable, no partial sum of the
momentum indices should vanish modulo $N$.
In addition to specifying the momentum index of each constituent,
one may specify one Cartan index of a single constituent;
together this information completely determines the Cartan and KK indices
of all constituents around the cycle.
The tree-level mass of such a closed string state is just
\begin{equation}
    M_{\rm tot} = \mW \sum_{i=1}^P |k_i| \,.
\label{eq:Mtree}
\end{equation}

\medskip
\noindent\textbf{``Near extremal'' states}:
An interesting subset of states are those with non-zero compact momentum
$P_3$ and whose tree-level mass equals the minimal value consistent with
this compact momentum,
\begin{equation}
    M = |P_3| \,.
\label{eq:nearextermal}
\end{equation}
This implies that the momentum indices of all constituents have the same sign.
One simple case, satisfying the constraint (\ref{eq:ring-constraint})
(plus non-decomposability),
are ``pearl necklace'' bound states containing $N$ $W$-bosons,
all with momentum indices equal to unity, $k_i = 1$,
or all equal to minus one, $k_i = -1$.
For these states
$P_3 = \mW \sum_i k_i = \pm N \, \mW = \pm 2\pi/L$
and the (tree level) rest mass $M = |P_3| = N \mW$.
The middle example in Fig.~\ref{fig:glueballs} illustrates this type of
pearl necklace state (with $P_3 = -2\pi/L$) in the case of $N \eq 4$.
Such a state 
is created by the $N$-body operator
\begin{equation}
    A^{i_1 i_2 \cdots i_N} \,
    (\phi_{-1}^{\, 1})^\dagger_{i_1} \,
    (\phi_{-1}^{\, 2})^\dagger_{i_2} \cdots
    (\phi_{-1}^{\, N-1})^\dagger_{i_{N-1}} \,
    (\phi_{-1}^{\,N})^\dagger_{i_N} \,,
\label{eq:AN}
\end{equation}
where the coefficients $\{ A^{i_1 \cdots i_N} \}$
(defining a rank-$N$ 2D spatial tensor) determine the spin wavefunction.

There are also near-extremal states with fewer constituents.
One can imagine fusing together any neighboring pair of constituents in the
operator (\ref{eq:AN}) and replacing them with a single $W$-boson having
the same Cartan charges and compact momentum as the pair.
Or doing the same fusing process with a neighboring triplets of constituents,
etc.
The resulting states are also near-extremal, and are created by
$N{-}1$ or $N{-}2$ body operators such as
\begin{subequations}%
\begin{align}
    &A^{i_1 i_2 \cdots i_{N-1}} \,
    (\phi_{-1}^{\, 1})^\dagger_{i_1} \,
    (\phi_{-1}^{\, 2})^\dagger_{i_2} \cdots
    (\phi_{-2}^{\, N-1})^\dagger_{i_{N-1}} \,,
\\
\noalign{\hbox{or}}
    &A^{i_1 i_2 \cdots i_{N-2}} \,
    (\phi_{-1}^{\, 1})^\dagger_{i_1} \,
    (\phi_{-1}^{\, 2})^\dagger_{i_2} \cdots
    (\phi_{-3}^{\,N-2})^\dagger_{i_{N-2}} \,.
\end{align}
\end{subequations}
Continuation of this fusing process leads to near-extremal states with any number
of constituents from $N$ down to 1.
Three and two body examples are
\begin{subequations}%
\begin{align}
    &A^{i_1 i_2 i_3} \,
    (\phi_{-1}^{\, 1})^\dagger_{i_1} \,
    (\phi_{-1}^{\, 2})^\dagger_{i_2} \,
    (\phi_{-(N-2)}^{\,3})^\dagger_{i_3} \,,
\\
\noalign{\hbox{and}}
    &A^{i_1 i_2} \,
    (\phi_{-1}^{\, 1})^\dagger_{i_1} \,
    (\phi_{-(N-1)}^{\, 2})^\dagger_{i_2} \,,
\end{align}
\end{subequations}
while the endpoint of this process is a neutral ``heavy photon'' state
created by a one-body operator such as
\begin{equation}
    A^{i} \, (\phi_{-N}^{\,1})^\dagger_{i} \,.
\label{eq:A1}
\end{equation}
More generally,
ignoring spin and center degeneracies
there are $\binom{N}{P{-}\delta^P_1}$
distinct categories of near-extremal states
containing $P$ constituents associated with different contiguous
fusing of the fields in the $N$-body operator (\ref{eq:AN}),
or altogether $2^N{-}N$ types of non-decomposable near-extremal states
having the same value of $P_3 = \pm N \mW$.

\medskip
\noindent\textbf{``Non-extremal'' states}:
Bound states containing constituents with oppositely signed momentum indices
are ``non-extremal.''
Such states have rest masses which exceed their compact momentum,
$M > |P_3|$, by an $\O(\mW)$ amount or more.
This includes all bound states of $W$-bosons having vanishing total
compact momentum, $P_3 = 0$,
such as the lightest glueballs (\ref{eq:lightglue}).

\medskip
\noindent\textbf{Binding energies}:
Calculating the $\O(\lambda \mW)$ binding energies of multi-body glueball states
requires one to find eigenvalues of the first-quantized Hamiltonian which describes
the sector of the theory (\ref{eq:HNR}) with the chosen number of constituents.
For ``closed string'' bound states composed of $P \le N$ $W$-bosons, this is
\begin{equation}
    \hat H
    =
    \sum_{i=1}^P
    \left[
	\frac {\mathbf p_i^2}{2m_i}
	+ \frac {\lambda \mW}{4\pi^2} \ln (\mu |\mathbf x_i{-}\mathbf x_{i-1}|)
    \right] ,
\end{equation}
with the understanding that $\mathbf x_0 \equiv \mathbf x_P$.
The scaling relation (\ref{eq:rescale2}) allows one to remove the dependence
on $\lambda$, but eigenvalues will be non-trivial functions of constituent
mass ratios,
\begin{equation}
    E_{\rm binding}
    =
    \frac{P\lambda\mW}{8\pi^2}
    \left[
	f(\{m_i/m_j\}) - \ln (\lambda \widetilde m \mW /\mu^2)
    \right] ,
\label{eq:gluebinding}
\end{equation}
where $f$ is a dimensionless $\O(1)$ function (depending on the chosen
energy level as well as mass ratios),
and $\widetilde m$ is the harmonic mean of the constituent masses.

For modest values of $P$ (three or four),
an accurate variational calculation should be feasible despite the fact that
computational effort will rise as a rather high power of the number
of single particle states included in the truncated basis.
We leave such calculations to future work.

An interesting limiting case partially amenable to analytic analysis
concerns low-lying states with large orbital angular momentum, $\ell \gg 1$,
and constituents all having the same mass $m$.
Such states include rotating ``pearl necklace'' configurations in which
each constituent contributes equally to the total orbital angular momentum.
A semiclassical analysis of such states is straightforward.
The classical Hamiltonian (for fixed $\ell$)
has a local minimum in which the constituents
lie at the vertices of a regular $P$-sided polygon whose circumscribed
circle has radius $r = 2 \pi \ell/(P \sqrt{\lambda m \mW})$,
rotating at angular velocity
$\Omega = \ell/(P \, m r^2) = P \lambda\mW / (4 \pi^2 \ell)$.
Semiclassical quantization of vibrations about this configuration leads to
energy levels whose binding energies (ignoring center of mass motion) are
given by
\begin{equation}
    E_{\rm binding}
    = \frac{P \lambda \mW}{4\pi^2}
    \left[
	\half +
	\ln \Big(\frac{4 \pi \ell \mu \sin(\pi/P)}{P \sqrt{\lambda m \mW}}\Big)
    \right]
    + \sum_{i=-(P-2)}^{P-2} (n_i {+} \half) \, \omega_{|i|}
    + \O(\ell^{-2}) \,,
\label{eq:large-lglue}
\end{equation}
where the $P{-}1$ vibrational frequencies $\{\omega_i\}$
are $\O(\lambda\mW/\ell)$.%
\footnote
    {%
    One mode, here labeled $i=0$, is a uniform ``breathing'' mode
    with $ \omega_0 = \sqrt 2 \, \Omega $.
    All other modes (present only for $P > 2$)
    are higher frequency doubly-degenerate asymmetric stretching modes.
    For $P \eq 2$, the form (\ref{eq:large-lglue}) agrees as it must
    with the prior results (\ref{eq:2bodyglue}) and (\ref{eq:2bodylargel}).
    }

The result (\ref{eq:large-lglue}) grows logarithmically with increasing
angular momentum $\ell$, with a coefficient of $P \lambda \mW/4\pi^2$
proportional to the number of constituents.
This linear increase with $P$ implies that these semiclassical
``pearl necklace'' states are not the minimal energy states with a
given large orbital angular momentum.
``Core-halo'' states will exist in which $P{-}1$ constituents are clumped
together in a region of size $\sqrt {P/\lambda m \mW}$ while a single
constituent circles at a distance of order $\O(\ell/\sqrt{\lambda m\mW})$
and contributes (nearly) all the orbital angular momentum.
The binding energy of such states will increase with $\ell$ just like
the two-body case, namely
$
    E_{\rm binding} \sim (\lambda \mW/2\pi^2) \, \ln \ell
$
as $\ell \to \infty$.
Computing the sub-dominant $\ell$-independent contribution coming from the
core wavefunction requires a full quantum calculation.

\subsubsection{Mesons}

Largely identical considerations apply to multi-body mesons.
Focusing, once again, on the case of $\nf \eq N$,
bound states containing a quark and antiquark having
half-integer compact momentum indices $k_q$ and $k_{\bar q}$, plus
$P$ $W$-bosons with momentum indices $\{k_1,{\cdots},k_P\}$,
will have total compact momentum
\begin{equation}
    P_3 = \mW \big(k_q - k_{\bar q} + \sum_{i=1}^P k_i \big) \,.
\end{equation}
For the state not to be decomposable into a glueball-meson molecule,
no partial sum of the $W$-boson momentum indices should vanish modulo $N$.
With tree-level mass $M_{\rm tot} = \mW (|k_q| + |k_{\bar q}| + \sum_i |k_i|)$,
it is immediate that $M_{\rm tot} \ge |P_3|$.
Any of the multi-body ``closed string'' glueball states discussed above
may be converted into an ``open string'' meson state by replacing any
one of the $W$-boson constituents by a $q \bar q$ pair collectively
having the same Cartan charges and compact momentum.
As an example, one analogue of the near-extremal $N$-body glueball operator
(\ref{eq:AN}) is the near-extremal meson operator
\begin{equation}
    B^{s_{\bar q} s_q \, i_1 i_2 \cdots i_{N-1}} \,
    (\chi_{+1/2}^{\,1})^\dagger_{s_{\bar q}}
    (\phi_{-1}^{\, 1})^\dagger_{i_1} \,
    (\phi_{-1}^{\, 2})^\dagger_{i_2} \cdots
    (\phi_{-1}^{\, N-1})^\dagger_{i_{N-1}} \,
    (\psi_{-1/2}^{\,N})^\dagger_{s_q} \,,
\label{eq:multibodymeson}
\end{equation}
(with $s_q$ and $s_{\bar q}$ denoting two-component spinor indices of the quark
and antiquark, respectively),
in which $N{-}1$ $W$-bosons are inserted between the quark and antiquark.

The $\O(\lambda \mW)$ binding energies of (non-decomposable)
multi-body meson states containing
$P$ $W$-bosons are given by eigenvalues of the first-quantized Hamiltonian
\begin{align}
    \hat H
    &=
    \sum_{i=0}^{P+1} \frac {\mathbf p_i^2}{2m_i}
    +
    \frac {\lambda \mW}{4\pi^2}
    \Bigl[
	- \tfrac 1N \ln (\mu |\mathbf x_0{-}\mathbf x_{P+1}|)
	+ \sum_{i=1}^{P+1} \ln (\mu |\mathbf x_i{-}\mathbf x_{i-1}|)
    \Bigr] ,
\end{align}
where $\mathbf x_0 \equiv \mathbf x_{\bar q}$ and
$\mathbf x_{P+1} \equiv \mathbf x_q$ refer to the antiquark and quark,
respectively,
and likewise for the momenta $\mathbf p_0$ and $\mathbf p_{P+1}$
and masses $m_0 \equiv m_{\bar q}$ and $m_{P+1} \equiv m_q$.
The resulting energy levels have the form
\begin{equation}
    E_{\rm binding}
    =
    (P+(1{-}\tfrac 1N)) \, \frac{\lambda\mW}{8\pi^2}
    \left[
	f(\{m_i/m_j\}) - \ln (\lambda \widetilde m \mW /\mu^2)
    \right] ,
\end{equation}
with $f$ some $\O(1)$ function,
differing from the glueball case (\ref{eq:gluebinding}) just in the prefactor.

Just as with closed-string glueballs, it is interesting to consider
open-string mesons with large orbital angular momentum, $\ell \gg 1$.
Among such states are semiclassical ``rotating wire'' states.
The classical Hamiltonian (for fixed orbital angular momentum $\ell$)
has local minima in which all constituents are arrayed along a straight
line which rotates uniformly with some angular velocity $\omega$,
with the positions of constituents along this line adjusted so that
the sum of forces (falling with inverse separation)
acting on each constituent provides the required centripetal acceleration,
and the common angular velocity $\omega$ is suitably adjusted to yield the
chosen angular momentum $\ell$.
Solving for this minimum analytically, for arbitrary $P$, is not easy,
but a numerical determination for chosen values of $P$ is straightforward.
Semiclassical quantization of such a stationary configuration will lead
to energy levels which, as in the glueball case (\ref{eq:large-lglue}),
grow logarithmically with increasing $\ell$, with a coefficient
which increases with the number of constituents.
Hence, for the same reasons discussed above,
lower energy ``core-halo'' mesonic states will exist in which all but
one constituent are clumped together and collectively carry little or no
angular momentum while a single constituent (which may be either a quark
or a $W$-boson) circles the core at a large $\O(\ell/\sqrt{\lambda m \mW})$
distance and carries (nearly) all the orbital angular momentum.


\subsubsection{Baryons}\label{sec:baryons}

Baryonic bound states containing quarks with no additional $W$-bosons
(``non-exotic baryons'')
may be formed from a collection of $N$ quarks,
each having a distinct color (Cartan) index.
Focusing, once again, on the case of $\nf = N$,
the momentum indices $\left\{ k_{1},{\cdots}, k_{N}\right\} $ of the quarks
are arbitrary half-integers
(with $k_i$ the momentum index of the quark with Cartan index $i$).
The total compact momentum $P_3=\mW \sum_{i}k_{i}$
and the tree-level mass $M_{\text{tot}}=\mW\sum_{i=1}^N|k_{i}|$.

Note that, for large values of $N$, baryons which are
composed of the lightest quark constituents with
$\O(1)$ momentum indices
will have a total mass $M_{\rm tot}$ which scales linearly with $N$.
Such baryons contain quarks of (nearly) all $N$ different flavors.
Baryons which are solely composed of quarks of a single
flavor will have a total mass which is at least $\O(N^2)$,
because the momentum indices of quarks must, in this case,
all be distinct and hence will, at a minimum,
have magnitudes ranging from $\half$ up to $\lfloor N/2 \rfloor$.

The strength of the attractive logarithmic interaction between two quarks
of differing colors is ${1}/{N}$, so the
first-quantized non-relativistic Hamiltonian for non-exotic baryons is
\begin{equation}
    \hat{H}
    =
    \sum_{i=1}^{N} \frac{\p_{i}^{2}}{2m_{i}}
    +
    \frac 1N \sum_{i<j=1}^{N}
	\frac{\lambda m_{W}}{4\pi^{2}} \,
	\ln\left(\mu|\x_{i}{-}\x_{j}|\right) ,
\label{eq:bar ham}
\end{equation}
with $m_i = \mW |k_i|$ the $i$'th constituent quark mass.

For the lightest class of baryons, each quark has
momentum index $\pm \half$ and the minimal
constituent mass $m_i = m_q \equiv \half \mW$.
Such states are created by operators of the form
\begin{equation}
    C^{s_{1}s_{2}\cdots s_{N}}
	\left(\psi_{\pm 1/2}^{1}\right)_{s_{1}}^{\dagger}
	\left(\psi_{\pm 1/2}^{2}\right)_{s_{1}}^{\dagger}
	\cdots
	\left(\psi_{\pm 1/2}^{N}\right)_{s_{N}}^{\dagger} \,,
\end{equation}
with $s_{i}$ denoting the two-component spinor index of the $i$th
quark.
(Fig.~\ref{fig:baryons} illustrates one such state for $N\eq4$.)
For simplicity of presentation, we will
focus our discussion on this lightest class of baryons.

For baryons with equal mass constituents,
the Hamiltonian (\ref{eq:bar ham}) is completely symmetric
under permutations of constituents.
The rescaling relation (\ref{eq:rescale3}) implies that
\begin{equation}
    \hat{H}
    \cong
    \frac{\lambda m_{W}}{4\pi^{2}}
    \biggl(
	\half \sum_{i=1}^{N} {\p_{i}^{2}}
	+
	\frac 1{2N} \sum_{i\ne j=1}^{N}
	\ln|\x_{i}{-}\x_{j}|
    \biggr)
    -
    (N{-}1) \,  \frac{\lambda m_{W}}{16\pi^{2}} \,
    \ln \biggl(\frac{\lambda \mW m_q}{4\pi^2 \mu^2}\biggr) \,.
\label{eq:bar ham2}
\end{equation}
The spectrum of this Hamiltonian was already discussed
in Sec.~\ref{sec:2body} in the special case of $N \eq 2$.
We now examine the opposite extreme, $N \gg 1$.

As discussed by Witten \cite{Witten:1979kh},
a Hartree approximation to the many-body wavefunction
is asymptotically accurate as $N \to \infty$.
The appropriate $N$-body Hartree wavefunction for the
ground state is just a product of identical one-body wavefunctions,
\begin{equation}
    \Psi(\x_{1},{\cdots},\x_{N})
    =
    \prod_{i=1}^{N} \psi(\x_{i}) \,,
\label{eq:Hartree}
\end{equation}
with the one-body wavefunction $\psi(\x)$
determined by minimizing the expectation value of
the Hamiltonian (subject to the normalization constraint
$\int d^2x \> |\psi(\x)|^2 = 1$).%
\footnote
    {%
    A better approximation would project this state onto
    vanishing center-of-mass momentum.
    However, such projection only affects $\O(1)$ contributions
    to the total energy of the state, which we neglect.
    }
The resulting ground state baryonic mass grows linearly with $N$
and is given by
\begin{equation}
    E_{\rm baryon}/N
    =
    M_{1/2}(\mW)
    + \frac {\lambda\mW}{4\pi^2}
	\Bigl(
	    \bar\epsilon
	    - \tfrac 14 \ln \frac{\lambda m_q}{4\pi^2\mW}
	\Bigr)
    + \O(1/N) \,,
\end{equation}
where
\begin{equation}
    \bar\epsilon
    \equiv
    \min_{\psi}
    \epsilon[\psi] \,,\qquad
    \epsilon[\psi]
    =
    \mathcal T[\psi] + \mathcal V[\psi] \,.
\end{equation}
Here,
\begin{subequations}\label{eq:TV}%
\begin{align}
    \mathcal T[\psi]
    &\equiv
    \half \int d^2\x \> |\nabla\psi(\x)|^2
    \Big/ \mathcal N[\psi] \,,
\label{eq:T}
\\
    \mathcal V[\psi]
    &\equiv
    \half \int d^2\x \, d^2\x' \>
    \ln|\x{-}\x'| \, |\psi(\x)|^2 \, |\psi(\x')|^2
    \Big/ \mathcal N[\psi]^2 \,,
\label{eq:V}
\end{align}
\end{subequations}
with
$
    \mathcal N[\psi] \equiv \int d^2\x \> |\psi(\x)|^2
$.
The ground state wavefunction which minimizes $\epsilon[\psi]$
satisfies the Hartree equation,
\begin{equation}
    \left[ -\half \nabla^2 + U(\x) \right] \psi(\x) = \lambda \, \psi(\x) \,,
\label{eq:Hartree eq}
\end{equation}
with the self-consistent potential
\begin{equation}
    U(\x) \equiv \int d^2\x' \> \ln|\x{-}\x'| \, |\psi(\x')|^2
    \Big/ \mathcal N[\psi] \,.
\label{eq:U}
\end{equation}
This wavefunction is guaranteed to be nodeless,
and hence is spherically symmetric, $\psi(\x) = \psi(|\x|)$.
After angular averaging of the logarithm, the potential (\ref{eq:U})
becomes a convolution with the radial Green's function,
\begin{equation}
    U(|\x|) \equiv
    \int_0^\infty r' \, dr' \> \ln \!\big(\! \max(|\x|,r')\big) \> \psi(r')^2
    \Big/
    \int_0^\infty r' \, dr' \> \psi(r')^2  \,.
\label{eq:U2}
\end{equation}

We minimize the functional $\epsilon[\psi]$ numerically,
using pseudospectral methods \cite{boyd2001chebyshev}.
We write
$
    \psi(r) = e^{-\mu r/2} f(r)
$
and then represent the function $f$ as an order $M{-}1$ polynomial
determined by its values $\{ f_k \}$ on the Gauss-Laguerre grid points
$\{ r_k \}$ which are the roots of the Laguerre polynomial $L_M(\mu r)$.
This is equivalent to, but much more computationally convenient than
using the coefficients $\{ c_k \}$ in the orthogonal polynomial  expansion
$
    f(r) = \sum_{k=0}^{M-1} c_k \, L_k(\mu r)
$.
The radial integrals in expressions (\ref{eq:TV})--(\ref{eq:U2}) are evaluated
using $M$-point Gauss-Laguerre quadrature.
Radial derivatives become dense $M\times M$ matrices acting on the
$M$-component vector $\vec f \equiv ( f_k )$, and the Hartree equation
(\ref{eq:Hartree eq}) becomes an $M$-dimensional linear eigenvalue equation.
Starting with a simple pure exponential initial guess for $\psi(r)$,
we compute the Hartree potential (\ref{eq:U2}),
solve for the lowest eigenvalue of the Hartree equation
(\ref{eq:Hartree eq}), and iterate
these two steps until convergence.%
\footnote
    {%
    Demanding stationarity of $\epsilon[\psi]$ under a rescaling
    $\psi(\x) \to \xi \psi(\xi \x)$ at $\xi = 1$ shows that
    $\mathcal T[\psi] = \tfrac 14$ at extrema of $\epsilon$.
    This is the analogue of the usual virial theorem for our
    logarithmic potential.
    Choosing the scale $\mu = \sqrt 2$ in our spectral representation
    gives our initial guess this correct value of $\mathcal T$.
    }

Due to the non-analyticity in the Green's function (\ref{eq:U2}), the
truncation error only falls with increasing basis size as $\O(1/M)$.
Six points suffice for 5\% accuracy,
thirty points yield better than 1\%,
and several hundred are needed to achieve 0.1\% accuracy.
For large $M$, the spectral matrices become quite ill-conditioned and
extended precision arithmetic with roughly $2M$ digits is needed to
avoid precision loss.
A very stable extrapolation in $1/M$ yields the result,
\begin{equation}
    \bar\epsilon
    =
    0.449558 \,.
\end{equation}

The degeneracy of this lightest baryon level, before taking into
account splittings due to higher order radiative corrections,
is $4^N$, growing exponentially as $N$ increases.
(For each quark, there is one factor of two for the choice of spin and
another factor of two from the compact momentum $k = \pm \half$.)

To compare our $N \,{=}\, 2$ and $N {\gg} 1$ results for ground state baryons
in a coupling independent fashion,
consider the binding energy scaled by $N{-}1$,
with the exactly known $\lambda \ln\lambda$ contribution removed,
\begin{equation}
    \delta E_{\rm binding}(N)
    \equiv
    \frac 1{N{-}1}
    \left[ E_{\rm baryon} - N M_{1/2}(\mW) \right]
    + \frac {\lambda \mW}{16\pi^2} \ln \frac {\lambda}{8\pi^2} \,.
\end{equation}
Our results,
\begin{equation}
    \frac
	{\delta E_{\rm binding}(\infty)}
	{\delta E_{\rm binding}(2)}
    =
    \frac{\bar \epsilon}{\half (\epsilon_{00} + \ln 2)}
    =
    1.0298 \,,
\label{eq:baryonbinding}
\end{equation}
show stunningly little dependence on $N$.
It would be interesting to see if this near-constancy
is a coincidence, or remains true for other values of $N$.

\begin{figure}
\begin{centering}
\includegraphics[scale=0.65]{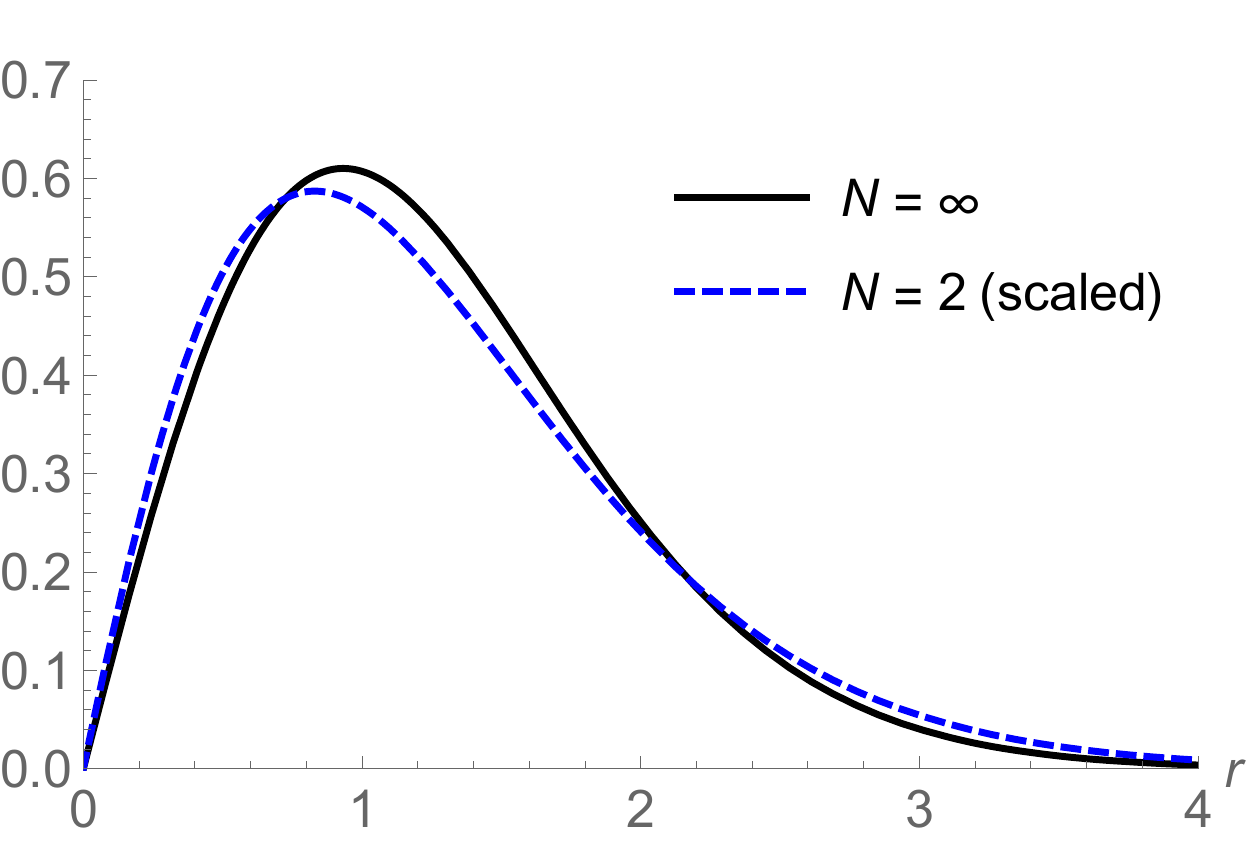}
\par\end{centering}
\caption
    {%
    Single particle radial probability density $|\x|\,p(\x)$ of
    ground state baryons at $N \,{=}\, \infty$
    as a function of $r \equiv \sqrt\lambda \, \mW |\x|/(4\pi)$
    (solid curve)
    overlaid with the corresponding density for $N\,{=}\,2$ baryons
    as a function of
    $r' \equiv \sqrt{2\lambda} \, \mW |\x|/(4\pi)$
    (dashed curve).
    \label{fig:baryon pdf}
    }
\end{figure}

At large $N$, the probability density to find a quark at position $\x$
relative to the baryon center of mass equals the square of the Hartree
single particle wavefunction, $p(\x) = |\psi(\x)|^2$.
To compare this with the corresponding distribution in $N\,{=}\,2$
ground state baryons,
recall that the Hamiltonian for relative motion (\ref{eq:Hrelative})
was expressed in terms of the separation between constituents,
so the corresponding distribution relative to the center of mass is
$p(\x) = 4|\psi_{\rm rel}(2\x)|^2$.
One finds that the single particle distribution is more
highly concentrated at $N\,{=}\,2$ than at $N\,{=}\,\infty$.
The mean square deviations differ by just about a factor of two,
\begin{equation}
    \langle \x^2 \rangle
    =
    \frac {8\pi^2}{\lambda \mW^2} \,
    \times
    \begin{cases}
	1.0907 \,, & N = 2;
    \\
	2.0294 \,, & N = \infty.
    \end{cases}
\end{equation}
Fig.~\ref{fig:baryon pdf} compares the $N\,{=}\,\infty$
single particle radial probability density $|\x| \, p(\x)$ with the
corresponding $N\,{=}\,2$ distribution when distance is rescaled by a
factor of $\sqrt 2$, that is
$
    \half |\x| \, p(\x/\sqrt 2)
$.
As one sees from the figure, with this rescaling the two radial
distributions are very similar.

Above the baryon ground state level there is a manifold
of vibrationally excited baryon levels.
For $N \gg 1$, energy levels in which a small number of
quarks are excited may be computed using a product
wavefunction with a few of the factors in
the ground state wavefunction (\ref{eq:Hartree}) replaced by
excited single particle wavefunctions.
Low lying levels with a single excited quark may be labeled
by the number of radial nodes $n$ and orbital angular momentum $\ell$
of the excited quark, and have excitation energies
\begin{equation}
    \Delta E_{n,\ell}
    =
    \frac{\lambda \mW}{4\pi^2}
	\left( \lambda_{n,\ell} - \lambda_{0,0} \right) ,
\end{equation}
where $\lambda_{n,\ell}$ is an eigenvalue of the Hartree equation
(\ref{eq:Hartree eq}) containing the mean field generated by
all the unexcited quarks.
The subtraction of $\lambda_{0,0}$ accounts for the decrease
in the number of quarks in the lowest single particle level.
Table~\ref{tab:excited baryons} lists the eigenvalues
$\lambda_{n,\ell}$ for the lowest few levels.
Excitation energies to baryon levels with multiple excited quarks are,
up to $1/N$ corrections, just the sum of the individual excitation
energies (provided the number of excited quarks is a negligible
fraction of $N$).

\begin{table}
\begin{center}
\begin{tabular}{c|c|c|c|c}
$n$ & $|\ell|=0$ & 1 & 2 & 3
\\
\hline
0
& 0.64911 & 1.1367 & 1.5182 & 1.8152
\\ 1
& 1.4448 & 1.7124 & 1.9450 & 2.1457
\\ 2
& 1.9018 & 2.0805 & 2.2458 & 2.3964
\\ 3
& 2.2169 & 2.3503 & 2.4780 & 2.5979
\\ 4
& 2.4569 & 2.5632 & 2.6669 & 2.7663
\end{tabular}
\end{center}
\caption
    {%
    Eigenvalues $\lambda_{n,\ell}$ of the Hartree equation
    (\ref{eq:Hartree eq}),
    with the self-consistent potential for the lowest baryon level,
    for indicated values of the radial quantum number $n$
    and orbital angular momentum $\ell$.
    \label{tab:excited baryons}
    }
\end{table}

Lastly, in the same manner discussed above for mesons,
it is also possible to form exotic baryons containing $N$ quarks plus
one or more $W$-bosons.
For the bound state to be non-decomposable into baryon-glueball molecules,
no partial sum of the $W$-boson momentum indices should vanish.
Such states can be progressively built from non-exotic baryons by
replacing a quark
with a quark plus one or more $W$-boson(s) which collectively have
the same Cartan charge and compact momentum as the removed quark.
One example of such a state is shown in Fig. \ref{fig:baryons}.
By suitably repeating this process one may, for example,
build baryons in which all $N$ quarks have the same color
while $N{-}1$ $W$-bosons mediate attractive interactions between
these quarks.

\section{Decay processes}
\label{sec:decays}

Higher order perturbative interactions turn most of the
hadronic states discussed in the previous sections into narrow resonances.
Examining the systematics of the various decay processes is our next topic.
First, however, we detail those states which \emph{cannot} decay.

\subsection{Stable states}
\label{sec:stable states}

In the light sector of the quarkless theory,
individual dual photons are exactly stable.
Each dual photon has a non-zero center charge $p = 1,{\cdots},N{-}1$,
and is the lightest state with that value of center charge.%
\footnote
    {%
    Recall that a $p \,{=}\, 0$ dual photon was artificially added
    to the light sector effective theory (\ref{eq:Slight}) to
    simplify the presentation, but this extra degree of freedom
    exactly decouples from all physical degrees of freedom.
    The physical particles of the $SU(N)$ gauge theory
    do not include a $p\,{=}\,0$ dual photon.
    }
To see this, note that the mass formula \eqref{massspectrum}
is a subadditive function of the center charge,
$m_{p_1+p_2} < m_{p_1} + m_{p_2}$.
This implies that any splitting of a dual photon into two or more
photons with the same total center charge is kinematically
forbidden.
The formation of $k$-body light sector bound states discussed in
Sec.~\ref{sec:NR_EFT_light} does not affect this conclusion,
as the $k$-body binding energies are exponentially small
compared to the relevant differences in photon masses.
The two-body bound state of dual photons with center charges
$1$ and $N{-}1$,
whose binding energy is given by Eq.~(\ref{eq:B2rel}),
is the lightest center charge zero excitation
and is likewise exactly stable.

If $\theta = 0$ then the theory is CP invariant.%
\footnote
    {%
    This paragraph assumes that $N \ge 3$.
    Because $SU(2)$ is pseudo-real, charge conjugation
    is a distinct symmetry in $SU(N)$ pure YM theory only for $N > 2$.
    }
Individual dual photons are CP odd.
The lightest CP even states with non-zero center charge $p$ are
bound states of two dual photons with charges $q$ and $p{-}q$
and minimal total mass $M_p = \min_q (m_q + m_{p-q})$.
Specifically, these are the $(q,p{-}q)$ bound states with
\begin{align}
    q =
    \begin{cases}
    1 \,, & \mbox{for } p = 2,{\cdots},\lfloor \frac N2 \rfloor,
    			\mbox{ or $p = N{-}1$};
    \\
    N{-}1 \,, & \mbox{for } p = \lfloor \frac {N+1}{2} \rfloor
			    ,{\cdots},N{-}2,
    			\mbox{ or $p = 1$}.
    \end{cases}
\label{eq:stables}
\end{align}
Similarly,
the lightest CP odd state with vanishing center charge is a
bound state of three dual photons with charges $( 1,1,N{-}2 )$
(or their conjugates).
These bound states are necessarily stable at $\theta = 0$.
Moreover, the charged two particle bound states (\ref{eq:stables})
remain absolutely stable at $\theta \ne 0$
for purely kinematic reasons.
These bound states are heavier than a single dual photon of the
same total center charge, but are lighter than all other multiparticle
bound states of the given charge, and hence have no allowed decay channels
which can conserve both energy and momentum.

Turning now to the theory with quarks,
as discussed in Sec.~\ref{sec:quarks}
with $\nf \le N$ massless quark flavors,
$\nf{-}1$ of the dual photons become exactly massless
and are the Goldstone bosons of spontaneously broken
$U(1)^{\nf-1}_A$ symmetry.
When $\nf \,{=}\, N$, this means
all $N{-}1$ dual photons are massless.
These massless Goldstone bosons are stable.

In the heavy sector, exactly stable states are those
protected by conservation of the $U(1)^{\nf}_V$  flavor charges
(\ref{eq:N^A})
and/or compact momentum (\ref{eq:P3}).
With $\nf \,{=}\, N$, mesons composed of a quark and antiquark
having the minimal mass, $m_q = m_{\bar q} = \half \mW$,
and opposite compact momentum indices, $k_q = -k_{\bar q} = \pm \half$,
have flavor charges $(+1,-1)$ under two different $U(1)$ flavor subgroups
and non-vanishing total compact momentum $P_3 = \pm \mW$.
Such mesons (with vanishing vibrational and rotational excitations)
are the lightest states with these flavor quantum numbers, and hence
are stable.%
\footnote
    {%
    More precisely, such mesons with opposite spins and total $S_z = 0$
    are stable.
    As noted in Sec.~\ref{sec:2body}, hyperfine interactions
    shift the $S_z = \pm 1$ mesons up in energy relative to the $S_z = 0$
    states.
    A light $S_z = \pm 1$ meson can decay to its
    corresponding $S_z = 0$ partner via emission of a dual photon ---
    the QCD analog of 21 cm radiation from hydrogen.
    }
These mesons are the small-$L$ avatars of charged pions and kaons
(in the chiral limit).

Baryons (or antibaryons) composed of $N$ quarks (or antiquarks)
all with mass $m_q = \half \mW$ are the lightest states with
non-vanishing baryon number, and a subset of these states
(those with minimal energy after including hyperfine interactions)
are stable.
Whether there are additional bound, and hence stable,
di-baryons or higher multi-baryon states is an interesting
open question.

Whether the heavy photons created by our EFT operators
$\vec\phi_{\pm N}^{\,aa}$ are stable is also an interesting open question.
These states have $P_3 = \pm N \mW$ and tree-level mass $M = N\mW$.
This is the same value of $P_3$ and the same tree-level mass as a
flavor singlet meson containing a quark and antiquark with
$k_q = -k_{\bar q} = \pm N/2$,
or of a collection of $N$ lightest mesons each with identical values
of $P_3 = \pm 1$ and flavor charges summing to zero,
or a variety of other ``near-extremal'' flavor singlet
multi-constituent states.
Whether heavy photons decay into flavor singlet mesons,
or collections of flavored mesons, or vice-versa, depends
on which of these near-extremal states have the lowest energy.
To determine this
one must, at a minimum, take into account the leading
$\O(\lambda \mW)$ perturbative energy shifts.
These include the binding energies
computed in Sec.~\ref{sec:mesons} for two-body mesons.
But $\O(\lambda \mW)$ energy shifts also include
corrections to the tree-level constituent rest masses.
Evaluation of such corrections requires an improved
one-loop matching of the EFT parameters to the underlying
4D gauge theory, and this matching calculation has not yet been completed.
Consequently, we are not yet able to determine which transitions
among near-extremal states are kinematically allowed.

\subsection{Light sector resonances}
\label{sec:light resonances}

Light sector bound states other than those discussed above
(which are stable due to the absence of any symmetry and
kinematically allowed decay channels)
will decay via emission of one of more dual photons.
Such decays are induced by the cubic and higher order terms in the
expansion of the effective Lagrangian (\ref{eq:Slight}) about its minimum.
The relative decay widths of all of these states are doubly
exponentially small.
Not only are the non-linear couplings within the dual photon sector
(\ref{eq:photon_rel_EFT}) exponentially small, $\O(e^{-4\pi^2/\lambda})$,
more importantly the binding momentum (\ref{eq:B2}) is so tiny that
the probability for two constituents of a bound state to be within
a Compton wavelength of each other is comparable to the
relative binding energy (\ref{eq:B2rel}).
Consequently, the logarithm of the relative decay width
is exponentially large and negative,
\begin{align}
    -\ln (\Gamma/m_\gamma) = \O(e^{4\pi^2/\lambda})
\end{align}
(neglecting powers of $\lambda$).
We have not attempted to compute any such decays quantitatively.

\subsection{Heavy sector resonances}
\label{sec:heavy resonances}

The primary decay processes for heavy sector resonances are direct
analogues of familiar processes in QED and atomic physics:
radiative decays and particle-antiparticle annihilations.
The key differences are the reduced dimensionality, additional
conserved quantities (compact momentum and center charge),
and multiple $U(1)$ gauge groups.
There are also more unusual decay processes
involving splitting or joining of $W$-boson constituents
within hadrons.
These include,
in particular, transitions among ``near-extremal'' states
whose tree-level masses are identical.
As noted above, understanding such processes requires a higher order
determination of rest masses in the non-relativistic EFT.
We leave explorations of such transitions to future work, and
focus here on radiative and annihilation processes, specifically
in two-body states.

\subsubsection{Radiative decays}

The relevant photon momenta for radiative decays will be in the range
$m_\gamma \ll p \ll \mW$, so the non-perturbative physics of the
light sector may be wholly ignored and photons treated as massless.
Excitation energies of low-lying heavy sector states
are $\O(\lambda \mW)$.
Photons of such energies have wavelengths parametrically large
compared to the characteristic $\O(\lambda^{-1/2} \,\mW^{-1})$ size of
these states.
Consequently, the usual multipole expansion of the photon field applies.
The fastest radiative decays will be electric dipole transitions.
Adapting the standard logic for hydrogenic decays to our 2D multi-photon
situation, one finds that the total dipole transition rate from
some initial state $|I\rangle$ to lower energy final states
$\{|F\rangle\}$ is given by
\begin{equation}
    \Gamma_{\rm tot}
    =
    \tfrac {\pi}{2} \, \kappa
    \sum_F \> \Delta E_{IF}^2 \>
    \big| \langle F | \x | I \rangle \big|^2 ,
\label{eq:Gammarad}
\end{equation}
where $\Delta E_{IF} \equiv E_I {-} E_F$ and $\kappa$ equals to the
strength of the logarithmic potential binding the constituents,
so $\kappa=\lambda \mW/(2\pi^2)$ for glueballs and
$(1{-}\frac 1N)\, \lambda \mW/(4\pi^2)$ for mesons.

\begin{figure}
\centering{}
\includegraphics[scale=0.48]{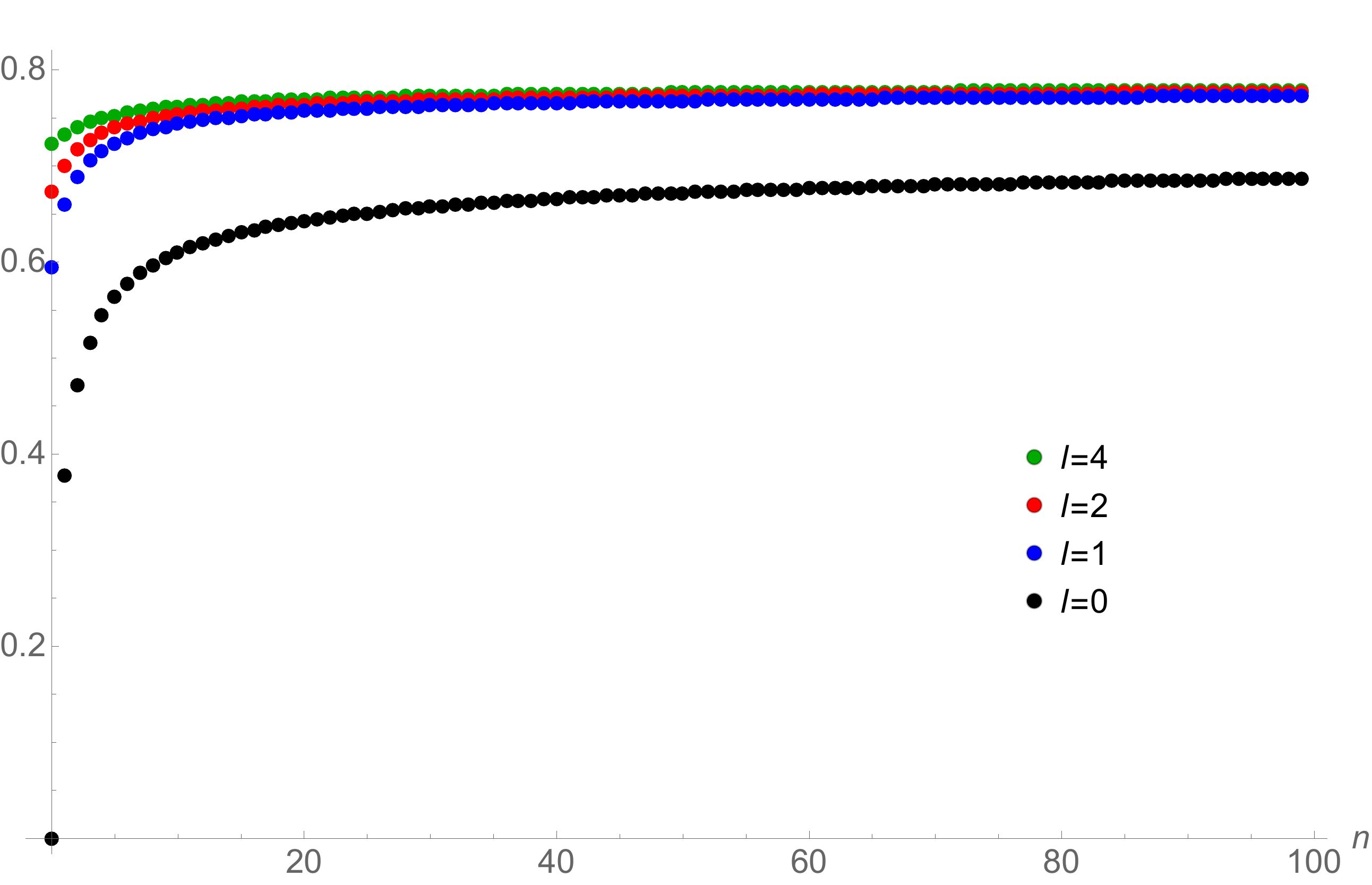}
\caption
    {%
    [Color online]
    Total radiative decay rates of two-body bound states
    in units of $\kappa^2/m$,
    as a function of the
    level number for the first 100 states
    with $\ell = 0,1,2$, and 4.
    Higher rows of points correspond to larger values of $\ell$.
    \label{fig:tot decay rates}
    }
\end{figure}

Parametrically, dipole decay rates for low-lying states are
$\O(\lambda^2 \mW)$.
To obtain quantitative results, including state dependence,
one must evaluate the precise dipole matrix elements.
We evaluated these matrix elements,
for level numbers $n$ up to 100,
using radial wavefunctions computed using pseudo-spectral methods
(as briefly described in footnote \ref{fn:numerics} and
Sec.~\ref{sec:baryons}),
with up to several hundred grid points.
Figure~\ref{fig:tot decay rates} shows the resulting
total dipole decay rates,
in units of $\kappa^2/m$ (with $m$ the reduced mass
of the two-body bound state),
for orbital angular momentum $\ell = 0$, 1, 2 and 4.
As seen in the figure,
decay rates at fixed $\ell$
grow with increasing level number $n$ and appear to asymptote to a
finite limit.
At fixed level number $n$, decay rates also grow with increasing
$\ell$, and quickly appear to reach a limiting value.
Our numerical results are consistent with a limiting value of
$\frac {\pi}{4} \, \kappa^2/m$
in either case,
with subleading $\O(1/\ell)$ corrections if $\ell$ increases at fixed $n$,
and $\O(n^{-1/2})$ corrections if $n$ increases at fixed $\ell$,
although this inverse power of $n$ is not well-constrained by our data
on the first 100 levels.

Consider states with positive orbital angular momentum, $\ell > 0$.
The interleaving of energy levels,
$
    \epsilon_{n,|\ell|} < \epsilon_{n,|\ell|+1} < \epsilon_{n+1,|\ell|}
$,
implies that the $|0,\ell\rangle$ minimal energy states
(for a given angular momentum)
decay down to the $|0,0\rangle$ ground level by sequential
$|0,\ell\rangle \to |0,\ell{-}1\rangle$ transitions,
with each emitted photon carrying off one unit of angular momentum.
States with non-zero angular momentum and non-minimal energy,
$n > 0$ and $\ell > 0$,
have multiple possible dipole allowed final states,
including both $\Delta\ell \eq {+}1$ and $\Delta\ell \eq {-}1$ transitions.
Examining transition rates to specific final states,
one finds that the total decay rates for states with $n,\ell > 0$
are highly dominated by decays to the nearest lower levels,
either
$
    |n,\ell\rangle \to |n,\ell{-}1\rangle
$
or
$
    |n,\ell\rangle \to |n{-}1,\ell{+}1\rangle
$.
Of these two decay channels,
the decay decreasing $\ell$ is significantly more likely than
the decay increasing $\ell$.
All other decays channels are smaller by one or more orders of magnitude.
(The predominance of transitions decreasing $|\ell|$ over those
increasing $|\ell|$ is visible in
Fig.~\ref{fig:tot decay rates} as the
smaller values of the $\ell \eq 0$ points compared to $\ell \eq 1$.)
Consequently, an excited state $|n,\ell\rangle$ with $n \gg 1$
will cascade stepwise down to $n \eq 0$, with $\ell$ undergoing
a random walk biased toward $\ell = 0$.

For high angular momentum, $\ell \gg 1$,
one may regard the $n \eq 0$ eigenstate as a quasiclassical circular orbit.
In two dimensions, the power radiated by an
electric dipole of magnitude $eR$ rotating at frequency $\omega$ is
\begin{equation}
    P = \tfrac 18 \, e^2 \, R^2 \, \omega^3 \,.
\end{equation}
For our high-$\ell$ bound states
with $e^2 = 2\pi\kappa$,
$R = \ell (\kappa m)^{-1/2}$, and orbital frequency
$\omega = \kappa /\ell$,
this gives
$
    P = \frac{\pi}{4} \, \kappa^3 / (m \ell)
$.
The power radiated must equal the photon frequency times the decay rate,
so this classical result implies an $\ell$-independent
asymptotic decay rate,
\begin{equation}
    \Gamma
    = \tfrac{\pi}{4} \, \kappa^2 / m
    \times \big( 1 + \O(\ell^{-1}) \big) \,.
\end{equation}
Decay rates from states with fixed $n$
nicely converge to this value as $\ell$ increases.

\subsubsection {Annihilation decay}

In addition to radiative decays,
two-body bound states having $\ell\,{=}\,0$ and composed of
particle-antiparticle pairs
can annihilate into two or more light sector photons.
This is a short-distance process,
represented by higher dimension operators in our non-relativistic EFT.
Annihilation rates are parametrically smaller than dipole-allowed
radiative transition rates, and hence only significant for the
lowest $\ell \,{=}\, 0$ energy levels.
Constituents with masses of order $\mW$ have Compton wavelengths
which are comparable (for small $N$), or larger (for large $N$),
than the compactification size $L$.
Consequently, annihilation rates are most easily calculated
using a dimensionally reduced relativistic EFT,
having the form (\ref{eq:2Wlagr}) for $W$-boson bound states
or $2{+}1$ dimensional QED for mesons.
The annihilation rate may be expressed as
\begin{equation}
    \Gamma_{\rm annih}
    =
    (\lim_{v\to 0} \, \sigma v) \, |\psi(0)|^2 \,,
\end{equation}
where $\sigma v$ is the flux-weighted cross-section
in two spatial dimensions
(a quantity with dimensions of length)
and $\psi(\x)$ is the wavefunction for relative motion,
so $|\psi(0)|^2$ is the 2D probability density for coincident
constituents.
Parametrically,
$\sigma v \sim \lambda^2/\mW$
for CP even states which can annihilate to two photons having
momenta of order $\mW$,
while $|\psi(0)|^2 \sim \kappa m$
since this is the inverse mean square size of the lowest
$\ell\,{=}\,0$ two-body bound states.
Hence
\begin{equation}
    \Gamma_{\rm annih}
    =
    \O({\lambda^3\, \mW}) \,,
\end{equation}
which is one power of $\lambda$ smaller than radiative decay rates.

Evaluating the cross section in the relativistic $2{+}1$D
relativistic EFT, we find%
\footnote
    {%
    We consider decays from bound states
    with vanishing total compact momentum and
    equal mass constituents.
    Higher KK modes (i.e., heavy photons) may be neglected.
    For $WW$ annihilation,
    each $W$-boson couples to two different $U(1)$ photons
    and consequently there are three different processes which
    contribute ($\gamma_A \gamma_A$, $\gamma_B\gamma_B$,
    and $\gamma_A\gamma_B$).
    Evaluating the leading order seagull, $t$, and $u$-channel
    diagrams and taking the non-relativistic limit yields
    the result shown.
    }
\begin{equation}
    \sigma_{WW\to2\gamma}
    =
    \frac{11 \pi}{64 \, v} \, \frac{\kappa^2}{m^3}
    \left[
	1 + \O(\mathbf p^{2}/m^2 )
    \right]
\label{eq:WW cross section}
\end{equation}
for annihilation of $W$-bosons with mass $m$
and interaction strength $\kappa = \lambda \mW / (2\pi^2)$,
and
\begin{equation}
    \sigma_{q\bar{q}\to2\gamma}
    =
    \frac{5\pi}{128\, v}
    \frac{\kappa^2}{m^3}
    \left[
	1 + \O(\mathbf p^{2}/m^2)
    \right]
\end{equation}
for $q\bar q$ annihilation with mass $m$
and interaction strength $\kappa = (1{-}\frac 1N) \, \lambda\mW/(4\pi^2)$.

For the lowest $n\,{=}\,\ell\,{=}\,0$ level of our two-body
logarithmic quantum mechanics,
the probability at the origin is
\begin{equation}
    |\psi(0)|^2 = 2.68915 \, \kappa \, \widetilde m
\end{equation}
with $\widetilde m$ the reduced mass of the two constituents.
Consequently, for the lightest CP-even glueballs and mesons
(with constituent masses equal to $\mW$ and $\half \mW$, respectively)
we find
\begin{equation}
    \Gamma_{\rm annih}
    =
    \begin{cases}
    {5.80815} \, \mW
    \left(\frac{\lambda}{4\pi^2} \right)^3 , & \mbox{glueballs;}
    \\
    {0.660017} \, \mW
    \left(\frac{\lambda}{4\pi^2}\right)^3
    \left( 1{-}\tfrac 1N \right)^3 , & \mbox{mesons.}
    \end{cases}
\label{eq:Gammaann}
\end{equation}

\section{Discussion}
\label{sec:discussion}

\subsection {Adiabatic continuation}

Recent studies have shown that it is possible to engineer circle
compactifications of 4D $SU(N)$ YM theory and QCD in such a way
that symmetry realizations for large and small circle sizes
coincide \cite{Unsal:2008ch,
Shifman:2008ja,Shifman:2008cx,Shifman:2009tp,
Cossu:2009sq,Myers:2009df,Simic:2010sv,Unsal:2010qh,Vairinhos:2011gv,
Thomas:2011ee,Anber:2011gn,
    Poppitz:2012sw,
    Poppitz:2012nz,Argyres:2012ka,Argyres:2012vv,
    Anber:2013doa,Cossu:2013ora,
    Anber:2014lba,Bergner:2014dua,Li:2014lza,
    Anber:2015kea,Anber:2015wha,Misumi:2014raa,
    Cherman:2016hcd}.
Available evidence is consistent with the natural
conjecture that the weakly coupled small-$L$ regime is
smoothly connected --- that is, without intervening phase transitions
--- to the strongly coupled large-$L$ regime.
The small circle regime offers a rare luxury:
controlled analytic calculations in a phase of the
theory with confinement and chiral symmetry breaking.
Taking advantage of this tractability, we have studied the behavior
of glueballs, mesons, and baryons, with a focus on the spectrum of
resonances and their decays.

\begin{figure}
\vspace*{-2.0em}
\begin{center}
\includegraphics[scale=0.45]{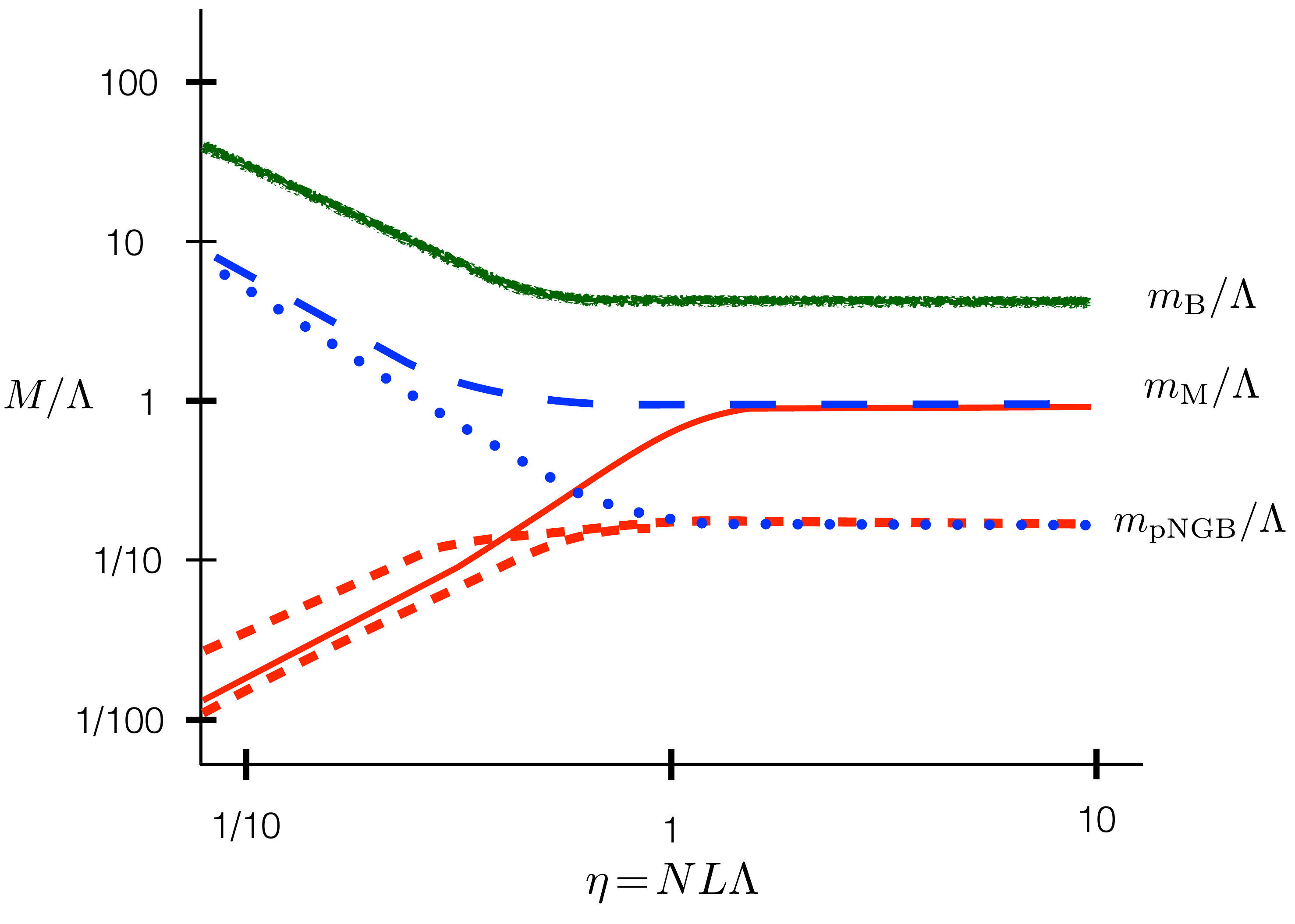}
\end{center}
\vspace*{-1.5em}
\caption
    {%
    [Color online.] A sketch of a possible interpolation of
    the spectrum
    as the circle size $L$ is varied in adiabatically compactified
    QCD with $n_f \eq N$ and $0 < m_q \ll \Lambda$.
    In this log-log cartoon, masses are in units of $\Lambda$
    and the abscissa $\eta \equiv NL\Lambda$.
    The short-dashed red and dotted blue curves correspond
    to the neutral and charged pNGBs, respectively.
    At small $L$, the neutral pNGBs are dual photons,
    while the charged pNGBs are non-relativistic quark-antiquark bound states.
    The large splitting in their masses
    at small $L$ is due to the partial breaking
    of flavor symmetry to its Cartan subgroup by our flavor-twisted
    boundary conditions.
    The solid red curve represents the lightest flavor singlet meson which,
    at small $L$, is a bound state of dual photons.
    The long-dashed blue curve represents glueballs and
    other mesons (both flavor singlet and non-singlet) which are
    not pNGBs and which, at small $L$, are bound states of $W$-bosons.
    The fuzzy green curve at the top of the figure represents the
    evolution of the mass of a baryon from small to large~$L$.
    \label{fig:sketch1}
    }
\end{figure}

Our results are broadly consistent with the conjecture of continuity
between small and large $L$.  Much physics in
adiabatically compactified theories depends on the circle size $L$ through the
parameter $\eta = N L\Lambda$.
To place our small $\eta$ results into perspective, first recall
that when  $\eta \gg 1$, the dynamics of QCD-like theories are
insensitive to the scale $L$.
(Finite volume effects vanish at least as fast as $L^{-2}$.)
With fundamental representation fermions ($\nf \lesssim N$)
with a common mass $m_q \ll \Lambda$, at large $L$
there are multiple characteristic scales for the masses of particles:
the pseudo-Nambu-Goldstone (pNGB) mass scale
$m_{\mathrm{pNGB}} \sim \sqrt{m_q \Lambda}$,
the glueball and meson mass scale $m_{M} \sim \Lambda$,
and the baryon mass scale $m_{B} \sim N \Lambda$.%
\footnote
    {%
    If $\nf \ll N$ and $m_q/\Lambda \ll 1/N\ll 1$,
    then there is an additional scale
    $\Lambda N^{-1/2}$ associated with the mass of the $\eta'$ meson
    \cite{Witten:1980sp}.
    }

In the weakly coupled regime $\eta \ll 1$, we find a similar
picture, but with particle masses depending on $L$ through the
combination $\eta = NL\Lambda$.
In adiabatically compactified QCD with, e.g., $\nf = N$
earlier work \cite{Cherman:2016hcd} found that
the pNGB masses lie in the range
$
    m_{\mathrm{pNGB}}
    \sim \left[ \O(1/N) \mbox{--} \O(1) \right]
    \times \eta \sqrt{m_q \Lambda}
$
at small $\eta$ (if double trace deformations stabilize the color-flavor center symmetry).
Our results in this paper show explicitly that
$m_{M}\sim \Lambda \eta^{-1}$ and  $m_{B} \sim N \Lambda\eta^{-1}$.
This is clearly similar to the large $L$ pattern, apart
from the natural appearance of dependence on
the parameter $\eta$ when $L$ is small.%
\footnote
    {%
    The dependence of pNGB masses (\ref{eq:pNGBmass0}) at small $L$
    on the charge of the particle under cyclic flavor permutations
    may, at first sight, seem surprising.
    But such dependence is also present when $L$ is large but finite in
    adiabatically compactified theories with flavor twisted boundary
    conditions,
    as seen explicitly in the results of Ref.~\cite{Cherman:2016vpt}.
    }
In Fig.~\ref{fig:sketch1}
we sketch a possible simple interpolation of the spectra of
light and heavy states as $L$ is varied.

\begin{figure}
\begin{centering}
\includegraphics[scale=0.45]{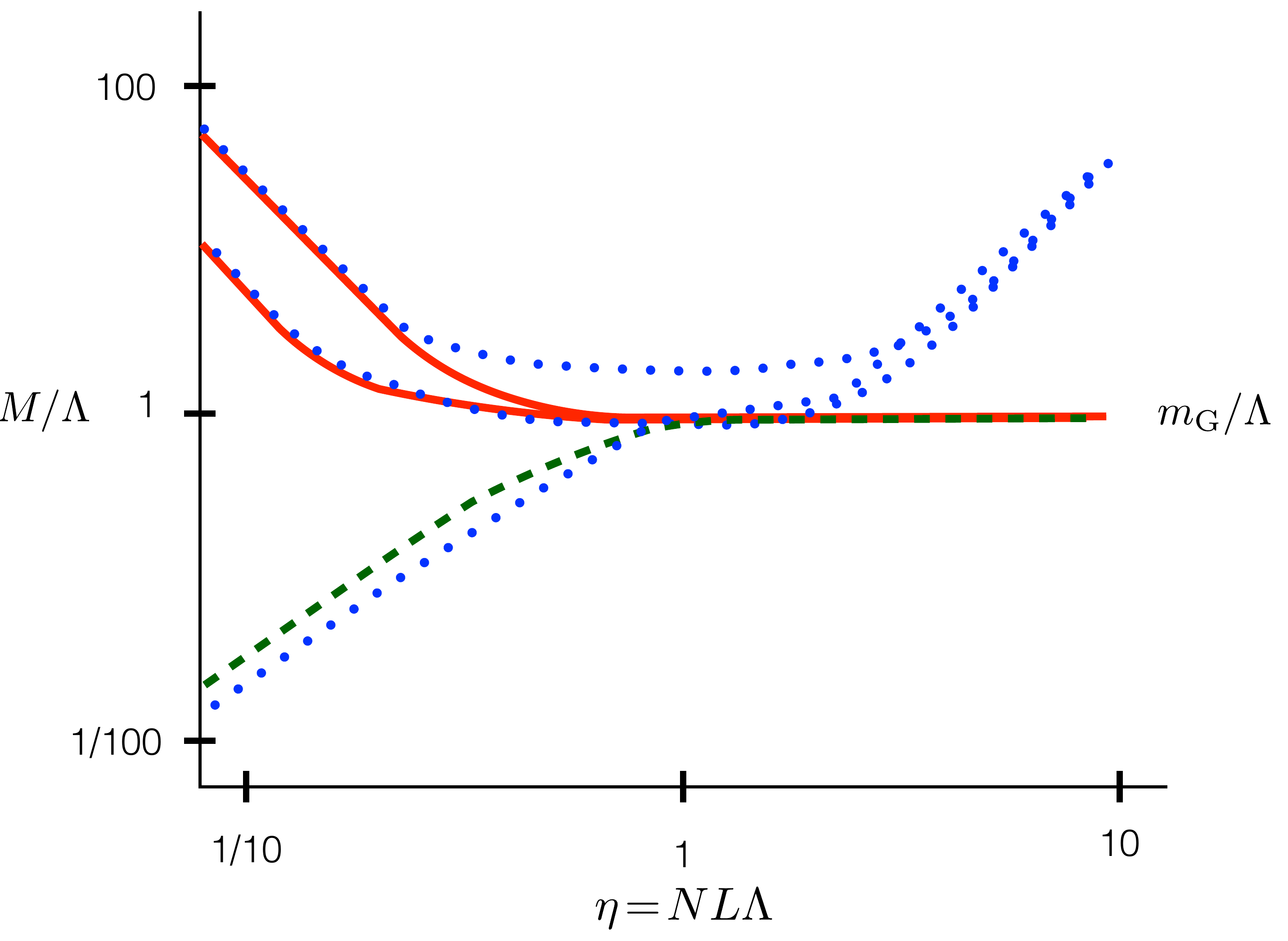}
\par\end{centering}
\caption
    {%
    [Color online.] A sketch of a possible interpolation of
    the spectrum
    as the circle size $L$ is varied in adiabatically compactified
    Yang-Mills. The illustration is for $N=2$ for simplicity.
    The dashed green curve shows the lightest glueball,
    with vanishing center charge, which is a bound state of two dual
    photons at small $L$.
    The dotted blue curve with a positive slope at small $\eta$ corresponds
    to the lightest topologically non-trivial
    ``glueball'' with non-zero center charge.
    This state is a dual photon at small $L$ and evolves into a state
    with a linearly growing energy, $m \sim \sigma L$ at large $L$,
    where $\sigma$ is the string tension.
    The two solid red curves correspond to center-neutral $W$-boson
    bound states, which evolve into
    ordinary glueballs at large $L$.
    The dotted blue curves with negative slope at small $\eta$ correspond
    to $W$-boson states with non-zero center charge, which evolve into
    wrapped-flux states with a linearly diverging mass at large $L$.
    \label{fig:sketch2}
    }
\end{figure}

The situation at $\nf \eq 0$ is depicted in Fig.~\ref{fig:sketch2}.
At small $L$,
instead of light pNGB mesons there are now light glueball states
involving dual photons and their bound states,
with masses
$
    m_{\rm light}
    \sim
    \left[ \O(1/N) \mbox{--} \O(1) \right]
    \times \Lambda \, \eta^{5/6}
$
(if double trace deformations stabilize center symmetry).
The $N{-}1$ dual photons are charged under the center symmetry,
indicating that they are topologically non-trivial excitations
containing flux wrapping the compactified direction.
These states cannot be created by topologically trivial local operators
(acting on the vacuum) and will have masses which do not
asymptote to finite limits at large $L$
but rather grow linearly, $m \sim \sigma L$,
with $\sigma$ the decompactified YM string tension.
The bound state of two dual photons with vanishing total center charge
is the lightest topologically trivial glueball at small $L$,
and can smoothly connect to the lightest glueball at large $L$.
In the heavy sector at small $L$, $W$-boson bound states form
nearly degenerate multiplets containing all values of center charge.
Within each such multiplet, the vanishing center charge state
can evolve into an ordinary topologically trivial glueball at large $L$,
while the remaining states with non-zero center charge will have
linearly diverging masses at large $L$.

Finally, when $1 \le \nf < N$, the overall picture is the same as
the sketch shown in Fig.~\ref{fig:sketch1}, except that
the light sector at small $L$ now contains $\nf{-}1$
pseudo-Nambu-Goldstone bosons, with masses vanishing at $m_q = 0$,
as well as non-pNGB states,
namely the remaining $N{-}\nf$ dual photons and their bound states.
These non-pNGB states have masses on the order of
$m_{\rm light} \sim \Lambda \, \eta^b$ with an exponent
$b > 0$ depending on $\nf/N$.
Whether these states should be described as glueballs or mesons,
or some admixture, is not clear.
There is no symmetry which clearly delineates a distinction.
It should be possible to clarify the situation by computing
the amplitudes with which these states are created by
local fermion bilinears or Polyakov loop operators, but such
an analysis has not yet been performed.
In any case, these states can smoothly evolve into ordinary glueballs
and mesons as $L \to \infty$.
The same is true of the glueballs and mesons in the heavy sector
at small $L$.
Due to string breaking by dynamical quarks,
none of these states will have masses which diverge as $L \to \infty$.

\subsection {Large $N$ behavior}

Our analysis has been carried out with $N$ arbitrary but fixed.
The usual large $N$ limit involves sending $N$ to infinity while
holding fixed the 't Hooft coupling $\lambda$
(or equivalently the strong scale $\Lambda$).
If the compactification size $L$ is also held fixed, then the
large $N$ limit takes the compactified theory out of the regime
$\eta = NL\Lambda \ll 1$ where a weak coupling analysis is possible
and into the strongly coupled domain, $\eta \gg 1$,
where large $N$ volume independence applies
\cite{EguchiKawaiOriginal,Bhanot:1982sh,GonzalezArroyo:1982ub,GonzalezArroyo:1982hz,Perez:2014sqa,Kovtun:2007py}.
Our small $\eta$ analysis adds nothing to the understanding of this limit.

However, it is interesting to consider an alternate $N \to \infty$ limit
in which $\eta = NL\Lambda$ is held fixed.
This is the key parameter which controls the physics of
adiabatically compactified QCD-like theories.
Viewing $\Lambda$ as a fixed physical scale,
fixing $\eta$ requires reducing the compactification size as $N$ increases,
$L \propto 1/N$,
or equivalently holding fixed $\mW \equiv 2\pi/(NL)$.
If $\eta$ is fixed at a small value, then a weak coupling analysis
remains valid for all $N$.

\subsubsection{Heavy sector}

Starting with the heavy non-relativistic sector, our results show
that the glueball and meson spectra remain stable as $N \to \infty$
(regardless of whether $\nf = N$, or $\nf \ll N$).
For example,
the value of $N$ simply never enters the result (\ref{eq:2bodyglue})
for two-body glueball binding energies,
while the only $N$ dependence in meson binding energies (\ref{eq:mesonmass})
comes from the quark-antiquark interaction strength proportional
to $1{-}\frac 1N$.
So masses of glueballs and mesons become $N$-independent at large $N$.
The lightest baryon masses, as one would expect, grow linearly with $N$,
but (based only on results at $N\eq 2$ and $N \gg 1$)
the ground state baryon binding energy per quark (\ref{eq:baryonbinding})
and the shape of the single particle distribution
(Fig.~\ref{fig:baryon pdf})
are quite insensitive to $N$.

Similarly, the only $N$ dependence in the glueball and meson
radiative decay (\ref{eq:Gammarad})
and annihilation rates (\ref{eq:Gammaann})
arises from the same $1{-}\frac 1N$ interaction strength factor for mesons.
Given this, one might guess that glueball and meson scattering
amplitudes would also have finite, non-zero large $N$ limits
--- but this is not entirely correct.
If one ignores higher order radiative corrections then,
for example,
two-body mesons (at small $\eta$) may be labeled by the Cartan charge
of their constituents.
The amplitude for the elastic scattering process
$
    M^a + M^b \to M^a + M^b
$
arising from the exchange of one or more Cartan photons
will include a trivial factor of
$
    \delta_{ab}
$
expressing the fact that both mesons must contain constituents
charged under the same $U(1)$ factor if they are to scatter via
photon exchange.
When radiative corrections are included,
the actual mass eigenstates are linear combinations of the
fixed Cartan charge states which (for $\nf = N$) have
definite center charge (or more precisely, definite color-flavor
center charge, as discussed in Ref.~\cite{Cherman:2017tey}),
$
    \widetilde M^p = N^{-1/2} \sum_a \omega^{-a p} M^a
$.
The resulting scattering amplitude for
$
    \widetilde M^p + \widetilde M^q \to \widetilde M^{p'} + \widetilde M^{q'}
$,
is $\O(1/N)$ for all center charges satisfying $p{+}q=p'{+}q'$,
instead of $\O(1)$ for coinciding Cartan charges
and zero otherwise.

The same argument applies to glueballs.
Consider, for simplicity,
glueballs which are bound states of two $W$-bosons,
with either $\nf \eq 0$ or $\nf \eq N$
(so the compactified theory has either an ordinary,
or intertwined color-flavor center symmetry).
As discussed in Sec.~\ref{sec:multiglue},
glueballs in our small $\eta$ regime,
before diagonalizing center symmetry,
may be labeled by a \emph{single} Cartan index
plus the ordered compact momenta of their $W$-boson constituents.
(Subsequent Cartan indices are determined by the
mass formula \eqref{eq:W-masses}, which in turn
is a consequence of the adjoint Higgs mechanism operative at small
$\eta$.)
The transformation to a mass eigenstate basis with definite
center charge involves exactly the same discrete Fourier
transform as for mesons,
$
    \widetilde G^p = N^{-1/2} \sum_a \omega^{-a p} \, G^a
$.
The resulting $2\leftrightarrow 2$ scattering amplitude for
$
    \widetilde G^p + \widetilde G^q \to \widetilde G^{p'} + \widetilde G^{q'}
$
is suppressed by $1/N$
for all center charges satisfying $p{+}q=p'{+}q'$.

More generally, scattering amplitudes at small $\eta$
involving $K$ external particles (incoming plus outgoing)
scale as $\O(N^{1-\frac 12 K})$.
This holds for processes involving any combination of
light dual photons and heavy sector bound states
(either mesons or glueballs) with $\O(1)$ constituents,
provided at least one of the particles in the scattering process
is a heavy sector bound state.
(Scattering involving only dual photons is discussed below.)
This relation shows that decay amplitudes into two particle final states
are $\O(N^{-1/2})$,
so decay rates to exclusive two particle final states are
suppressed by $1/N$.
That may appear inconsistent with the $\O(1)$ total radiative and
annihilation rates computed in Sec.~\ref{sec:decays},
but inclusive decay rates sum over all accessible final states.
Because the splittings between states with differing center charge
are parametrically smaller than heavy sector binding or rest energies
(by powers of $\lambda$ for heavy states, or $m_\gamma/\mW$ for
light dual photons),
inclusive $1 \to 2$ decay rates
pick up a factor of $N$ from summing over
all possible center charges of the final state particles
consistent with the initial state center charge.%
\footnote
    {%
    This assumes the decay channel is not parametrically
    close to threshold, so that the decay kinematics is insensitive to
    the splittings between final state particles with differing
    center charges.
    }
The same logic shows that while fully exclusive
$2 \leftrightarrow 2$ scattering rates are $\O(N^{-2})$,
inclusive $2 \leftrightarrow 2$ scattering rates for mesons and glueballs
are $\O(N^{-1})$ as $N \to \infty$.

    Meson-baryon scattering amplitudes scale as $\O(N^0)$,
since a quark (or antiquark) with any given Cartan index
can interact with the quark having the same Cartan index in the baryon.
The same scaling holds for glueball-baryon scattering (for both heavy
sector bound state glueballs, and light dual photons).
Baryon-baryon scattering amplitudes are $\O(N)$, since every quark in
one baryon can interact via an unbroken $U(1)$ gauge group with one
of the quarks in the other baryon.

These large $N$ scaling relations at small $\eta$ may be compared
with conventional large $N$ behavior when
$\Lambda$ and $L$ are held fixed, and hence $\eta \to \infty$.
It will be interesting to compare with conventional behavior in
both the 't Hooft ($\nf$ fixed as $N\to\infty$) and
Veneziano ($\nf/N$ fixed as $N \to \infty$) limits.
In all cases, meson and glueball spectra are stable as $N \to \infty$,
while the lightest baryon masses grow linearly with $N$.
One unusual consequence of our flavor-twisted boundary conditions,
at small $\eta$, is that baryons composed of only a single flavor of quark
(or more generally $\O(1)$ different flavors)
have masses which grow quadratically with $N$.

In the standard 't Hooft large $N$ limit,
glueball scattering amplitudes scale as
$\O(N^{2-K_g})$, with $K_g$ the number of external glueballs
(incoming plus outgoing)
\cite{Witten:1979kh}.
For processes involving mesons,
possibly with additional glueballs, the scaling of scattering
amplitudes becomes $\O(N^{1-K_g -\frac 12 K_m})$,
where $K_m$ is the number of external mesons.
Hence,
meson decay widths are $\O(N^{-1})$ and
glueball decay widths to either two glueball, or two meson final states
are $\O(N^{-2})$.
Rates for two glueballs to scatter into two glueballs,
or into two mesons, are $\O(N^{-4})$, while
$2 \leftrightarrow 2$ meson scattering rates are $\O(N^{-2})$.
Baryon-baryon scattering amplitudes are $\O(N)$ while
baryon-meson scattering amplitudes are $\O(1)$ \cite{Witten:1979kh}.

In the Veneziano large $N$ limit, the additional factors of
$\nf \propto N$ in sums over final states
(assuming a common quark mass for all flavors)
make both meson and glueball decay rates $\O(1)$.
Hence, except for the lightest states in each symmetry channel,
mesons and glueballs remain resonances, with finite lifetimes,
as $N \to \infty$.
The inclusive rate for two mesons to scatter into two mesons is $\O(N^{-1})$,
while two glueballs can scatter into two mesons with an
$\O(N^{-2})$ inclusive rate, parametrically faster than the $\O(N^{-4})$
rate for pure glueball scattering.

Comparing these conventional large $\eta$ scaling relations
with our small $\eta$ results,
one sees that for mesons
our $\O(N^0)$ total decay rates,
$\O(N^{-1})$ inclusive two particle scattering rates, and
$\O(N^{-2})$ exclusive two particle rates all coincide with
the behavior of mesons in the Veneziano limit.
The scaling of our baryon-baryon and baryon-meson scattering 
amplitudes is the same as in conventional large $N$ limits.
But the fact that, at small $\eta$, glueball processes have
the \emph{same} large $N$ scaling as mesons is quite peculiar.

Two significant features contribute to this change in behavior
of glueball dynamics between large and small $\eta$.
First is the adjoint Higgs mechanism induced by the center-symmetric
holonomy at small $\eta$.
This suppresses fluctuations in off-diagonal components of the $SU(N)$
gauge field,
so that only the $N{-}1$ gluonic degrees of freedom play a singificant role
in resonance formation, scattering, and decay.
In contrast, at large $\eta$ there are huge fluctuations in the holonomy
and all $N^2$ gluonic degrees of freedom contribute to every glueball
operator.
This leads to the familiar $1/N^2$ suppression factors in
exclusive decay rates and $2 \to 2$ scattering amplitudes of glueballs.
A second essential difference at large and small $\eta$ is the contribution
of states with non-zero center charge.
At $\nf = 0$, such states have linearly diverging energy as $\eta \to \infty$
(as shown in Fig.~\ref{fig:sketch2}), and play no role in scattering processes
involving $\O(N^0)$ energies.
But at small $\eta$ these topologically non-trivial states become nearly
degenerate in energy with vanishing center charge states,
and dominate inclusive scattering and decay rates at large $N$.

\subsubsection{Light sector}

Turning now to the light sector,
when $\nf \ll N$, the smallest non-zero dual photon mass
is $\O(m_{\gamma}/N)$.  Holding $\eta$ fixed as $N \to \infty$ implies that the
light scale $m_\gamma$ is also held fixed.
Consequently, the lightest (non-Goldstone boson) mass vanishes
as $N \to \infty$.  

The interpretation and consequences of the vanishing of the mass
of the lightest non-Goldstone boson excitations in the small-$\eta$
large $N$ limit were the focus of Ref.~\cite{Cherman:2016jtu}.
At very low energies, small compared to $m_{\gamma}$,
the theory does not flow to a trivial fixed point.
Rather, to all orders in the semi-classical expansion
the low energy theory becomes gapless as $N \to \infty$.
The low energy dynamics at $N \eq \infty$ is most naturally written
as a \emph{four-dimensional} theory, despite the fact that the
``parent'' UV theory was compactified on a tiny circle.
The fourth dimension in the low energy, large $N$ dynamics is emergent,
appearing only on length scales large compared to $m_{\gamma}^{-1}$.

The results in this work are consistent with this picture, but
do not shed much additional light on the origin or interpretation
of this unexpected phenomena.
%
The quartic interactions of dual photons,
shown in Eq.~(\ref{eq:quarticphoton}), may be interpreted
in the large $N$ emergent dimension description
as momentum-dependent interactions with vertex factors
proportional to $1/N$ times the product of photon momenta
in the emergent dimension.
Consequently, for $\O(N^0)$ momenta (in the original spatial dimensions),
dual photon scattering amplitudes scale as $\O(N^{-1})$ at large $N$,
the same as for heavy sector glueballs.

As shown in Eqs.~(\ref{eq:non-id binding}) and (\ref{eq:id binding}),
the dual photon binding energies (and momenta) discussed in
Sec.~\ref{sec:NR_EFT_light} vanish exponentially as $N \to \infty$.
So these bound states play no significant role at large $N$,
and the emergence of the extra dimension in the light sector of the
theory happens just as described in Ref.~\cite{Cherman:2016jtu}.
To understand how, e.g., the glueballs arising from
$W$-boson bound states fit into the large $N$ emergent dimension picture,
recall that the emergent dimension appears as an $N$-site discretized
circle with lattice spacing $m_{\gamma}^{-1}$ \cite{Cherman:2016jtu}.
A continuum 4D description is only relevant for physics with momenta
small compared to $m_{\gamma}$.
But at small $\eta$, the $\O(\mW)$ $W$-boson masses, their
$\O( \lambda \mW)$ binding energies,
and the $\O(\lambda^2 \mW)$ radiative corrections to binding energies
are all large compared to $m_{\gamma}$.
So the large $N$ bound state dynamics does not involve the
low energy emergent dimension,
and must be treated using a 3D effective field theory,
as done in the present paper.

\subsection{Outlook}

The analysis and results of this paper raise a number of questions
which would be interesting to study in future work.
First, as noted near the end of Sec.~\ref{sec:stable states},
we have not performed the matching calculation necessary to
determine the $\O(\lambda)$ corrections to the rest mass
parameters of the 3D non-relativistic EFT.~
Differences in the short distance corrections to the EFT rest masses
are needed to determine the relative stability of meson, glueball,
and heavy photon resonances whose leading order masses are identical.
For example, the lightest glueball resonances with mass near $4\mW$
might be composed of two $W$-bosons each with (tree level) mass $2 \mW$,
or from four of the lightest $W$-bosons each with mass $\mW$.
Such glueball states are nearly degenerate with heavy photons
having a tree-level mass of $4\mW$.
The results of a one-loop matching calculation of EFT rest energies
would enable one to determine the relative ordering
of these states.
In particular, this would allow one to answer the interesting question
of which near-extremal states are absolutely stable by virtue of
minimizing the ratio of mass to compact momentum, $M/|P_3|$.

Second, as emphasized in Sec.~\ref{sec:bound_states},
the bound state spectra for glueballs, mesons, and baryons
have an exponentially rising (Hagedorn) density of states.
It is interesting that this Hagedorn scaling emerges as a consequence
of a logarithmic potential within the domain of validity of
non-relativistic quantum mechanics,
in contrast to the common lore that Hagedorn scaling is characteristic
of relativistic string dynamics.
In any case, the implications of Hagedorn scaling in the density
of states for the thermodynamics of adiabatically compactified QCD
deserve further study.
Previous work \cite{Simic:2010sv,Anber:2011gn,Anber:2013doa}
considered the $SU(2)$ deformed Yang-Mills theory
(see also Refs.~\cite{Dunne:2000vp,Kovchegov:2002vi}),
and argued that a thermal confinement-deconfinement transition
occurs near the temperature
\begin{align}
\label{eq:betac}
    \beta^{-1}_c \simeq \frac{\lambda  \mW}{ 4 \pi^2} \,.
\end{align}
The picture behind this conclusion is that in the regime%
\footnote
    {%
    This temperature range is similar to, but slightly more restrictive
    than the condition for the validity of our non-relativistic EFT
    analysis, and is needed to justify the treatment of  the
    monopole-instanton gas as two-dimensional.
    }
$\zeta^{1/3} \ll \beta^{-1} \ll \mW$,
the dilute monopole-instanton
gas representation of the 3D Euclidean vacuum
effectively reduces to a dilute two-dimensional gas of magnetically
charged particles subject to binary logarithmic interactions.
At the same time, there is also a thermal gas of electrically charged
particles, namely $W$-bosons.
The thermal phase transition is believed to be driven by a
competition between the effects of these electrically and magnetically
interacting gases.
However, in
Refs.~\cite{Dunne:2000vp,Kovchegov:2002vi,Simic:2010sv,Anber:2011gn,Anber:2013doa}
the electrically-charged component of the gas was treated classically,
and the existence of Hagedorn behavior in the density of states was
not taken into account.
It would be interesting to revisit these
calculations in light of our results here,
and clarify whether the temperature \eqref{eq:betac}
is indeed a correct estimate of the phase transition temperature.


Next, it would be very interesting if lattice gauge theory simulations
could be performed in both pure Yang-Mills and QCD exploring the cross-over
regions in Figs.~\ref{fig:sketch1} and \ref{fig:sketch2}, along the lines of
Refs.~\cite{Vairinhos:2011gv,Bergner:2014dua}.
This would require simulations in a variety of lattice volumes
with one dimension having double trace center stabilizing terms
and flavor-twisted boundary conditions on quarks.

Last, and perhaps most interesting from a phenomenological perspective,
is the possibility of studying multi-baryon states at small $L$.
To motivate this, recall that in the real world there is a wide
separation between ``nuclear'' excitation scales relevant
in multi-baryon systems and the energy scales characteristic of
single baryons.
For example, the saturation binding energy per nucleon of nuclear matter,
roughly $14\,\mathrm{MeV}$, is tiny compared to the
$\approx 300\,\mathrm{MeV}$ energy required
to excite a single nucleon beyond its ground state.
Or, one may compare nuclear binding scales to nucleon masses
of nearly a GeV.
Both comparisons indicate a wide separation between nuclear and
single-baryon energy scales.
Moreover, lattice simulations indicate that the nuclear/hadronic
scale separation persists even as quark masses are varied
\cite{Beane:2014ora,Parreno:2016fwu,Wagman:2017tmp},
and that it also persists when $N \eq 2$ instead of 3 \cite{Detmold:2014kba},
suggesting that this scale separation
is robust feature of QCD.
This scale separation is vital for essentially all phenomenological
understanding nuclear physics, including
the modeling of nuclei as a collection of individual nucleons.

The puzzle is that there is no fundamental explanation for this important
exper\-imentally-observed scale separation from QCD.
For example, this scale separation is not an automatic consequence of
either the large $N$ or chiral limits.
The adiabatic small-$L$ regime
allows one to use straightforward numerical and analytic methods
to study multi-baryon systems for any quark mass and any number of colors.
Further exploration of QCD phenomenology on a small circle may thus
yield useful insights into the long-standing and important puzzle
of the separation between nuclear and hadronic energy scales in QCD.

\acknowledgments
    {%
    We are grateful to M.~\"Unsal for helpful discussions.
    This work was supported, in part, by the U.~S.~Department of Energy
    via grants DE-FG02-00ER-41132 (A.C.) and  DE-SC0011637 (K.A.~\& L.Y.)
    and by a Discovery Grant of the National Science and Engineering
    Research Council of Canada (E.P.).
    L.~Yaffe thanks the University of Regensburg and the Alexander von
    Humboldt foundation for their generous support and hospitality during
    completion of this work.
    }

\appendix

\section {Non-relativistic EFT derivation}\label{app:NReft}

We denote $SU(N)$ indices by $a,b,c,d$, etc., each running from
1 to $N$, and define the set of $N\times N$ basis matrices $\{E^{ab}\}$
by $(E^{ab})_{cd} \equiv \delta_{ac} \, \delta_{bd}$.
We use an $N$-dimensional basis for the root vectors $\beta_{ab}$ ($a\ne b$).
The positive roots are $\beta_{ab} = (0,...,0,1,0,...,0,-1,0,...,0)$, $a < b$,
with $1$ and $-1$ in the $a$-th and $b$-th position, respectively;
the negative roots are $\beta_{ba} = - \beta_{ab}$, $a < b$.
The indices $\mu = 0,1,2$ denote the noncompact spacetime directions and
$x^3 \equiv x^3 + L$ is the coordinate of the compact direction.
The circumference $L \equiv 2 \pi R$.
The Cartan generators are denoted by $H^{a} \equiv E^{aa}$.
The overall $U(1)$ photon coupling to $\sum_a H^a$
decouples from the $SU(N)$ dynamics and is introduced solely for
the convenience of working with an $N$-dimensional weight basis.
Since all weight vectors are orthogonal to the vector $(1,1,1,1,...,1)$,
the static interactions discussed below in Appendix \ref{app:lightsector}
only involve $SU(N)$ charges which are neutral with respect to this
overall $U(1)$.

Until otherwise specified [just before Eq.~(\ref{eq:Lmass1})],
we write Euclidean space expressions in this appendix.
The Yang-Mills Lagrangian
$
    L = \frac{N}{4 \lambda} \, {\rm tr} \, F_{\alpha\beta}^2
$,
with
$F_{\alpha\beta}$ Hermitian.
The 't Hooft coupling
$\lambda \equiv N g^2(\mW)$, where the scale
$\mW \equiv 1/(NR)$ denotes the lightest $W$-boson mass.
We  decompose the gauge field into components along the compact
and noncompact directions,
\begin{subequations}\label{eq:expansion}%
\begin{align}
A_3 &=  \sum\limits_{1\le a \le N}   A_3^a(x^\mu) \: H^a \,,
\\
A_\mu &=  \sum\limits_{1\le a \le N} A_\mu^a (x^\mu, x^3) \: H^a  +\sum\limits_{1 \le a < b \le N}    W_{ \mu}^{ab}(x^\mu, x^3) \: E^{ab} + W_{ \mu}^{ab *}(x^\mu, x^3) \;  E^{ba}   \,.
\end{align}
\end{subequations}
The expansion (\ref{eq:expansion}) is written in the unitary gauge, where
the only nonzero gauge field components along the $S^1$ direction
are the Cartan components and they have no $x^3$-dependence.
The $N$ real fields $A_\mu^a$
describe 3D photons in
the Cartan subalgebra, while the $\half (N^2{-}N)$ complex fields
$W^{ab}_\mu$
($a < b$)
in the off-diagonal elements describe charged $W$-bosons.

Next, we expand $A_3^a$ around the  center symmetric expectation
value (\ref{eq:Omega}) of the  holonomy,
$
    A_3^a \equiv  \rho^a/(N R) + \phi^a
$,
so that $\phi^a$ represents
the  fluctuations of the holonomy.%
\footnote
    {%
    Here, $\rho^a = \half (N{+}1)-a$ are the components of the
    Weyl vector in our basis.
    The expectation value $\langle A_3^a \rangle = \rho^a/(NR)$ corresponds
    to $\mathbb Z_N$ symmetric eigenvalues of the holonomy and produces
    vanishing traces in the fundamental representation,
    $\langle {\rm tr}_F  \, \Omega^k \rangle= 0$ for $k=1,...,N{-}1$.
    }
Plugging the expansion (\ref{eq:expansion})
into the Yang-Mills Lagrangian one obtains, up to quadratic order in
the $W$-boson fields,
\begin{align}
\label{eq:2Wlagr}
    {\cal L}_{2W}
    &=
    \frac{ N }{ 4 \lambda  }
    \Biggl\{
	\sum_{1 \le a \le N}
	    {F}_{\mu\nu}^a \; F^{\mu\nu \; a}
	    +  2 \,\partial_\mu  {\phi}^a \, \partial^\mu \phi^a
	    +  2 \,\partial_3 {A}_\mu^a \, \partial^3 A^{\mu \, a}
\nonumber\\ &\qquad{}
    + \!\! \sum\limits_{1 \le a < b \le N}
	2\, \Big\vert
	    \partial_{\mu}  W_{\nu}^{ab}
	    + i ( A_{\mu}^{a} {-} A_{\mu}^{b} ) \, W_{\nu}^{ab}
	    - (\mu \leftrightarrow \nu) \Big\vert^2
\nonumber \\ & \quad\qquad\qquad {}
	+ 4\, \Big\vert
	    \big(
		{-} i \partial_3 + \tfrac{a{-}b}{R N} + {\phi_b {-} \phi_a }
	    \big) \,
		W_\mu^{ab}
	    \Big\vert^2
	+ 2i \left(
	    {F}_{\mu\nu}^a{-}{F}_{\mu\nu}^b
	    \right)
	    W_{[\mu}^{ab} \, W_{\nu]}^{ab *}
	\Biggr\} \,.
\end{align}
The second line shows explicitly that
the $W$-boson field $W^{ab}_\mu$ has charge
$+1$ and $-1$ under the $a$-th and $b$-th Cartan $U(1)$ gauge groups,
respectively.
Hereafter, we neglect the fluctuations $\phi^a$ of the holonomy;
as explained in Sec.~\ref{sec:NR_EFT}, they play no
role in the dynamics to the order that we  study.
(These neutral fluctuations are gapped by the perturbative
center-stabilization mechanism.)

Next,
we derive the leading-order EFT valid for momenta $p \ll \mW$
(but large compared to the non-perturbatively induced mass gap
(\ref{eq:mgamma}), $p \gg m_\gamma$).
This EFT describes the interactions
of charged massive $W$-bosons with the (perturbatively) massless
Cartan photons and with the ``heavy photons," modes in the Kaluza-Klein (KK)
tower containing the Cartan photon fields.
As a final step before considering the $p \ll \mW$ non-relativistic
limit, we rewrite the Lagrangian (\ref{eq:2Wlagr}) in a mass
(or KK) eigenstate basis.
The KK expansions are defined as usual, e.g.,
\begin{align}
  \label{eq:KKexpansions}
   {A}_\mu^a (x^\nu, x^3)
   &=
   \sum_{n = - \infty}^\infty
   e^{i  x^3 {n }/{R}} \,
   {A}_\mu^{a, n}(x^\nu) \,,
\end{align}
with $ {A}_\mu^{a, -n}(x^\nu) = ( {A}_\mu^{a, n}(x^\nu))^* $,
and similarly for the $W^{ab}_\mu$ fields
(without a corresponding reality condition).
Inserting these expansions into the 4D Lagrangian (\ref{eq:2Wlagr}),
integrating over $x_3$
(and neglecting holonomy fluctuations),
leads to an effective three-dimensional Lagrangian
\begin{equation}
    {\cal L}^{3D} = {\cal L}_2 + {\cal L}_3  + {\cal L}_3^{\prime} +\cdots \,,
\label{eq:L3D}
\end{equation}
in which we separate, for convenience, quadratic, cubic, and higher order terms.
The quadratic part is given by
\begin{align}
\label{eq:3dfreeL}
    {\cal L}_2
    &=
    \frac{ N L }{ 4 \lambda}
    \!\sum\limits_{n=-\infty}^\infty
    \!\left(
	\sum\limits_{1 \le a \le N}
	\vert  {F}_{\mu\nu}^{a, n}\big\vert^2
	+ 2 \big\vert  m_n^{aa} \, {A}_\mu^{a,n}\big\vert^2
	+ \!\!
	\sum\limits_{1 \le a < b \le N}  \!\!
	    2\, \big\vert
		\partial_{[\mu} W_{\nu]}^{ab,n}
		\big\vert^2
	+ 4 \big\vert  m_n^{ab} \, W_\mu^{ab,n}  \big\vert^2
	\right),
\end{align}
with the KK masses
\begin{equation}
\label{wmass}
    m^{ab}_n \equiv \mW |a-b + n N|\,,
    \qquad \mW \equiv (NR)^{-1} \,.
\end{equation}
The cubic terms contain the  coupling of the Cartan photons to the
charge currents of the $W$-bosons,
\begin{align}
\label{eq:3dcubicL1}
    {\cal L}_3
    =
    \frac{ N L}{ 4 \lambda}
    \sum\limits_{m,n=-\infty}^\infty \>
    \sum\limits_{1 \le a < b \le N}
	2 i \partial_{[\mu} W_{\nu]}^{ab,n*}
	\big(A_{[\mu}^{a,n-m} - A_{[\mu}^{b, n-m}\big) \,
	W_{\nu]}^{ab,m}
	+ \mbox{(h.c.)}
    ,
\end{align}
as well as their  magnetic-moment coupling to the spin of the $W$-bosons,
\begin{align}
\label{eq:3dcubicL2}
    {\cal L}_3^{\prime}
    =
    \frac{ N L }{ 4 \lambda}
    \sum\limits_{m,n=-\infty}^\infty \>
    \sum\limits_{1 \le a < b \le N}
    2i \left( {F}_{\mu\nu}^{a,n-m} -  {F}_{\mu\nu}^{b,n-m} \right)
    W_{[\mu}^{ab,m} \, W_{\nu]}^{ab,n*}  \,.
\end{align}
Quartic terms in the Lagrangian, if needed, can be worked out similarly.

We shall eventually return to our Lagrangian of interest, ${\cal L}^{3D}$,
but first we discuss the construction of a non-relativistic effective
field theory (NR EFT) in
the simpler case of a single massive charged vector boson.
To this end, let $W_\mu$ denote a 3D complex vector field
with $U(1)$ gauge symmetry,
$W_\mu \rightarrow e^{i \alpha} W_\mu$,
and Lagrangian%
\footnote
    {%
    At this point, we revert to Minkowski space expressions
    using a $(-,+,+)$ metric signature.
    \label{metric}
    }
\begin{align}
\label{eq:Lmass1}
    L
    =&
    -\frac{1 }{ 4 e^2} \, \big(\partial_{[\mu} A_{\nu]}\big)^2
    -\tfrac{1 }{ 2}  \big\vert \partial_{[\mu}  W_{\nu]}\big\vert^2
    -  M^2 \, W_\mu W^{\mu \, *}
\nonumber \\ & {}
    + i \, A^\mu
	\big(
	    W^{\nu \,*}  \, \partial_{[\mu} W_{\nu]}
	    - W^\nu \, \partial_{[\mu} W_{\nu]}^*
	\big)
    - \tfrac{1 }{ 2} \, \big|A_{[\mu} W_{\nu]}\big|^2
    - \tfrac{i }{ 2} \,  \partial^{[\mu}  A^{\nu]} \,W_{[\mu} \, W_{\nu]}^*
    \,.
\end{align}
This charged vector boson Lagrangian
contains precisely the kinds of terms appearing in the Lagrangian
(\ref{eq:L3D})--(\ref{eq:3dcubicL2}) of our
full theory.
We use  $e^2$ to denote the coupling constant of the massless photon.
The leading-order correspondence with our full theory is
\begin{equation}
    e^2 =\frac{ \lambda }{ N L} = \frac{\lambda \mW }{ 2 \pi} \,.
\label{eq:ee2lam}
\end{equation}
Note that the vector field $W_\mu$ has a conventional normalization,
but we have chosen to scale the charge $e$ out of covariant derivatives
and define the photon field $A_\mu$ as having dimension 1.

A 3D massive vector field has two polarization states.
Define polarization vectors $e_\mu^i(\k)$, $i=1,2$, obeying
$e^i_\mu(\k) \, e^{j\, \mu}(\k) = \delta^{ij}$ and
$k^\mu \, e^{i}_ {\mu}(\k) = 0$, for on-shell momenta
$k_\mu \equiv (|\k|,\k)$.
Explicitly,
\begin{align}\label{eq:vectors}
    e^1_{ \mu}(\k) &\equiv
    \Big(0, \frac{\widetilde\k }{ |\k|}\Big) \,,
    \qquad
    e^2_{\mu}(\k) \equiv
    \Big(\frac{|\k| }{ M}, \frac{\k }{ |\k| } \frac{\omega_k }{ M}\Big) \,,
\end{align}
where
$\omega_k \equiv \sqrt{\k^2 + M^2}$ and
$ (\widetilde\k)_i \equiv \epsilon_{ij} (\k)_j$
(we use $i,j=1,2$ to denote spatial indices and take
$\epsilon_{12} = - \epsilon_{21}=1$).
The free mode expansion of the second quantized field is
\begin{align}
\label{eq:massive1}
    W_\mu(t, \x)
    &=
    \int \frac{d^2 k }{ (2\pi)^2 \, \sqrt{2\omega_k}} \>
    \sum_{i=1}^2
    \left[
	e^{i(\omega_k t - \k\cdot \x)} \, e_\mu^i(\k) \, a^{i}(\k)^\dagger
	+ e^{- i(\omega_k t - \k\cdot \x)} \, e_\mu^i(\k) \, b^{i}(\k)
    \right] ,
\nonumber \\ &{}
    \equiv W_\mu^+(t, \x) + W_\mu^-(t, \x) \,,
\end{align}
where
$
    [a^{i} (\k), a^{j}(\p)^\dagger ] =
    [b^{i} (\k), b^{j}(\p)^\dagger ] =
    (2 \pi)^2  \delta^{ij} \, \delta^{2} (\p{-}\k)
$,
and all other commutators vanish.
It is convenient to denote by $W^\pm_\mu$
the positive frequency $(\propto e^{i \omega_k t})$
and negative frequency $(\propto e^{- i \omega_k t})$ parts,
respectively.
The $U(1)$ charge operator
$
    Q \equiv \int d^2 x  \, 2\> {\rm Im}
    \big( W^{\nu \, *} \partial_0 W_\nu \big)
$,
after normal ordering, becomes
$
    Q = \int \frac{d^2 k }{ (2 \pi)^2}
    \sum_{i}
    \big[ a^{i}(\k)^\dagger a^i(\k) - b^{i}(\k)^\dagger b^i (\k) \big]
$,
from which it is evident that the operators
$a^{i}(\k)^\dagger$ ($a^{i}(\k)$) are creation (annihilation)
operators of positively charged vector bosons while the operators
$b^{i}(\k)^\dagger$ ($b^{i}(\k)$) create (annihilate)
negatively charged antiparticles.
Polarization index $i{=}1$ ($i{=}2$) refers to particles with
transverse (longitudinal) polarization, respectively.
The free Hamiltonian
$
    P_0= \int \frac{d^2 k }{ (2 \pi)^2} \>
    \omega_k \sum_{i}
    \big[ a^{i}(\k)^\dagger a^i (\k) + b^{i}(\k)^\dagger b^i (\k) \big]
$
and has eigenvalue $\omega_k$ for all four single-particle states
of a given spatial momentum $\k$.

Apart from explaining the physical content of the massive vector boson
theory, the mode expansion (\ref{eq:massive1}) provides an easy way
to see that an effective theory describing the dynamics of non-relativistic
vector bosons can be expressed solely in terms of the spatial components
$W_i$ of the vector field $W_\mu$.
For small momenta, $|\k| \ll M$,
the longitudinal polarization vector
$
    e^2_{ \mu}(\k)
    =
    \big(0, \frac{\k }{ |\k|}\big)
    + {\cal O}\big(\frac{|\k| }{ M}\big)
$,
with  only spatial components to leading order.
Since the  transverse polarization vector $e_\mu^1(\k)$ is purely spatial,
in the non-relativistic limit the time component $W_0$ can be eliminated,
leading to an effective theory for a spatial vector field.

One may construct the Lagrangian of this effective non-relativistic theory
by writing all terms consistent with the symmetries and matching the
coefficients to terms in the relativistic theory to the desired order
in the small coupling and small momentum expansion
(treating $\frac{|\bnabla|}{ M} \sim \frac{|\k|}{ M} \ll1$,
where $\bnabla$ is a spatial gradient.)
To carry out this procedure, we introduce two different two-component
complex fields, $\vec{\phi}_+(t, \x)$ and $\vec{\phi}_-(t, \x)$.
In a second-quantized non-relativistic theory,
these fields (and their Hermitian conjugates)
annihilate (or create) particles of charges $+1$ and $-1$,
respectively.
The two-component vector represents the direction in the
two-dimensional polarization space.
To leading order in the derivative and small-coupling expansion,
the fields $\vec\phi_\pm$ can be considered as scalars
under $SO(2)$ spatial rotations,
with an emergent $SO(2)$ ``flavor" symmetry acting as rotations
in the polarization space.
Magnetic moment interactions explicitly break this
$SO(2) \times SO(2)$ symmetry down to the diagonal $SO(2)$.
(This is completely analogous to the approximate spin rotation symmetry
in light atoms and molecules when spin-orbit interactions can be neglected.)

Temporarily ignoring the gauge field $A_\mu$,
to lowest non-trivial order in powers of $\frac{\bnabla}{ M}$,
the Lagrangian of the NR EFT is
\begin{align}
\label{eq:Lmass2}
    L_{\textrm{NR}}
    &=
    \vec\phi_+^{\; \dagger} \,i\partial_t \, \vec\phi_+
    + \vec\phi_-^{\; \dagger} \,i\partial_t \, \vec\phi_-
    - M\, |\vec\phi_+|^2- M\, |\vec\phi_-|^2
    - \frac{|\bnabla \vec\phi_{+}|^2}{2 m}
    - \frac{|\bnabla \vec\phi_{-}|^2}{2 m} \,.
\end{align}
and the corresponding Hamiltonian is
\begin{align}
\label{eq:Lmass3}
    H = \int d^2 x \>
    \vec{\phi}_+ (\x)^{\dagger} \cdot
	\big( M - \tfrac{\bnabla^2 }{2m} \big)
    \vec{\phi}_+(\x)
    +
    \vec{\phi}_- (\x)^{\dagger} \cdot
	\big( M - \tfrac{\bnabla^2 }{2m} \big)
    \vec{\phi}_-(\x) \,.
\end{align}
The conserved charge
$
    Q =
    \int d^2x \>
    (\vec{\phi}_+)^{\dagger} \cdot \vec{\phi}_+
    +
    (\vec{\phi}_-)^{\dagger} \cdot \vec{\phi}_-
$.
Mode expansions of the non-relativistic fields read
\begin{align}
    \phi_+^{i}(t, \x)^\dagger
    =
    \int \frac{d^2 k }{ (2 \pi)^2 } \>
    e^{i \varepsilon_k t - i \k \cdot  \x} \, a^{i}(\k)^\dagger \,,
    \quad
    \phi_-^{i}(t, \x)^\dagger
    =
    \int \frac{d^2 k }{ (2 \pi)^2 } \>
    e^{i \varepsilon_k t  - i \k \cdot \x} \, b^{i}(\k)^\dagger \,,
\label{eq:NRfieldops}
\end{align}
where
$
    \varepsilon_k \equiv M + \k^2/(2m)
$,
and $a^{i}(\k)^\dagger$ and $b^{i}(\k)^\dagger$
are the same creation operators appearing
in the relativistic expansion (\ref{eq:massive1})
(and its Hermitian conjugate).
The fields (\ref{eq:NRfieldops})
satisfy non-relativistic canonical commutation relations,
$
    [\phi^i_+(t, \x), \phi_+^{j}(t, \y)^\dagger] =
    [\phi^i_-(t, \x), \phi_-^{j}(t, \y)^\dagger] =
    \delta^{ij} \, \delta^{2}(\x{-}\y)
$,
with other commutators vanishing.

To fix parameters in the NR EFT
one demands that physical quantities, computed in the EFT
and in the IR limit of the full theory,
agree with each other order by order in the low energy and
weak coupling expansions.
At low orders, the matching is rather straightforward.
In the free theory (\ref{eq:Lmass2}),
single-particle states have energies
$\varepsilon_\k = M + \frac{\k^2 }{2 m}$ and charges $\pm 1$.
This agrees with the energy and charge of low momentum states
in the relativistic theory (\ref{eq:Lmass1}) provided
both the rest mass parameter $M$, and the kinetic mass $m$, appearing
in the non-relativistic theory (\ref{eq:Lmass2}) equal,
at lowest order, the physical mass $M$ of the original theory.

Note that if one ignores the explicit polarization vector dependence in
the relativistic expression (\ref{eq:massive1}),
then the operator $\phi_+^{i}(t,\x)^\dagger$ corresponds,
in the non-relativistic limit and after a trivial rescaling by $\sqrt{2 M}$,
to the positive-frequency part $W^+_i$ of $W_\mu$,
while $\phi_-^{i}(t,\x)^\dagger$ corresponds
to the positive frequency part $(W^-_i)^\dagger$ of the conjugate
field $W_\mu^\dagger$.

We now proceed to write down the NR EFT Lagrangian describing the theory
(\ref{eq:Lmass1})
to leading order in the small-$\lambda$ and derivative expansion.
We choose to work in Coulomb gauge for the photon field $A_\mu$.
The time component $A_0$ is not an independent field but is determined
by the charge distribution of the $W$-bosons via Gauss' law.
We denote the vector boson charge density by
\begin{equation}
\label{eq:wcharge}
    n(t, \x)
    =
    i W^{\nu \,*}  \, \partial_{[0} W_{\nu]}
    - i W^\nu \, \partial_{[0} W_{\nu]}^*
    \,,
\end{equation}
(neglecting higher order ``seagull'' contributions).
Varying the action, the Lagrangian (\ref{eq:Lmass1}) gives
$
    A_0 (t, \x) = e^2 \int d^2 y \, G(\x {-} \y) \, n(t, \y)
$,
where the two-dimensional Laplacian
Green's function $G$ was defined in Eq.~(\ref{eq:G}).
Using this result to eliminate $A_0$ from the action, one obtains
the Coulomb energy,
$
    V_C \equiv
    -\frac {e^2}{2} \int d^2 x \, d^2 y \,
    n(t,\x) \, G(\x{-}\y) \, n(t, \y)
$,
as a contribution to (minus) the Lagrangian.
Ignoring, for the moment, interactions
mediated by spatial components of the photons
as these are higher order in the non-relativistic limit,
our effective theory (\ref{eq:Lmass2}) changes,
to leading order, only by the inclusion of the
Coulomb energy in the action,
\begin{align}
\label{eq:Lmass5}
    S_{\textrm{NR}}
    =&\int dt \, d^2 x \>
    \bigg(
	\vec\phi_+^{\, \dagger} \,i\partial_t \, \vec\phi_+
	+ \vec\phi_-^{\, \dagger} \,i\partial_t \, \vec\phi_-
	- M\, |\vec\phi_+|^2
	- M\, |\vec\phi_-|^2
	- \frac{|\bnabla \vec\phi_{+}|^2}{2 M}
	- \frac{|\bnabla \vec\phi_{-}|^2}{2 M}
    \bigg)
\nonumber \\ &{}
    + \frac {e^2}{2} \int dt \, d^2 x \, d^2 y \>
	n(t, \x) \, G(\x{-} \y) \,n(t, \y) \,,
\end{align}
where
$
    n(t, \x)
    =
    \vec\phi_+ (t, \x)^\dagger \cdot \vec\phi_+ (t, \x)
    -\vec\phi_-(t, \x)^\dagger \cdot \vec\phi_- (t, \x)
$
is the non-relativistic limit of the vector boson charge density
(\ref{eq:wcharge}).
The corresponding Hamiltonian is just
\begin{align}
    H &=
    \int d^2 x \>
    \vec{\phi}_+ (\x)^{\dagger} \cdot
	\big( M - \tfrac{\bnabla^2 }{2M} \big)
    \vec{\phi}_+(\x)
    +
    \vec{\phi}_- (\x)^{\dagger} \cdot
	\big( M - \tfrac{\bnabla^2 }{2M} \big)
    \vec{\phi}_-(\x)
\nonumber \\ &\quad {}
    - \frac {e^2}{2} \int d^2 x \, d^2 y \>
	n(\x) \, G(\x{-} \y) \, n(\y) \,.
\label{eq:Hmass5}
\end{align}
The action (\ref{eq:Lmass5}) or Hamiltonian (\ref{eq:Hmass5})
include all leading-order terms in the non-relativistic ($v/c \ll 1$)
limit using the systematic power counting rules that we discuss next.
(In what follows,  $c\equiv 1$.)

Higher order terms which can appear in the NR EFT may be classified
and ordered using a suitable power counting scheme for the operators
and their matrix elements,
evaluated in characteristic bound states.%
\footnote
    {%
    These are determined by solving the two-particle
    Schr\"odinger equation which results from
    projecting the Hamiltonian (\ref{eq:Hmass5})
    into the two-particle Hilbert space.
    }
This approach is now well-established for 3+1 dimensional Coulombic
systems \cite{Lepage:1992tx}.
Compared to such systems,
several important differences arise in our 2+1D theory.
The first is that a particle of mass $M$ moving in a non-relativistic orbit
due to a central force $F \sim  e^2 /r$
moves at speed $v \sim e/\sqrt M$,
for any orbit radius,
rather than $v \sim e/\sqrt {Mr}$ as in a three dimensional
$F \sim e^2/r^2$ central force field.
The second is the appearance of $e^2 \ln (e^2/M)$ terms,
non-analytic in the coupling,
in the ground state energy
[as seen in Eq.~(\ref{eq:logE})],
owing to the scaling properties of the logarithmic potential.
Ignoring such logarithmic factors,
the appropriate power counting is similar to that detailed in
Ref.~\cite{Lepage:1992tx}:
the size of bound states is of order
$a_0 \sim (e^2 M)^{-1/2}$ and their characteristic binding energy
$\Delta E \sim e^2$.
For estimating the parametric dependence of matrix elements of
arbitrary operators that may arise in the NR EFT Hamiltonian,
evaluated in low-lying bound states of the lowest-order
theory (\ref{eq:Hmass5}),
we take%
\footnote
    {%
    These power counting rules for $\vec\phi_\pm$ follow, e.g.,
    by demanding that $\int d^2x \> \phi^\dagger \phi \sim 1$ in a
    bound state of size $a_0$.
    For the remaining assignments, the arguments are the same
    as given in Ref.~\cite{Lepage:1992tx};
    see also Ref.~\cite{Kinoshita:1995mt}.
    }
the fields $\vec\phi_\pm$ to scale as $\sqrt{e^2 M}$,
time derivatives $\partial_t \sim e^2$,
spatial derivatives $\nabla \sim \sqrt{e^2 M}$,
and Coulomb-gauge scalar and vector potentials
$e A_0 \sim e^2$ and
$e \mathbf{A} \sim e^4 / \sqrt{e^2 M}$.
Thus, the field strengths scale as
$ e \mathbf {E} \sim e^2 \sqrt{e^2 M}$
and
$e B \sim e^4$.
(Here, and below, we have rescaled the Maxwell action for the photon by $e^2$,
to give the gauge field a conventional perturbative normalization.)
Using these parametric estimates,
it follows that all terms in the lowest-order NR Hamiltonian (\ref{eq:Hmass5}),
excepting the rest-energy terms, are of order $e^2$, as required.

To assess the relative importance of higher order terms,
we begin with  the magnetic moment coupling of the
vector bosons,
the last term in the NR Lagrangian (\ref{eq:Lmass1}).
Writing the leading terms consistent with the
symmetries of the theory which couple the field strength tensor
$F_{ij}$ to the NR vector fields $\phi^i$ and $\phi^{i \, \dagger}$,
one finds, to leading order in  $1/M$, that there is a
unique such term,
\begin{align}
\label{eq:Lmagn1}
    L_{\rm NR}^{\rm mag}
    = -
    \frac{i }{ 2 M} \, e F_{ij} \,
    \big(
	\phi^{i \, \dagger}_+ \phi^j_+
	- \phi^{i \, \dagger}_- \phi^j_-
    \big) ,
\end{align}
whose coefficient follows by matching to the relativistic form
(\ref{eq:Lmass1}) using relations
(\ref{eq:vectors}), (\ref{eq:massive1}), and (\ref{eq:NRfieldops}).
The above power counting rules show that
magnetic moment interactions will
shift bound state energy levels by an amount of order $e^4/M$,
or a relative $\O(e^2/M)$ correction to binding energies.

Given our original choice (\ref{eq:vectors}) of polarization vectors,
the NR fields $\phi_\pm^i$, $i=1,2$ annihilate vector bosons which are
linearly polarized, either transverse or parallel to their momenta,
respectively.
However, using operators that create particles in eigenstates of $S_z$, the spin of the vector boson field  $S_z = \int d^2 x (\epsilon_{ij} \dot{W}_i W^*_j + {\rm h.c.})$,
is typically more convenient when
discussing bound states in a central potential.
Such operators can be obtained by redefining our NR field
operators as follows:
\begin{equation}
\begin{split}
\label{eq:Lzoperators}
    \phi_\pm^{1}
    =
    \tfrac{1 }{ \sqrt{2}}
    \big(
	\phi_\pm^{\uparrow} + \phi_\pm^{\downarrow}
    \big) \,,
\qquad
\qquad
    \phi_\pm^{1\; \dagger}
    =
    \tfrac{1 }{ \sqrt{2}}
    \big(
	\phi_\pm^{\uparrow \; \dagger  } + \phi_\pm^{\downarrow \; \dagger}
    \big) \,,
\\
 \phi_\pm^{2}
   =
    \tfrac{1 }{\sqrt{2} i}
     \big(
	 \phi_\pm^{\uparrow} - \phi_\pm^{\downarrow}
     \big) \,,
\qquad
\qquad
    \phi_\pm^{2 \; \dagger}
    =
    \tfrac{i }{ \sqrt{2} }
    \big(
	\phi_\pm^{\uparrow \; \dagger } - \phi_\pm^{\downarrow \; \dagger }
    \big) \,.
\end{split}
\end{equation}
The new operators
$\phi_\pm^{\uparrow \, \dagger}$ and $\phi_\pm^{\uparrow}$
obey canonical commutation relations and create or destroy vector bosons
with $S_z = 1$,
similarly,
$\phi_\pm^{\downarrow \, \dagger}$ and $\phi_\pm^{\downarrow}$
create or destroy $S_z = -1$ states.
Using these redefined fields,
the magnetic
moment coupling (\ref{eq:Lmagn1}) becomes
\begin{align}
\label{eq:Lmagn2}
  L_{\rm NR}^{\rm mag}    = -
    \frac{ e  B }{ 2 M} \,
    \big(
	\phi_+^{\uparrow\;  \dagger} \phi_+^{\uparrow}
	- \phi_+^{\downarrow \; \dagger} \phi_+^{\downarrow}- \phi_-^{\uparrow\;  \dagger} \phi_-^{\uparrow}
	+ \phi_-^{\downarrow \; \dagger} \phi_-^{\downarrow}
    \big) \,,
\end{align}
showing that the magnetic moment couplings  split, at order
$e^4/M$, the level degeneracy of $\uparrow \uparrow$ and
$\downarrow \downarrow$ bound states.
In particular, the magnetic
moment interaction term (\ref{eq:Lmagn2}) leads to the spin-spin
hyperfine interaction potential (local in 2D), discussed in
Sec.~\ref{sec:glueballs}.

Coefficients of further operators in the EFT can be found by matching scattering
amplitudes between the full and effective theories, as done in
continuum NRQED in Ref.~\cite{Kinoshita:1995mt}.
(See Ref.~\cite{Lepage:1992tx} for matching in NRQCD using lattice
gauge theory.)
The resulting terms are dimensionally reduced versions of
ones listed in the above references and include,
for example,
$
    \frac{e}{M^2}
    \big[
    C_1 \,
    \mathbf\nabla \cdot \mathbf E \; \phi^{j \; \dagger}_+ \phi^j_+
    +
    C_2 \,
    ( \partial_i E_j - \frac{1}{2} \delta_{ij}\mathbf\nabla \cdot\mathbf{E}) \,
    \phi^{i \; \dagger}_+ \phi^j_+
    + \cdots
    \big]
$,
whose coefficients can
be found by matching scattering amplitudes in external static
electric fields.
The contribution of these operators to the bound state
energies also scale as $e^{4}/M$.
Additionally,
there are a number of possible contact terms involving four non-relativistic
fields, schematically of the form
$e^2 (\phi^\dagger \phi)^2$,
that also contribute $\O(e^4/M)$ energy shifts.
There are, of course, also corrections arising from
higher orders in the expansion of the relativistic dispersion relation
of the form
$\phi_+^{i \; \dagger} \frac{\nabla^4 }{ M^3} \phi^i_+ + \cdots$.
According to the power counting rules,
these also contribute to bound state energies
at order $e^4/M$.
We have not systematically enumerated all possible higher order
terms in the NR EFT and leave their detailed study and matching for
future work.

To conclude this Appendix, we invite the reader to consider the
transition from the non-relativistic effective theory (\ref{eq:Lmass5})
for our toy single vector boson model (\ref{eq:Lmass1}),
to the effective theory (\ref{eq:Sheavy}) describing our full theory
(\ref{eq:L3D})--(\ref{eq:3dcubicL2}).
The transition from the toy NR EFT (\ref{eq:Lmass5}) to
our full EFT (\ref{eq:Sheavy}) is largely
one of bookkeeping due to the proliferation of fields in the full
theory.
In brief,
in the NR EFT (\ref{eq:Sheavy}), the fields
$\vec\phi^{ab}$ with $a>b$ correspond to the field $\vec\phi_+$
of the toy model,
while the fields $\vec\phi^{ab}$ with $a<b$ correspond to the
$\vec\phi_-$ field of the single complex vector model.
The charge densities (\ref{eq:na})
are the multi-field generalizations of toy model
charge density (\ref{eq:wcharge}).
The Hamiltonian (\ref{eq:Hmass5}) is easily seen
to give rise to the complete form (\ref{eq:HNR})
(with the same normal ordering issues discussed in Sec. \ref{sec:NR_EFT}).

\section{Light sector details}
\label{app:lightsector}

We start with the quadratic part (\ref{eq:3dfreeL}) of the 4D action
to remind the reader about the 3D photon-scalar duality and the
normalization of the dual photon field used in dual
description (\ref{eq:Slight}).
The Cartan generators in the fundamental representation have eigenvalues
given by the $N$ weight vectors $\bnu_A$, $A=1,...,N$.
In our basis and choice of normalization,
the highest weight is
$\bnu_1 =(1{-}\frac 1{N}, -\frac 1{N}, ...,-\frac 1{N})$
and coincides with the first fundamental weight vector $\bmu_1$ of
$su(N)$.
Consider a static quark, or fundamental representation probe charge,
placed at the origin of $\mathbb R^2$ and having
some weight vector $\bnu$ characterizing its color charge.
This adds a source to the 3D (Minkowskian, c.f. footnote \ref{metric})
Lagrangian for the static (KK index $n=0$) Cartan components
of the gauge field,
$
    - \frac{NL}{4 \lambda} F_{\mu\nu}^a F^{\mu\nu \; a}
    + A_0^a (\x) \, \nu^a \, \delta^{2}(\x)
$
(where a sum on $a$ is implied, and
$\nu^a$ is the $a$-th component of the quark's weight).

The resulting $A_0^a$ equation of  motion,
$
    \frac{NL}{\lambda} \, \nabla^2 A_0^a(\x) = \nu^a \, \delta^{2}(\x)
$,
implies Gauss' law,
$
    \oint_C dl \, \hat{n}^i \big( \frac{NL}{ \lambda} F_{i0}^a \big)
    = \nu^a
$,
where the curve $C$ encircles the origin (counterclockwise)
and $\hat{n}$ is its outward normal.
An $N$-component dual photon field $\bsigma$ may be introduced via
the relation
$
    \frac{NL}{ \lambda} \, F_{i0}^a =
    \frac{1}{2\pi} \, \epsilon_{ij} \, \partial_j  \sigma^a
$
(with $\epsilon_{12} \equiv 1$).
The choice
of coefficient ensures that
$
    \oint_C dl \, \hat{n}^i \, \epsilon_{ij} \,
    \partial_j \, \sigma^a = 2\pi \nu^a
$,
i.e., the monodromy of the dual
photon field is  $2\pi$ times the charge (the weight vector $\bnu$).
To be consistent with probes in all fundamental representations,
the dual photon field is defined to be periodic with a periodicity
of $2\pi$ times the $su(N)$ weight lattice,
generated by the fundamental weights $\{ 2 \pi \bmu_A \}$.
The $2{+}1$D Lorentz invariant form of the above
duality relation is
$
    F_{\mu\nu}^a = \frac{\lambda}{ 2\pi NL} \,
    \epsilon_{\mu\nu\lambda} \, \partial^\lambda \sigma^a
    = \frac{\lambda \mW}{ 4 \pi^2} \,
    \epsilon_{\mu\nu\lambda} \partial^\lambda \sigma^a
$
(with $\epsilon_{0ij}\equiv -\epsilon_{ij}$).
To implement the duality, we replace the Maxwell part of the
quadratic action (\ref{eq:3dfreeL}) by
$
    - \frac{NL }{ 4 \lambda} \,
    F_{\mu\nu}^a F^{\mu\nu \, a}
    + \frac{1 }{ 4 \pi} \, \epsilon_{\mu\nu\lambda} \,
    F^{\mu\nu \, a} \, \partial^\lambda \sigma^a
$.
Treating $\sigma^a$ and $F_{\mu\nu}^a$
as independent integration variables and integrating out the
field strength $F_{\mu\nu}^a$,
the resulting kinetic term for the dual photon is
$
    \frac{\lambda }{ 8 \pi^2 NL } \,
    (\partial_\lambda \sigma^a)^2
    =
    \frac{\lambda \mW }{ 16 \pi^3} \,
    (\partial_\lambda \sigma^a)^2
$, as shown in the light sector action (\ref{eq:Slight}).

The Coulomb energy $V_C$ of two static charges with weights
$\blambda_1$ and $\blambda_2$, separated by a distance $r$,
can also be obtained from the above expressions.
One finds
$
    V_C
    =
    -  \frac{\lambda \mW }{ 4 \pi^2} \,(\blambda_1 \cdot \blambda_2) \, \ln r
$.
The weights for $W$-bosons are root vectors, and since roots have length
two, the interaction energy of oppositely charged static $W$-bosons
is  $\frac{\lambda \mW }{ 2 \pi^2}  \log r$,
as shown in Eqs.~(\ref{eq:pair}) and (\ref{eq:kappaW}).
For a fundamental quark and an antiquark of opposite weights, we have
$
    - \blambda_1 \cdot \blambda_2
    = \bnu \cdot \bnu = 1 - \frac{1}{ N}
$,
hence they experience attraction of that strength, as shown in
(\ref{eq:kappaq}).
On the other hand, a quark with weight $\blambda_1 =\bnu$
and antiquark with weight $\blambda_2 = - \bnu'$,
with $\bnu\ne \bnu'$, experience repulsion since
$
    - \blambda_1 \cdot \blambda_2
    = \bnu \cdot \bnu'
    =  -\frac{1}{ N}
$,
as shown in Fig.~\ref{fig:mesons}.
Likewise, it follows that quarks (or antiquarks) of different
weights attract with strength $\frac {1}{ N}$,
as shown in Fig.~\ref{fig:baryons}.

Finally,
a magnetic monopole-instanton of magnetic charge $\alpha$
(one of the affine roots), is represented in the dual description
by insertions of $e^{i \balpha \cdot \bsigma(x)}$
($x \in \mathbb R^3$).
Hence the interaction action between two monopole-instantons  of charges
$\balpha_1$ and $\balpha_2$ can be obtained as
$
    \langle
	e^{i \balpha_1 \cdot \bsigma(x_1)}
	e^{i \balpha_2 \cdot \bsigma(x_2)}
    \rangle =
    \exp \big[
	{- \frac{2\pi^2}{\lambda \mW} \,
	    \frac{{\balpha_1 \cdot \balpha_2} }{|x_1 {-} x_2|}}
    \big]
$,
where the expectation value was calculated with
the free field portion of the $\bsigma$-field Lagrangian (\ref{eq:Slight}).
A remark relevant for the thermal case is that,
when reduced to two dimensions,
the corresponding correlator becomes
$
    e^{ \frac{4 \pi^2 T }{ \lambda \mW} \,
	{{\balpha_1 \cdot \balpha_2} \, \ln (|x_1 {-} x_2| T)}}
$
for $|x_1 {-} x_2| \gg 1/T$.

\section {Symmetry transformations}
\label{app:symms}

Let us choose to work in $A_3 = 0$ gauge,
where the holonomy $\Omega$ is an independent degree of freedom.
Regarding $A_\mu(x)$ as anti-Hermitian, and viewing the quark field $q$ as an $N \times \nf$ matrix of spinors, we will define
$\Omega_F = \mathrm{diag}(\xi^{1/2},\xi^{3/2},{\cdots},\xi^{N-1/2})$.
Our boundary conditions (in both index-free and component forms) are
\begin{subequations}
\begin{align}
    A_\mu(x_3{+}L) &= \Omega \, A_\mu(x_3) \, \Omega^\dagger \,,
&
    A_\mu(x_3{+}L)^{ab} &\simeq \omega^{a-b} \, A_\mu(x_3)^{ab} \,,
\label{eq:BCA}
\\
    q(x_3{+}L) &= \Omega \, q(x_3) \, \Omega_F^\dagger \,,
&
    q(x_3{+}L)^{aA} &\simeq
    \omega^{-\frac N2 +(a-\frac 12)} \,
    \xi^{-(A-\frac 12)} \,
    q(x_3)^{aA} \,,
\label{eq:BCq}
\end{align}
\end{subequations}
where $\simeq$ means when $\Omega$ has the form (\ref{eq:Omega}).

\subsection*{Mode expansions}

Suppose that $\Omega$ has the form (\ref{eq:Omega}) with negligible fluctuations, let $\y \equiv (y_1,y_2)$ denote the non-compact spatial coordinates, and ignore interactions.
Then:
\begin{align}
    A_\mu(t,\y,x_3)^{ab}
    &=
    \frac 1L \sum_{n\in\mathbf Z}
    \int \frac{d^2p}{(2\pi)^2 \sqrt{2 \omega}}
    \Bigl[
    e^{-i (\omega t - \p \cdot \y - k_n^{ab} x_3)}
    e^i_\mu(\vec p) \,
    \phi_i(\p)_n^{ab}
\nonumber\\ & \hspace*{1.45in} {}
    -
    e^{i (\omega t + \p \cdot \y + k_n^{ab} x_3)} \,
    e^i_\mu(-\vec p)^* \,
    \big(\phi_i(-\p)_{-n}^{ba} \big)^\dagger
    \Bigl] \,,
\end{align}
where
$\mu = 0,1,2$,
the compact momentum
$
    k_n^{ab} \equiv \frac{2\pi}{NL} (a-b+nN)
$,
the 3D spatial momentum
$
    \vec p \equiv (p_1,p_2,k_n^{ab})
$,
and the frequency
$
    \omega \equiv (\p^2 + (k_n^{ab})^2)^{1/2}
$
(with dependence on $\p$, $n$, $a$ and $b$ implicit).
The polarization vectors $\{ e_\mu^i(\vec p)\}$, $i = 1,2$, satisfy
$2{+}1$D transversality,
$p^\mu e_\mu^i = 0$, with $p^0 \equiv \omega$.
This expansion satisfies BCs (\ref{eq:BCA}),
anti-Hermiticity and transversality of $A_\mu$,
and the 4D free wave equation $\square A_\mu = 0$.

The corresponding mode expansion for the quarks is
\begin{align}
    q(t,\y,x_3)^{aA}
    &=
    \frac 1L \sum_{n\in\mathbf Z+\frac 12}
    \int \frac{d^2p}{(2\pi)^2 \sqrt{2 \omega}} \,
    \Bigl[
	e^{-i (\omega t - \p \cdot \y - k_n^{aA} x_3)} \,
	u_s(\vec p)  \, \psi_s(\p)_n^{aA}
\nonumber\\ & \hspace*{1.65in} {}
	+
	e^{i (\omega t + \p \cdot \y + k_n^{aA} x_3)} \,
	u_{-s}(-\vec p) \, \big(\chi_s(-\p)_{n}^{aA} \big)^\dagger
    \Bigl] \,,
\end{align}
where the quark compact momentum is
$
    k_n^{aA}
    \equiv
    \frac{2\pi}{NL}
    [(a{-}\frac 12)-\frac {N}{\nf}(A{-}\frac 12)+ n N]
$,
the 3D spatial momentum
$\vec p \equiv (p_1,p_2,k_n^{aA})$,
and the frequency
$
    \omega \equiv (\p^2 + (k_n^{aA})^2)^{1/2}
$.
The free particle spinors $u_s(\vec p)$ have
helicity $s=\pm 1$ and satisfy
$\gamma_\alpha p^\alpha u_s(\vec p) = 0$
with $p^\alpha \equiv (\omega,\vec p)$.
In a chiral basis,
$
    \gamma_0 \equiv
    \left(\begin{smallmatrix} \phantom- 0 & 1 \\ -1 & 0 \end{smallmatrix}
    \right)
$,
$
    \gamma_i \equiv
    \left(\begin{smallmatrix} 0 & \sigma_i \\ \sigma_i & 0 \end{smallmatrix}
    \right)
$,
$
    \gamma_5 \equiv -i \gamma_0\gamma_1\gamma_2\gamma_3 =
    \left(\begin{smallmatrix} 1 & \phantom- 0 \\ 0 & -1 \end{smallmatrix}
    \right)
$,
one has
$
    u_+(\vec p) =
    \left(\begin{smallmatrix} \xi_+(\hat p) \\ 0\end{smallmatrix}\right)
$
and
$
    u_-(\vec p) =
    \left(\begin{smallmatrix} 0 \\ \xi_-(\hat p) \end{smallmatrix}\right)
$,
where $\xi_\pm(\hat p)$ are two-component spinors satisfying
$\hat p \cdot \vec\sigma \> \xi_\pm(\hat p) = \pm \xi_\pm(\hat p)$
with phase convention
$\xi_\pm(\hat p)^* = \pm i \sigma_2 \, \xi_\mp(\hat p)$.
The free particle spinors satisfy
$\gamma_5 \, u_s(\vec p) = s \, u_s(\vec p)$ and
$u_s(\vec p)^* = C u_{-s}(\vec p)$ with
$C \equiv i \gamma_5 \gamma_2$ and
$C^\dagger \gamma^\alpha C = (\gamma^\alpha)^*$.
The above mode expansion satisfies the boundary conditions (\ref{eq:BCq})
and the massless Dirac equation $\gamma^\alpha \partial_\alpha q = 0$.

The coordinate space EFT operators are just 2D spatial Fourier transforms of
the momentum-space mode operators,
\begin{subequations}
\begin{align}
    \vec\phi(\y)_n^{ab}
    &\equiv
    \int \frac{d^2p}{(2\pi)^2} \> e^{i \p \cdot \y} \,
    \vec\phi(\p)_n^{ab} \,,
\\
    \psi_\pm(\y)_n^{aA}
    &\equiv
    \int \frac{d^2p}{(2\pi)^2} \> e^{i \p \cdot \y} \,
    \psi_\pm(\p)_n^{aA} \,,
\\
    \chi_\pm(\y)_n^{aA}
    &\equiv
    \int \frac{d^2p}{(2\pi)^2} \> e^{i \p \cdot \y} \,
    \chi_\pm(\p)_n^{aA} \,.
\end{align}
\end{subequations}

\subsection*{Axial $U(1)_A^{\nf}$}

Let $\bm\theta = \mathrm{diag}(\theta_1,{\cdots},\theta_{\nf})$.
The axial transformation is standard:
\begin{align}
    q(x) &\to e^{i\gamma_5 \bm\theta} q(x) \,,
&
    q(x)^{aA} &\to e^{i\gamma_5 \theta_A} q(x)^{aA} \,,
\label{eq:axialtrans}
\end{align}
with $\gamma_5 \equiv (\gamma_5)^\dagger$.
Non-invariance under the diagonal $U(1)_A$ only appears in the
non-perturbative light sector.
This transformation is produced by
\begin{align}
    \psi_\pm (\p)_n^{aA} &\to
    e^{\pm i \theta_A} \, \psi_\pm (\p)_n^{aA} \,,
&
    \chi_\pm (\p)_n^{aA} &\to
    e^{\pm i \theta_A} \, \chi_\pm (\p)_n^{aA} \,.
\end{align}
Building two-component operators,
$
    \psi(\p)_n^{aA} \equiv
    \left(\begin{smallmatrix}
	\psi_+(\p)_n^{aA} \\ \psi_-(\p)_n^{aA}
    \end{smallmatrix}\right)
$
and
$
    \chi(\p)_n^{aA} \equiv
    \left(\begin{smallmatrix}
	\chi_+(\p)_n^{aA} \\ \chi_-(\p)_n^{aA}
    \end{smallmatrix}\right)
$,
this transformation is equivalent to
\begin{align}
    \psi (\p)_n^{aA} &\to
    e^{i \theta_A \sigma_3} \, \psi (\p)_n^{aA} \,,
&
    \chi (\p)_n^{aA} &\to
    e^{i \theta_A \sigma_3} \, \chi (\p)_n^{aA} \,.
\end{align}

\subsection*{Charge conjugation}

Recall that $N$ is assumed odd.
Combine the basic charge conjugation transformation,
$A_\mu \to A_\mu^*$,
with global color and flavor permutations $V$ and $V_F$, respectively,
chosen to preserve the form (\ref{eq:Omega}) of $\Omega$
at the $\mathbb Z_N$ symmetric minimum
and the quark boundary conditions,
\begin{align}
    V &\equiv \| \delta_{a+b,N+1} \| = \| \delta_{\bar a,b} \| \,,
&
    V_F &\equiv \| \delta_{A+B,\nf+1} \| = \| \delta_{\bar A,B} \| \,,
\end{align}
where $\bar a \equiv N+1-a$, $\bar A \equiv \nf+1-A$.
Note that
$
    \Omega^* = V \, \Omega \, V^\dagger
$
and
$
    \Omega_F^* = V_F \, \Omega_F \, V_F^\dagger
$.
The action of charge conjugation is
\begin{subequations}\label{eq:chargeconj}%
\begin{align}
    \Omega &\to V \Omega^* V^\dagger
    \simeq \Omega \,,
\\
    A_\mu(x) &\to V A_\mu(x)^* V^\dagger \,,
&
    A_\mu(x)^{ab} &\to
    (A_\mu(x)^{\bar a,\bar b})^*
    =
    -A_\mu(x)^{\bar b,\bar a}
    \,,
\\
    q(x) &\to C (V q(x)^* V_F^\dagger) \,,
&
    q(x)^{aA} &\to C (q(x)^{\bar a \bar A})^* \,,
\end{align}
\end{subequations}
This transformation is produced by
\begin{align}
    \vec\phi(\p)_n^{a\, b} &\to -\vec\phi(\p)_n^{\bar{b} \, \bar{a}} \,,
&
    \psi_s(\p)_n^{aA} &\to \chi_{s}(\p)_{-n}^{\bar a \bar A} \,,
&
    \chi_s(\p)_n^{aA} &\to \psi_{s}(\p)_{-n}^{\bar a \bar A} \,.
\end{align}

\subsection*{$x_3$ reflection}

Let $y \equiv (x^0,x^1,x^2,L{-}x^3)$ denote the reflected coordinates.
Combine the basic reflection,
$A_\mu (x) \to A_\mu(y)$ (recall $A_3 \equiv 0$), with
the global color and flavor permutations $V$ and $V_F$
defined above.
Then the action of $x_3$ reflection is
\begin{subequations}\label{eq:reflection}
\begin{align}
    \Omega &\to V \Omega^\dagger V^\dagger
    \simeq \Omega \,,
\\
    A_\mu(x) &\to V A_\mu(y) V^\dagger \,,
&
    A_\mu(x)^{ab} &\to A_\mu(y)^{\bar a \bar b} \,,
\\
    q(x) &\to R_3 (V q(y) V_F^\dagger) \,,
&
    q(x)^{aA} &\to R_3 \, q(y)^{\bar a \bar A} \,,
\end{align}
\end{subequations}
where $R_3$ satisfies
$R_3^\dagger \gamma^\alpha R_3 = (1-2\delta^\alpha_3) \gamma^\alpha$ and
in our chiral basis $R_3 = \gamma_5\gamma_3$.
The free particle spinors satisfy $R_3 u_s(\vec p\,') = s u_{-s}(\vec p\,)$
where $\vec p\,'\equiv (p_1,p_2,-p_3)$.
This transformation is produced by
\begin{align}
    \vec\phi(\p)_n^{ab} &\to \vec\phi(\p)_{-n}^{\bar a \bar b} \,,
&
    \psi_s(\p)_n^{aA} &\to -s \, \psi_{-s}(\p)_{-n}^{\bar a \bar A} \,,
&
    \chi_s(\p)_n^{aA} &\to s \, \chi_{-s}(\p)_{-n}^{\bar a \bar A} \,.
\end{align}

\subsection*{$\mathbb Z_N$ center}

Assume here that either $\nf=0$, or $\nf = N$.
Combine the basic $\mathbb Z_N$ center transformation,
$\Omega \to \omega \, \Omega$,
with global color and flavor permutations $P$ and $P_F$
chosen to preserve the form (\ref{eq:Omega}) of $\Omega$
and the quark boundary condition,
\begin{align}
    P &\equiv \| \delta_{a,b-1} \| \,,\qquad
&
    P_F &\equiv \| \delta_{A,B-1} \| \,,
\end{align}
with color and flavor indices
regarded as defined modulo $N$.
Note that
$
    P^\dagger \Omega P
    =
    \omega \, \Omega
$
when $\Omega$ has the form (\ref{eq:Omega}),
and similarly
$
    P_F^\dagger \Omega_F P_F
    =
    \omega \, \Omega_F
$.
The action of a $\mathbb Z_N$ center transformation is
\begin{subequations}\label{eq:centertransA}
\begin{align}
    \Omega &\to \omega \, P \Omega P^\dagger
    \simeq \Omega \,,
\\
    A_\mu(x) &\to P A_\mu(x) P^\dagger \,,
&
    A_\mu(x)^{ab} &\to A_\mu(x)^{a-1,b-1} \,,
\\
    q(x) &\to P q(x) P_F^\dagger \,,
&
    q(x)^{aA} &\to q(x)^{a-1,A-1}
\end{align}
\end{subequations}
This transformation is produced by
\begin{align}
    \vec\phi_n^{\,ab}(\p) &\to \vec\phi_{n-\delta_a+\delta_b}^{\, a-1,b-1}(\p) \,,
&
    \psi_n^{aA}(\p) &\to \psi_{n-\delta_a+\delta_A}^{a-1, A-1}(\p) \,,
&
    \chi_n^{aA}(\p) &\to \chi_{n-\delta_a+\delta_A}^{a-1, A-1}(\p) \,,
\label{eq:centertransB}
\end{align}
where $\delta_a = 1$ if $a = 1$, otherwise 0.
This is equivalent to relations (\ref{eq:ZNa}) and (\ref{eq:ZNb}).

\newpage

\bibliographystyle{JHEP}
\bibliography{small_circle}

\end{document}